\documentclass[twocolumn]{aastex62}
\shorttitle{SGAS-HST}
\shortauthors{Sharon et al.}

\usepackage{amsmath}
\usepackage{natbib}
\usepackage{color}
\usepackage{url}

\definecolor{LightCyan}{rgb}{0.88,1,1}

\newcommand{\lenstool}{{\tt{Lenstool}}}

\newcommand{\galfit}{\texttt{GALFIT}}
\newcommand{\megasaura}{M\textsc{eg}a\textsc{S}a\textsc{ura}}
\newcommand{\megasauralong}{The Magellan Evolution of Galaxies Spectroscopic and Ultraviolet Reference Atlas}

\newcommand{\hst}{{\it HST}}
\newcommand{\HST}{{\it HST}}
\newcommand{\HSTlong}{{\it Hubble Space Telescope}}

\newcommand{\Swift}{{\it Neil Gehrels Swift Observatory}}
\newcommand{\spitzer}{{\it Spitzer}}
\newcommand{\Spitzer}{{\it Spitzer}}
\newcommand{\Spitzerlong}{{\it Spitzer Space Telescope}}
\newcommand{\chandra}{{\it Chandra}}

\newcommand{\zspec}{z_{spec}}
\newcommand{\zphot}{z_{phot}}

\newcommand{\Lya}{Ly$\alpha$}
\newcommand{\lya}{Ly$\alpha$}
\newcommand{\OII}{[O II]~3727, 3729~\AA}
\newcommand{\OIII}{[O III]~4959, 5007~\AA}
\newcommand{\ciiidoublet}{[C~III] 1907, C~III] 1909~\AA}
\newcommand{\HeI}{\hbox{{\rm He}\kern 0.1em{\sc i}}}
\newcommand{\kms}{km s$^{-1}$}
\newcommand{\msun}{M$_{\odot}$}
\newcommand{\multislit}{multi-slit}

\newcommand{\multiobject}{multi-object}

\newcommand{\as}{$^{\prime\prime}$ }

\begin{document}

\title{Strong Lens Models for 37 Clusters of Galaxies from the SDSS Giant Arcs Survey\footnote{Based on observations made with the NASA/ESA 
  {\it Hubble Space Telescope}, obtained at the Space Telescope Science
  Institute, which is operated by the Association of Universities for
  Research in Astronomy, Inc., under NASA contract NAS 5-26555. These
  observations are associated with programs GO-13003, GO-14622, GO-14230, GO-14896, GO-11974, GO-11100}}
\email{kerens@umich.edu} 
  
\author{Keren Sharon}
\affiliation{Department of Astronomy, University of Michigan, 1085 S. University Ave, Ann Arbor, MI 48109, USA}

\author{Matthew B. Bayliss}
\affiliation{Kavli Institute for Astrophysics \& Space Research, Massachusetts Institute of Technology, 77 Massachusetts Ave., Cambridge, MA 02139, USA}

\author{H{\aa}kon Dahle}
\affiliation{Institute of Theoretical Astrophysics, University of  Oslo,  P. O. Box 1029, Blindern, N-0315 Oslo, Norway }

\author{Samuel J. Dunham}
\affiliation{Department of Astronomy, Vanderbilt University, 6301 Stevenson Center Lane, Nashville, TN 37212, USA}
\affiliation{Department of Physics and Astronomy, University of Tennessee, Knoxville, TN 37996, USA}

\author{Michael K. Florian}
\affiliation{Observational Cosmology Lab Code 665, NASA Goddard Space Flight Center, Greenbelt, MD 20771, USA}

\author{Michael D. Gladders}
\affiliation{Department of Astronomy and Astrophysics, University of Chicago, 5640 South Ellis Avenue, Chicago, IL 60637, USA}
\affiliation{Kavli Institute for Cosmological Physics, University of Chicago, 5640 South Ellis Avenue, Chicago, IL 60637, USA}

\author{Traci L. Johnson}
\affiliation{Department of Astronomy, University of Michigan, 1085 S. University Ave, Ann Arbor, MI 48109, USA}

\author{Guillaume Mahler}
\affiliation{Department of Astronomy, University of Michigan, 1085 S. University Ave, Ann Arbor, MI 48109, USA}

\author{Rachel Paterno-Mahler}
\affiliation{University of California, Irvine, 4129 Frederick Reines Hall, Irvine, CA 92697, USA}

\author{Jane R. Rigby}
\affiliation{Astrophysics Science Division, Goddard Space Flight Center, 8800 Greenbelt Rd., Greenbelt, MD 20771, USA}

\author{Katherine E. Whitaker}
\affiliation{Department of Physics, University of Connecticut, Storrs, CT 06269, USA}
\affil{Cosmic Dawn Center at the Niels Bohr Institute, University of Copenhagen and DTU-Space, Technical University of Denmark}

\author{Mohammad Akhshik}
\affiliation{Department of Physics, University of Connecticut, Storrs, CT 06269, USA}

\author{Benjamin P. Koester}
\affiliation{Department of Physics, University of Michigan, 540 Church St, Ann Arbor, MI 48109, USA}

\author{Katherine Murray}
\affiliation{Space Telescope Science Institute, Baltimore, MD, USA}

\author{Eva Wuyts}
\affiliation{Armen TeKort Antwerp, Belgium}

\begin{abstract}
We present strong gravitational lensing models for 37 galaxy clusters from the SDSS
Giant Arcs Survey. We combine data from multi-band \textit{Hubble Space
Telescope} WFC3 imaging, with ground-based imaging and spectroscopy from Magellan, Gemini, APO, and MMT, 
in order to detect and spectroscopically confirm new multiply-lensed background sources behind the
clusters. We report spectroscopic or photometric redshifts of sources
in these fields, including cluster galaxies and background sources. 
Based on all available lensing evidence, we construct and present strong
lensing mass models for these galaxy clusters. 
\end{abstract}

\keywords{Gravitational lensing: strong --- Galaxies: clusters: individual: 
SDSS~J0004$-$0103, SDSS~J0108$+$0624,  SDSS~J0146$-$0929,  SDSS~J0150$+$2725,  SDSS~J0333$-$0651,  SDSS~J0851$+$3331,  SDSS~J0915$+$3826,  SDSS~J0928$+$2031,  SDSS~J0952$+$3434,  SDSS~J0957$+$0509,  SDSS~J1002$+$2031,  SDSS~J1038$+$4849,  SDSS~J1050$+$0017,  SDSS~J1055$+$5547, SDSS~J1110$+$6459, SDSS~J1115$+$1645,  SDSS~J1138$+$2754,  SDSS~J1152$+$0930,  SDSS~J1152$+$3313,  SDSS~J1156$+$1911,  SDSS~J1207$+$5254,  SDSS~J1209$+$2640,  SDSS~J1329$+$2243, SDSS~J1336$-$0331, SDSS~J1343$+$4155,  SDSS~J1420$+$3955,  SDSS~J1439$+$1208,  SDSS~J1456$+$5702,  SDSS~J1522$+$2535,  SDSS~J1527$+$0652, SDSS~J1531$+$3414, SDSS~J1604$+$2244, SDSS~J1621$+$0607, SDSS~J1632$+$3500, SDSS~J1723$+$3411, SDSS~J2111$-$0114, SDSS~J2243$-$0935 }

\section{Introduction}

Strong lensing models of galaxy clusters provide a detailed description
of the mass distribution for the inner $\sim 100$ kpc. 
This method is complementary to other mass measurements, as
it provides a high-resolution measurement of substructure at the core,
on spatial scales at which other proxies such as X-ray, the Sunyaev-Zel'dovich
(SZ) effect, dynamical masses, and weak lensing do not resolve.  

In addition, the high magnification by clusters allows their use  as ``cosmic
telescopes'' to study background galaxies. Some of the highest redshift
galaxies known were discovered as magnified sources in the fields of
massive strong-lensing clusters \citep[e.g.,][]{zheng12,coe13,zitrin14,watson15,salmon18}.
At intermediate redshifts, some of the best-studied star forming galaxies are those
that are magnified by gravitational lensing, which enables high
signal-to-noise measurements \citep[e.g.,][]{pettini00cb58,pettini02cb58,cabanac08,quider09,quider10,dessauges11,rigby18a}, and probes sub-galactic scales \citep[e.g.,][]{whitaker14}.
Magnifications caused by the gravity of  strong lensing clusters can exceed factors
of hundreds, if galaxies are fortuitously placed in close proximity to
a lensing caustic.  

To use galaxy clusters as telescopes, a detailed description of
their mass distribution is key. The lens model determines the lensing magnification, 
which in turn connects the intrinsic and observed physical properties of the
background sources: size, luminosity,
star formation rate, and stellar mass.   
Thus, lens models provide the context for these lensed galaxies. 
 
In the era of large optical surveys, the field of strong
gravitational lensing is growing beyond detailed studies of a
handful of strong lensing clusters. 
Mining the Sloan Digital Sky Survey (SDSS) has yielded hundreds of
new galaxy cluster scale lenses, which form the SDSS Giant Arcs Survey (SGAS). Extensive
followup observations of clusters from this survey included optical
ground-based imaging and spectroscopy with a variety of telescopes, as
well as two Large \Spitzerlong\ programs. 
For some of these clusters there are
X-ray and SZ observations. 
In a Large \HSTlong\ (\hst) Guest Observer (GO) program, GO-13003
(hereafter, SGAS-HST; PI Gladders), we obtained multi-band WFC3 images.  

In this paper, we present strong gravitational
lensing models of  37 galaxy clusters from SGAS, based on 
multi-band \hst\ data and followup spectroscopy. 

The \hst\ subsample of SGAS, hereafter SGAS-HST, was designed to enable detailed study of the background highly-magnified galaxies, most of which are observed at redshifts of $1<z<3$, the epoch when most of the stars in the Universe were formed \citep{madau14}.
These \hst\ observations span a large wavelength range in order to sample the spectral energy distribution of each background galaxy. The high spatial resolution of \hst\ reveals their morphology and stellar substructure, and combined with lensing magnification, probes sub-kpc scales \citep[e.g.,][]{johnson17,rigby17,johnson17L}.
The high spatial resolution of \HST\ is critical for computing reliable lens models. It provides robust identification 
of images and counter images of strongly-lensed galaxies, and reveals additional lensed galaxies beyond the main (typically bright) source in each field. 

This paper is organized as follows. In Section~\ref{s.sgas}, we describe the Sloan Giant Arcs Survey; 
in Section~\ref{s.observations} we present the \HST\
imaging and the ground based spectroscopy using the Gemini and Magellan telescopes; Section~\ref{s.lensing} 
describes the lens modeling process in general, followed by cluster-specific lens model details, 
which are given in the subsections. 
We conclude in Section~\ref{s.summary}. 
The reduced \hst\ images and lens model outputs 
are made available to the scientific community as high level data products with this publication.

Throughout this work, unless specifically noted otherwise, we assume a flat cosmology with $\Omega_{\Lambda} = 0.7$, $\Omega_{m}
=0.3$, and $H_0 = 70$ km s$^{-1}$ Mpc$^{-1}$. Magnitudes are reported
in the AB system.

\section{The Sloan Giant Arcs Survey}\label{s.sgas}
The Sloan Giant Arcs Survey (SGAS) is a survey for highly magnified
galaxies that have been gravitationally-lensed by foreground clusters of galaxies,
in the Sloan Digital Sky Survey \citep[SDSS;][]{sdss}. The work presented here is 
based primarily on imaging and photometric catalogs from Data Release 7 \citep[DR7;][]{sdssdr7}.  
A full discussion of the survey, selection criteria, and purity and
completeness estimate, will appear in a separate paper
({M. Gladders et al., in preparation}).   
Galaxy clusters and groups were selected from the photometry catalog using the cluster
red-sequence algorithm of \cite{gladdersyee00}.   
 Imaging data in $g$, $r$, $i$, and $z$, were downloaded and cropped
 to $4\farcm5 \times 4\farcm5$  in size, centered on the detected
 clusters, with some overlap around larger clusters or filaments. The
 SDSS data were combined into color images, using a custom selection
 of scale and contrast in order to maximize the visual detectability
 of faint arc-like features.  
The color images were visually inspected by four investigators for
evidence of gravitational lensing. Each inspector scored each image on a scale of 0 to 3, 
where the highest score was given to obvious strong lenses. The
process included a re-inspection rate of 10\%  of the images by each inspector in order to
quantify the selection efficiency of each inspector, and the overall
completeness of the sample.  
Candidate lenses were followed up by larger telescopes, with most of
the confirmation done at the 2.56-m Nordic Optical Telescope
(NOT). Spectroscopy of high-confidence arcs was obtained using Gemini North with the GMOS \multislit\ spectrograph
\citep{bayliss11,bayliss11L}. All candidate lenses with an average score above 2 have been followed up in this way with 100\% completeness; we reach 95\% followup completeness for lens candidates 
with an average score between 1 and 2.  

The multi-wavelength follow-up campaign of  the SGAS
sample includes imaging from space and ground: radio observations  
of the Sunyaev-Zel'dovich (SZ) Effect \citep{gralla11}, \Spitzer\
infrared imaging (Program ID \#90232, PI: Rigby; Program ID \# 70154, PI Gladders), wide field optical
imaging for weak lensing with Subaru \citep{oguri12}, \HST\
(GO-13003, PI: Gladders; see Section~\ref{s.observations}), UV imaging
with SOAR \citep{bayliss12}, and  X-ray imaging with the X-ray telescope of the \Swift\  (PI:
Sharon) and \chandra\ (Program ID \#19800436, PI: Bayliss). Spectroscopy of all the primary arcs is complete 
\citep{bayliss11}, as well as spectroscopic campaigns to obtain redshifts of
secondary arcs with Magellan (PI: Sharon) and Gemini (PI: Sharon; Johnson), see Section~\ref{s.spectroscopy}.  

Since its initial assembly in 2008 \citep{hennawi08}, the SGAS sample of strong lensing
clusters has been mined for bright lensed galaxies that have been studied
 in detail \citep{koester10,bayliss10,rigby11,rigby14,rigby15,gladders13,
 wuyts12a,wuyts12b,bayliss14.1050,dahle13,dahle15,sharon17,johnson17,johnson17L,rigby17,chisholm18}.  
Many of the brightest lensed arcs from this program have high-quality rest-frame 
ultraviolet spectroscopy, obtained as part of \megasaura: \megasauralong\ \citep{rigby18a}.
The study of these galaxies, which would otherwise be too faint for
detailed spectroscopy, is only possible due to the magnification boost
by the foreground cluster and fortuitous position of the background
source in a region of highest magnification.  
The large sample of strong lensing clusters also enabled research of
the mass distributions of the galaxy clusters themselves 
\citep[e.g.,][]{gralla11,oguri12,blanchard13,sharon14}
 and cosmic structure \citep{bayliss11L,bayliss14LOS}.

\section{Observations}\label{s.observations}

\subsection{\HST\ Imaging} \label{s.hst}
The SGAS-HST program (GO-13003, PI: Gladders) consists of 107 orbits
of \HST\ imaging during Cycle~20. Each cluster was  imaged with
four WFC3 filters over three orbits. 
Table~\ref{tab.observations} details
dates, filters, and exposure times of the \hst\ observations.
We supplement our observations with archival optical WFPC2 data from programs GO-11974 (PI: Allam) and GO-11100 (PI: Brada{\v c}), which were available for five fields. Subsequent \hst\ programs of some of these targets acquired grism spectroscopy or additional bands, including GO-14622 (PI: Whitaker), GO-14230 (PI: Rigby), and GO-14896 (PI: Bayliss).

The GO-13003 observations were obtained with both the IR and UVIS channels of the 
WFC3 instrument \citep{wfc3} in order to minimize overheads. 
The primary science objective of
GO-13003 is to measure the physical properties of
the lensed galaxies. To accommodate that, 
the filter choice for each cluster was driven by the redshift of the
primary arc, rather than the cluster redshift, for a long
baseline rest-frame wavelength coverage with broad filters for maximal
throughput.   
Each target was observed with the reddest filter available; the F160W on the
IR camera. On the blue side, all but two clusters were imaged with the
F390W filter on UVIS. The two intermediate bands were chosen to straddle the
4000\AA\ break at the redshift of the  primary arc.      

The observations using the two IR filters were executed within the same orbit, 
and data taken with the
UVIS filters were executed in a single orbit each, all within one visit. 
In all cases, observations were split into four sub exposures with sub-pixel
dithers. The number of sub exposures was
found to be the minimum needed to sample the point spread function (PSF), 
cover the chip gap, 
and enable removal of cosmic rays and other artifacts,
while keeping to a minimum the readout noise and losses due to
charge transfer efficiency (CTE).
Targets were centered on the center of Chip~1 (UVIS) and with a
channel pointing offset of $+23\farcs5$ (IR) to optimally overlap the
two fields of view, and avoid placing the center of the cluster on the
chip gap. Where the visibility window allowed, we selected orientation
angles that place the main arc in each cluster close to the readout
side of the UVIS chip in order to minimize CTE losses in our primary
science targets. 

In the UVIS channel, the declining CTE of
the detector can cause large flux losses and increases the 
correlated noise in the image. 
To mitigate, the bluest UVIS data were taken with post-flash,
to increase the image background and ensure that faint sources have
high enough counts \citep[see WFC3 Data Handbook,][]{rajan10}. 
Post-observation image corrections were applied to individual
exposures using the Pixel-based Empirical CTE Correction
Software\footnote{\url{http://www.stsci.edu/hst/wfc3/ins\_performance/CTE/}}
provided by STScI. 

The WFC3-IR observations were executed within a single orbit using a
sampling interval parameter SPARS25, four
images per filter with small box dithers for PSF reconstruction and
to cover artifacts such as the ``IR Blobs'' and ``Death Star'' \citep[WFC3 Data
Handbook;][]{rajan10}. 
We used a custom algorithm for a secondary correction of IR blobs that
were not fully accounted for in the reduction pipeline of the 
WFC3-IR images as follows. We generated a superflat image from all the IR imaging
in out program, and constructed an empirical model of the IR blobs using
\galfit\ \citep{peng10}. The IR blobs were then suppressed
from individual exposures prior to drizzling, by dividing the data by
the scaled model in a process similar to flatfielding. 

Individual corrected IR and UVIS images were combined using the AstroDrizzle package
\citep{gonzaga12} with a pixel scale of $0\farcs03$ pixel$^{-1}$, and drop size of
0.5 for the IR filters and 0.8 for the UVIS filters. We find that this
approach provides good recovery of the PSF in all 
bands and maximizes the sensitivity to low signal-to-noise
features such as faint lensed galaxies. Finally, all images were drizzled onto 
the same pixel frame. 

The typical point-source $5\sigma$ magnitude limits of the
entire survey are given in Table~\ref{tab.observations}.
We estimate the depth in each filter by randomly placing circular apertures with a diameter of 
$0\farcs7$ across the PSF-matched images. We eliminate apertures that overlap with objects, 
and extract from the remaining background apertures the mean and standard deviation of the noise distribution for 
each field and filter.  We next consider the distribution of depths for each filter.  
While the observing strategy was similar for most observations, different fields show some variation in their 
achieved depth, which is quantified as the standard deviation of the distribution (listed in Table~\ref{tab.observations}).

\begin{deluxetable*}{llllllllllllr} 
 \tablecolumns{11} 
\tablecaption{\HST-GO-13003 Observations  \label{tab.observations}} 
\tablehead{\colhead{Cluster} &
            \colhead{R.A.}     & 
            \colhead{Decl.}    & 
            \colhead{Date UT}    & 
            \multicolumn{9}{c}{Exposure time (s)} \\[-8pt]
\colhead{} & 
            \colhead{J2000}     & 
            \colhead{J2000}    &            
            \colhead{}    & 
                        \colhead{\scriptsize F390W}       & 
            \colhead{\scriptsize F475W}       & 
            \colhead{\scriptsize F606W}       & 
            \colhead{\scriptsize F775W}       & 
            \colhead{\scriptsize F814W}       & 
            \colhead{\scriptsize F105W}       & 
            \colhead{\scriptsize F110W}       & 
            \colhead{\scriptsize F125W}       & 
            \colhead{\scriptsize F160W}      \\[-16pt]
     }
\startdata 
SDSS J0004$-$0103  & 00:04:52 & $-$01:03:15.80  & 2013-09-24 & 2388 & \nodata & \nodata & \nodata & 2400 & \nodata & \nodata & 1112 & 1212 \\
SDSS J0108$+$0624  & 01:08:42 & $+$06:24:43.50  & 2013-08-03 & 2388 & \nodata & 2400 & \nodata & \nodata & 1112 & \nodata & \nodata & 1212 \\
SDSS J0146$-$0929  & 01:46:56 & $-$09:29:52.40  & 2013-09-29 & 2388 & \nodata & 2400 & \nodata & \nodata & 1112 & \nodata & \nodata & 1212 \\
SDSS J0150$+$2725  &01:50:01 & $+$27:25:36.20  & 2013-08-02 & 2272 & \nodata & 1456 & \nodata & 1980 & \nodata & \nodata & \nodata & 912 \\
SDSS J0333$-$0651  & 03:33:05 & $-$06:51:22.20  & 2013-07-17 & 2388 & \nodata & \nodata & \nodata & 2400 & \nodata & \nodata & 1112 & 1212 \\
SDSS J0851$+$3331  & 08:51:39 & $+$33:31:10.83  & 2013-02-26 & 2368 & \nodata & \nodata & \nodata & 2380 & \nodata & \nodata & 1112 & 1212 \\
SDSS J0915$+$3826  & 09:15:39 & $+$38:26:58.77  & 2013-04-03 & 2372 & \nodata & \nodata & \nodata & 2388 & \nodata & \nodata & 1112 & 1212 \\
SDSS J0928$+$2031  & 09:28:06 & $+$20:31:25.55  & 2013-02-01 & 2388 & \nodata & 2400 & \nodata & \nodata & \nodata & 1112 & \nodata & 1212 \\
SDSS J0952$+$3434  & 09:52:40 & $+$34:34:47.09  & 2013-09-28 & 2368 & \nodata & 2380 & \nodata & \nodata & 1112 & \nodata & \nodata & 1112 \\
SDSS J0957$+$0509  & 09:57:39 & $+$05:09:31.80  & 2013-03-14 & 2388 & \nodata & \nodata & \nodata & 2400 & \nodata & \nodata & 1112 & 1212 \\
SDSS J1002$+$2031  & 10:02:27 & $+$20:31:02.61  & 2013-05-15 & 2388 & \nodata & 2400 & \nodata & \nodata & 1112 & \nodata & \nodata & 1212 \\
SDSS J1038$+$4849  & 10:38:43 & $+$48:49:18.73  & 2013-05-16 & 2392 & \nodata & \nodata & \nodata & \nodata & \nodata & 1112 & \nodata & 1212 \\
SDSS J1050$+$0017  & 10:50:40 & $+$00:17:06.91  & 2013-04-19 & 2388 & \nodata & 2400 & \nodata & \nodata & \nodata & 1112 & \nodata & 1212 \\
SDSS J1055$+$5547  & 10:55:05 & $+$55:48:23.23  & 2013-12-07 & 2408 & \nodata & \nodata & 2424 & \nodata & \nodata & 1112 & \nodata & 1212 \\
SDSS J1110$+$6459  & 11:10:18 & $+$64:59:47.02  & 2013-01-08 & 2420 & \nodata & 2436 & \nodata & \nodata & 1112 & \nodata & \nodata & 1212 \\
SDSS J1115$+$1645  & 11:15:04 & $+$16:45:38.40  & 2013-10-28 & 2388 & \nodata & \nodata & \nodata & 2400 & \nodata & \nodata & 1112 & 1212 \\
SDSS J1138$+$2754  & 11:38:09 & $+$27:54:30.90 & 2013-03-25 & 2384 & \nodata & \nodata & 2396 & \nodata & \nodata & \nodata & 1112 & 1212 \\
SDSS J1152$+$0930  & 11:52:47 & $+$09:30:14.54  & 2013-05-17 & 2388 & \nodata & 2400 & \nodata & \nodata & 1112 & \nodata & \nodata & 1212 \\
SDSS J1152$+$3313  & 11:52:00 & $+$33:13:41.72  & 2013-07-07 & \nodata & 2380 & 2380 & \nodata & \nodata & \nodata & 1112 & \nodata & 1112 \\
SDSS J1156$+$1911  & 11:56:06 & $+$19:11:12.68  & 2012-12-04 & 2388 & \nodata & \nodata & \nodata & 2400 & \nodata & \nodata & 1112 & 1212 \\
SDSS J1207$+$5254  & 12:07:36 & $+$52:54:58.20  & 2012-11-29 & 2400 & \nodata & 2416 & \nodata & \nodata & 1112 & \nodata & \nodata & 1212 \\
SDSS J1209$+$2640  & 12:09:24 & $+$26:40:46.50  & 2013-05-05 & 2384 & \nodata & \nodata & \nodata & \nodata & 1112 & \nodata & \nodata & 1212 \\
SDSS J1329$+$2243  & 13:29:34 & $+$22:43:00.24  & 2013-05-16 & 2384 & \nodata & 2400 & \nodata & \nodata & 1112 & \nodata & \nodata & 1212 \\
SDSS J1336$-$0331  & 13:36:00 & $-$03:31:28.63  & 2013-05-08 & 2388 & \nodata & 2400 & \nodata & \nodata & \nodata & \nodata & 1112 & 1212 \\
SDSS J1343$+$4155  &13:43:33 & $+$41:55:04.08  & 2013-03-13 & 2380 & \nodata & \nodata & \nodata & \nodata & 1112 & \nodata & \nodata & 1112 \\
SDSS J1420$+$3955  & 14:20:39 & $+$39:55:04.96  & 2013-01-17 & 2372 & \nodata & 2388 & \nodata & \nodata & 1112 & \nodata & \nodata & 1112 \\
SDSS J1439$+$1208  & 14:39:10 & $+$12:08:24.75  & 2013-01-21 & 2384 & \nodata & \nodata & \nodata & 2400 & \nodata & \nodata & 1112 & 1212 \\
SDSS J1456$+$5702  & 14:56:01 & $+$57:02:20.60  & 2013-01-10 & 2404 & \nodata & \nodata & \nodata & 2424 & \nodata & \nodata & 1112 & 1212 \\
SDSS J1522$+$2535  & 15:22:53 & $+$25:35:29.36  & 2013-07-06 & 2384 & \nodata & 2396 & \nodata & \nodata & 1112 & \nodata & \nodata & 1212 \\
SDSS J1527$+$0652  & 15:27:45 & $+$06:52:31.79  & 2013-01-16 & \nodata & 2400 & 2400 & \nodata & \nodata & \nodata & 1112 & \nodata & 1212 \\
SDSS J1531$+$3414  & 15:31:11 & $+$34:14:24.91  & 2013-05-06 & 2256 & \nodata & 1440 & \nodata & 1964 & \nodata & \nodata & \nodata & 912 \\
SDSS J1604$+$2244  & 16:04:10 & $+$22:44:16.69  & 2013-06-26 & 2384 & \nodata & \nodata & 2400 & \nodata & \nodata & 1112 & \nodata & 1212 \\
SDSS J1621$+$0607  & 16:21:32 & $+$06:07:19.85  & 2013-04-29 & 2384 & \nodata & \nodata & 2400 & \nodata & \nodata & 1112 & \nodata & 1212 \\
SDSS J1632$+$3500  & 16:32:10 & $+$35:00:30.16  & 2013-10-12 & 2372 & \nodata & \nodata & \nodata & 2388 & \nodata & 1112 & \nodata & 1112 \\
SDSS J1723$+$3411  & 17:23:36 & $+$34:11:59.37  & 2013-03-14 & 2368 & \nodata & \nodata & 2380 & \nodata & \nodata & 1112 & \nodata & 1112 \\
SDSS J2111$-$0114  & 21:11:19 & $-$01:14:23.57  & 2013-06-13 & 2388 & \nodata & \nodata & \nodata & \nodata & \nodata & 1112 & \nodata & 1212 \\
SDSS J2243$-$0935  & 22:43:23 & $-$09:35:21.40  & 2013-08-31 & 2388 &
\nodata & 2400 & \nodata & \nodata & 1112 & \nodata & \nodata & 1212
\\
\hline
\multicolumn{4}{l}{Typical point source limiting magnitude ($5\sigma$)}& $26.39$& $26.38$& $26.11$& $25.65$& $25.49$& $25.85$& $26.17$& $25.81$& $25.66$\\
\multicolumn{4}{l}{Limiting magnitude uncertainty }& $\pm0.08$& $\pm0.03$& $\pm0.24$& $\pm0.23$& $\pm0.31$& $\pm0.22$& $\pm0.26$& $\pm0.31$& $\pm0.20$\\
\enddata 
 \tablecomments{Observing dates and exposure time of the clusters as part of \hst\ program GO-13003. Columns 6 through 13 are the cumulative exposure times in each of the WFC3 filters, as labeled. Exposure times are in seconds. The limiting magnitudes are measured within a
circular aperture of diameter $0\farcs7$.}
\end{deluxetable*} 

\subsection{Spectroscopy}\label{s.spectroscopy}
Several different programs conducted spectroscopy of SGAS fields, using 
multiple observatories, with the science goal determining the observation
strategy. 
These follow-up programs generally fall under two categories.
The first type of programs includes \multiobject\ spectroscopy using either a
fiber-fed spectrograph or slit masks with multiple
slits or slitlets placed primarily on arc candidates and cluster
galaxies. Some of the results from these programs are published in 
\citet{bayliss11,bayliss14LOS}. 
With these observations, we secured
new spectroscopic redshifts of lensed sources, and informed lens models
and statistical analysis of lensing cross section; spectroscopy of 
cluster-member galaxies were used to study the
dynamical properties of the lensing clusters. 
The second type of spectroscopic program collects deep, high spectral resolution spectroscopy of individual
lensed galaxies in order to study their physical properties, leveraging the high magnification by the foreground cluster 
\citep[e.g.,][Akhshik et al. in preparation, Florian et al. in preparation]{bayliss14.1050,wuyts12a,wuyts12b,rigby14,rigby15,rigby18a}. 
Where available, we use results from these programs. 

In this paper, we present new spectroscopic redshifts from Gemini and Magellan
\multislit\ spectroscopic programs. 
The primary goal of these programs was securing spectroscopic
redshifts of secondary arcs in SGAS-HST fields. 
By construction of the
\hst\ target list, the primary arcs in each field already had secure
spectroscopic redshifts. The secondary arcs have either not been
identified prior to the \hst\ observations, or previous attempts with Magellan or Gemini
were unsuccessful at yielding redshifts. 
Spectroscopic redshifts of secondary arcs are
critical for an accurate lens model, as the slope of the mass profile
is degenerate with the distance of the lensed galaxies used as
constraints.  Redshifts of multiple source planes help break
this degeneracy \citep[e.g.,][]{smith09,johnson16}. The
secondary arcs are typically fainter than the primary arcs in each
field. We therefore planned the observations to have longer exposures
than previous SGAS programs \citep{bayliss11,bayliss14LOS}. The observing strategy 
was optimized for securing spectroscopic redshifts
of the background galaxies. 
Multiobject\ slit masks primarily targeted the lensing features around the cluster core. 
We placed additional slits on cluster-member galaxies and other objects of interest.
The filters and gratings
were selected to provide maximal wavelength coverage, to increase the
redshift range in which redshifted emission lines can be observed.

Table~\ref{tab.clusterz} lists for each cluster the best available
spectroscopic redshift, from the new observations described here, 
supplemented with information from previous programs including 
\citet{bayliss11,bayliss14LOS}, public databases (e.g., SDSS), and the literature. 
Where spectroscopic redshifts
are available for several cluster-member galaxies, we define the cluster
redshift as the mean redshift of these galaxies. 
Otherwise, the redshift of the cluster is assumed to be the
redshift of the brightest cluster galaxy (BCG). In our 
lensing analysis, the redshift of each cluster is taken to be known
exactly without errors; the uncertainty has a negligible effect on the lensing analysis. 

The velocity dispersion of each cluster is measured from all available
spectroscopic redshifts. 
For clusters with a small number of galaxies with spectroscopic redshifts, 
$N_{\rm spec}< 15$, 
we use the Gapper estimator
\citep{beers90}.
Otherwise, we use the 
square root of the bi-weight sample variance \citep{beers90,ruel14}. 
The uncertainties on the velocity dispersion are
calculated as $\Delta\sigma_v = \pm0.91\sigma_v/\sqrt{N_{\rm spec}-1}$, following \cite{ruel14}. 
The velocity dispersions are listed in Table~\ref{tab.clusterz}. 
 
\subsubsection{Gemini/GMOS-North Spectroscopy}
Spectroscopic observations of SGAS clusters were obtained using the
Gemini North telescope as part of programs GN-2015A-Q-38 (PI: Sharon)
and GN-2015B-Q-26 (PI: Sharon) using the Gemini Multi-Object Spectrograph
\citep[GMOS;][]{GMOS}. 
 
All Gemini/GMOS-North observations were conducted in queue mode using the R400\_G5305 grating in 
first order and the G515\_G0306 long pass 
filter. Table~\ref{tab.specobs} lists the observation dates and
exposure times for the clusters that were targeted in these programs. 
We designed one \multiobject\ slit mask for each cluster, placing slits
on the high priority secondary arcs in each cluster for which spectroscopic
redshifts had not been previously measured. Other slits
were placed on cluster galaxies, selected by their color relative to
the cluster red sequence.  

We also report redshifts from program GN-2011A-Q-11 (PI: Gladders) of primary arcs and cluster members that were not published elsewhere. 
Those observations, as well as the data reduction procedures of the GMOS data, are described in 
\cite{bayliss14LOS}. 

\begin{deluxetable*}{lllll}
 \tablecolumns{5} 
\tablecaption{Summary of Spectroscopic Observations  \label{tab.specobs}} 
\tablehead{
  \colhead{Cluster}     &
  \colhead{Telescope/Instrument} &
            \colhead{UT Date}    & 
            \colhead{Exposure Time} &   
            \colhead{Program ID}    
        }
\startdata 
SDSS~J0851$+$3331 &Gemini/GMOS-North & 2015-04-18 & 2$\times$2400 s & GN-2015A-Q-38\\
SDSS~J0928$+$2031 &Gemini/GMOS-North & 2015-04-18,19 & 2$\times$2400 s & GN-2015A-Q-38\\
SDSS~J0952$+$3434 &Gemini/GMOS-North & 2015-10-23,2016-01-03 & 2$\times$2880 s & GN-2015B-Q-26\\
SDSS~J1002$+$2031 &Gemini/GMOS-North & 2016-01-04,07 & 2$\times$2880 s & GN-2015B-Q-26\\
SDSS~J1002$+$2031 &Gemini/GMOS-North & 2011-05-22 & 2$\times$2400 s & GN-2011A-Q-19\\
SDSS~J1110$+$6459 &Gemini/GMOS-North & 2016-01-08 & 2$\times$2880 s & GN-2015B-Q-26\\ 
SDSS~J1207$+$5254 &Gemini/GMOS-North & 2016-02-04 & 2$\times$2880 s & GN-2015B-Q-26\\
SDSS~J1209$+$2640 &Gemini/GMOS-North& 2016-02-09 & 2$\times$2880 s & GN-2015B-Q-26 \\
SDSS~J1329$+$2243& Gemini/GMOS-North& 2011-06-02 & 2$\times$2400 s & GN-2011A-Q-19 \\
SDSS~J1604$+$2244 &Gemini/GMOS-North& 2011-04-29 & 2$\times$2400 s & GN-2011A-Q-19 \\
SDSS~J1632$+$3500 &Gemini/GMOS-North& 2012-04-15 & 2$\times$2400 s & GN-2011A-Q-19 \\
SDSS~J0146$-$0929 & Magellan/IMACS MOS & 2013-11-06 & 3$\times$2400 s&\nodata\\
SDSS~J0150$+$2725 & Magellan/IMACS MOS & 2013-03-16 & 6$\times$2400 s (2 masks)&\nodata\\
SDSS~J1152$+$0930 & Magellan/IMACS MOS & 2014-04-24,25,26            &  10$\times$2400 s (2 masks)&\nodata\\
SDSS~J1336$-$0331 & Magellan/IMACS MOS & 2013-07-11, 2014-04-24 & 10$\times$2400 s  (2 masks)&\nodata\\
SDSS~J1336$-$0331 & Magellan/LDSS3 longslit  &2013-05-2 & 2400 s & \nodata\\
SDSS~J1439$+$1208 & Magellan/IMACS MOS & 2014-04-26 & 2$\times$2400+3300 s &\nodata\\
SDSS~J1439$+$1208 & Magellan/IMACS longslit & 2013-03-17       & 2$\times$1200 s  &\nodata\\
SDSS~J1621$+$0607 & Magellan/IMACS MOS & 2013-08-3              & 3$\times$2400 s &\nodata\\
                                  &                                    & 2014-04-24,25,26  & 2$\times$3000,  2$\times$2880,  2$\times$2700 s &\nodata\\
SDSS~J1621$+$0607 & Magellan/LDSS3 longslit & 2013-05-2& 1800+2400 s&\nodata\\
SDSS~J2111$-$0114 & Magellan/IMACS MOS & 2013-08-2,3,2014-04-24,25 & 2$\times$3600, 6$\times$2400 s &\nodata\\
SDSS~J2243$-$0935 & Magellan/IMACS MOS & 2013-07-11, 2013-11-6,7 & 4$\times$2400 s&\nodata\\
SDSS~J0333$-$0651& Magellan/FIRE & 2013-03-01 & 2400 s &\nodata\\
SDSS~J0928$+$2031& Magellan/FIRE & 2013-02-28 & 2400 s &\nodata\\
SDSS~J1439$+$1208 & Magellan/FIRE & 2013-02-28 & 9600 s &\nodata\\
SDSS~J0108$+$0624 & APO/DIS & 2012-01-20 & 2$\times$1200 s &\nodata\\
SDSS~J0150$+$2725 & APO/DIS & 2012-01-20 & 2$\times$3000 s &\nodata\\
SDSS~J1456$+$5702 & MMT/Blue Channel & 2014-05-04 & 2$\times$1200 s &\nodata\\
\enddata 
 \tablecomments{Spectroscopic observation dates and exposure
   times. UT date lists the date or dates of the
   observation. Number of exposures and integration time of each exposure are listed in the fourth column. Magellan observations are typically split over two or more sub-exposures. Program ID lists the proposal identifier in the Gemini
   archive.}
\end{deluxetable*} 

\subsubsection{Magellan Spectroscopy}
We followed up clusters that are observable from the south, with Decl.$<20^{\circ}$, with the 
Magellan 6.5m telescopes at Las Campanas Observatory, 
Chile. 
As with the Gemini observations described in the previous section, 
we obtained spectra of secondary arcs that were identified in the \hst\ data, and not targeted by \cite{bayliss11}. 
We observed eight cluster fields with the Inamori-Magellan Areal Camera \& Spectrograph (IMACS)  
on the Baade telescope, 
using \multislit\ spectroscopy. We designed one or two masks per field, placing $1\farcs0$ slitlets on the lensed galaxies at highest priority, and filling the rest of the 
$27\farcs2 \times 27\farcs2$ field of view with cluster-member galaxies and other sources. 
We observed one cluster field with $1\farcs0$ longslit. 
Observations were typically several exposures of 2400s each.

We report redshifts in two cluster fields that we observed with the Low Dispersion Survey Spectrograph 3 (LDSS3) 
camera on the Clay telescope using either the $1\farcs0$ or $1\farcs5$ longslit.  

We report new redshift measurements in three fields: SDSS~J0333$-$0651, SDSS~J0928$+$2031, and SDSS~J1439$+$1208, using the Folded-port InfraRed Echellette (FIRE) on the Baade telescope. 
The observations took place  on 2013 March 01.
with clear weather conditions and $0\farcs7$ seeing.  
We used the Eschelle disperser with $1\farcs0$ slit, typically integrating for 40 minutes. 

\subsubsection{Apache Point Observatory Spectroscopy}
Observations with the Dual Imaging Spectrograph (DIS) on the 3.5 m telescope at Apache Point Observatory (APO) on 2012 Jan 20 were used to measure the redshifts of two arcs in this paper. The observation strategy and data reduction procedures are identical to those detailed in \cite{bayliss12}. 

\subsubsection{Multiple Mirror Telescope Observatory Spectroscopy}
Observations with the 6.5~m Multiple Mirror Telescope (MMT) using the Blue Channel spectrograph were obtained on 2014 May 4 (PI: Bayliss) for one of the targets in this paper, the  giant arc in SDSS~J1456+5702. 
We used the 500-line grating in first-order, tilted to set the central wavelength to 5575\AA, using the $1\farcs25$ wide longslit aperture at a position angle (PA) of 87.3 degrees east of north.  The source was at airmasses between 1.18 and 1.38 during the science exposures, and conditions remained clear with consistently sub-arcsecond seeing. 
The science frames were followed by HeNeAr arc lamp wavelength calibration frames and quarts lamp flat calibration frames taken at the same PA. We bias-subtracted, flat-field corrected, and wavelength calibrated the data with standard IRAF routines in the iraf.noao.imred.ccdred and iraf.noao.onedspec packages. We then used  custom IDL code built using the XIDL package to subtract a two-dimensional model of the sky on a pixel-by-pixel basis. The science spectrum was boxcar extracted to capture from the full extended object profile of the giant arc along the slit, using a 3\arcsec\ wide aperture. The final combined spectrum covers a range in in wavelength of $\sim$4000-7175\AA, with a dispersion of 1.19\AA\ per pixel and a median spectral resolution, R $= \delta \lambda / \lambda \simeq 1150$. 

\begin{deluxetable*}{llllll} 
 \tablecolumns{6} 
\tablecaption{Cluster properties  \label{tab.clusterz}} 
\tablehead{\colhead{} &
            \colhead{Cluster}     & 
            \colhead{Redshift}    & 
            \colhead{$\sigma_v$}    & 
            \colhead{N$_{\rm spec}$}       & 
            \colhead{Ref}       \\
            \colhead{} &
            \colhead{}     & 
            \colhead{}    & 
            \colhead{\kms }       & 
            \colhead{}       & 
            \colhead{}             }
\startdata 
1&SDSS~J0004$-$0103&0.514$\pm$0.003&556$\pm$178&9&Carrasco+17\\
2&SDSS~J0108$+$0624&0.54778$\pm$0.0002&\nodata&1&SDSS DR12 BCG\\
3&SDSS~J0146$-$0929&0.4469$\pm$0.00006&\nodata&1&SDSS DR12 BCG\\
4&SDSS~J0150$+$2725&0.30619$\pm$0.00005&\nodata&1&SDSS DR12 BCG\\
5&SDSS~J0333$-$0651&0.5729$\pm$0.0006&\nodata&1&BCG spectrum from APO/DIS\\
6&SDSS~J0851$+$3331&0.3689$\pm$0.0007&890$\pm$130&41&Bayliss+14\\
7&SDSS~J0915$+$3826&0.3961$\pm$0.0008&960$\pm$120&39&Bayliss+14\\
8&SDSS~J0928$+$2031&0.1920$\pm$0.0009&965$\pm$209&19&This work: GMOS, SDSS DR12\\
9&SDSS~J0952$+$3434&0.357$\pm$0.003&706$\pm$214&10&This work: GMOS, SDSS DR12 \\
10&SDSS~J0957$+$0509&0.448$\pm$0.001&1250$\pm$290&25&Bayliss+14\\
11&SDSS~J1002$+$2031&0.319$\pm$0.006&1110$\pm$170&37&This work: GMOS, SDSS DR12\\
12&SDSS~J1038$+$4849&0.4308$\pm$0.0003&659$\pm$69&48&Irwin+17, Bayliss+14\\
13&SDSS~J1050$+$0017&0.5931$\pm$0.0008&560$\pm$80&32&Bayliss+14\\
14&SDSS~J1055$+$5547&0.466$\pm$0.003&678$\pm$220&13&Bayliss+11\\
15&SDSS~J1110$+$6459&0.656$\pm$0.006&1022$\pm$226&18&Johnson+17\\
16&SDSS~J1115$+$1645&0.537 &\nodata &1&BCG spec Stark+13\\
17&SDSS~J1138$+$2754&0.451$\pm$0.008&1597$\pm$380&11&Bayliss+11\\
18&SDSS~J1152$+$0930&0.517$\pm$0.009&1360$\pm$320&6&Bayliss+11\\
19&SDSS~J1152$+$3313&0.3612$\pm$0.0007&800$\pm$90&38&Bayliss+14\\
20&SDSS~J1156$+$1911&0.54547$\pm$0.00013&\nodata&1&SDSS DR12 BCG\\
21&SDSS~J1207$+$5254&0.275$\pm$0.003&790$\pm$410&4&SDSS DR12, Kubo+10\\
22&SDSS~J1209$+$2640&0.5606$\pm$0.0012&1155$\pm$217&25&This work, Bayliss+11\\
23&SDSS~J1329$+$2243&0.4427$\pm$0.0007&830$\pm$120&31&Bayliss+14\\
24&SDSS~J1336$-$0331&0.17637$\pm$0.00003&\nodata&1&SDSS DR12 BCG\\
25&SDSS~J1343$+$4155&0.418$\pm$0.006&1011$\pm$290&7&Bayliss+11\\
26&SDSS~J1420$+$3955&0.607$\pm$0.006&1095$\pm$175&13&Bayliss+11\\
27&SDSS~J1439$+$1208&0.42734$\pm$0.00013&\nodata&1&SDSS DR12 BCG\\
28&SDSS~J1456$+$5702&0.484$\pm$0.009&1536$\pm$320&10&Bayliss+11\\
29&SDSS~J1522$+$2535&0.602$\pm$0.006&1187$\pm$300&14&This work: GMOS, SDSS DR12\\
30&SDSS~J1527$+$0652&0.392$\pm$0.005&923$\pm$233&14&Bayliss+11\\
31&SDSS~J1531$+$3414&0.335$\pm$0.005&998$\pm$190&11&Bayliss+11\\
32&SDSS~J1604$+$2244&0.286$\pm$0.004&1040$\pm$340&9&This work: GMOS + SDSS DR12\\
33&SDSS~J1621$+$0607&0.3429$\pm$0.0005&721$\pm$97&48&This work, Bayliss+11\\
34&SDSSJ1632+3500&0.466$\pm$0.006&1212$\pm$350&11&This work: GMOS, SDSS DR12\\
35&SDSS~J1723$+$3411&0.44227$\pm$0.00009&\nodata&1&SDSS DR12 BCG\\
36&SDSS~J2111$-$0114&0.6363$\pm$0.0009&1090$\pm$146&48&Bayliss+11, Carrasco+16\\
37&SDSS~J2243$-$0935&0.447$\pm$0.005&996$\pm$200&20&Bayliss+11\\
\enddata 
 \tablecomments{Spectroscopic redshifts and velocity dispersions or the SGAS-HST lensing clusters. N$_{\rm spec}$ is the
   number of unique cluster member galaxies with spectroscopic
   redshifts that were used to compute the velocity
   dispersion. We used the Gapper estimator for clusters with N$_{\rm
     spec}<15$ and the bi-weight sample variance otherwise (see
   Section~\ref{s.spectroscopy}).
 References are: \citet{carrasco17,bayliss11,bayliss14.1050,irwin2015,johnson17,kubo10,stark13}.}
\end{deluxetable*}

\section{Strong Lensing Analysis}\label{s.lensing}
We compute strong lensing models using the public
software \lenstool\ \citep{jullo07}.  \lenstool\
assumes a  ``parametric'' solution to describe the mass distribution,
and solves for the best-fit set of parameters using Markov Chain Monte
Carlo (MCMC) sampling of the parameter space. The best fit model is
defined as the one that results in the smallest scatter between observed
and model-predicted image positions in the image plane. 

We model each cluster as a linear combination of 
 mass halos, representing the dark matter mass distribution
attributed to the cluster and to correlated large scale
structure, and individual cluster galaxies. 
The halos are modeled as pseudo-isothermal ellipsoidal mass
distribution \citep[PIEMD, also known as dPIE;][]{jullo07}, with
seven parameters: position $x$, $y$; ellipticity $e$; position
angle $\theta$; core radius $r_{core}$, cut radius $r_{cut}$, and a
normalization $\sigma_0$. We note that $\sigma_0$, the effective
velocity dispersion, is correlated with, but not identical to, the observed velocity
dispersion of the mass halo. The relation to the observed velocity dispersion,
$\sigma_v$, can be found in \cite{eliasdottir07} for the
PIEMD potential. 
The two mass slope parameters, $r_{core}$ and $r_{cut}$, define an
intermediate region $r_{core}\lesssim r \lesssim r_{cut}$ within which the
mass distribution is isothermal; the potential transitions smoothly to
a flat profile towards $r=0$, and asymptotes to zero at large radii. 

The normalization and radii parameters of the galaxy-scale halos are scaled
to their observed luminosity (see \citealt{limousin05} for a description of the scaling relations). 
The positional parameters ($x$, $y$, $e$, $\theta$) are fixed to 
their observed measurements, as measured with Source Extractor \cite{bertin96}.

Unless noted otherwise, all the parameters of cluster-scale halos are allowed to vary, with
the exception of $r_{cut}$, which is far outside of the strong-lensing
region and thus cannot be constrained by the lensing evidence. We therefore
fix this parameter at 1500 kpc. 

We estimate the statistical uncertainties on individual model parameters 
from the MCMC sampling of the parameter space. These
uncertainties do not include systematic uncertainties due to modeling
choices, structure along the line of sight, or correlated structure and
substructure, which have been  shown to cause systematic uncertainties 
\citep[e.g.,][]{acebron17,meneghetti17,zitrin15,cerny18,priwe17,johnson17}.
From the lensing potential we compute the mass
distribution, convergence, shear, magnification, and deflection maps. We refer the
reader to \citet{johnson14} for a concise description of these
derivatives, and their dependence on the cluster and source
distances. We estimate the statistical uncertainties of these lensing
outputs by computing a suite of models sampled from the MCMC chain.

\subsection{Identification of cluster galaxies}\label{s.galaxies}
Cluster members are identified as forming a tight ridge in
color-magnitude diagrams, often referred to as ``red sequence''
\citep[e.g.,][]{gladdersyee00,koester07}. The color of the red sequence depends on the
cluster redshift and the selection of filters. 
We note that some
fraction of the galaxy population is not quiescent \citep{butcher78,butcher84}, and thus the red-sequence method forms an incomplete cluster-member selection. 
However, this fraction decreases significantly in the inner 1~Mpc 
\citep[e.g.,][]{hansen09,porter08,bai07,saintonge08,loh08,fairley02}.
In cases where the lensing signal is clearly affected by
individual galaxies, even when they are not on the red sequence, we
include these galaxies in the model.   

\subsection{Lensed Sources}\label{s.lensed}
The SGAS clusters were selected based on visual inspection of SDSS
images of lines of sight towards galaxy clusters. Therefore, by construction of the sample,
each cluster has at least one bright arc detectable with shallow
ground-based imaging. Deeper followup
provided a multitude of fainter arc candidates, some of which were
confirmed by spectroscopy \citep{bayliss11}.  
In optical bands, the depth of the \HST\ imaging does not exceed that of
the ground data; however, the excellent resolution allows us,
in most cases, 
to unambiguously identify multiple images, detect substructure in
extended arcs, and discriminate among arc candidates. The deep \hst\ near-IR
imaging provides wide wavelength coverage with a lever arm on color
variance, and a window to the high-redshift universe as well as highly
extinguished star forming galaxies.  
In discovering new arc candidates, we are often assisted by the
preliminary lens model, which is based on the primary arc and other
known lensed galaxies. The initial model is often
successful in predicting the approximate location of possible counter
images. This is typical for lens models which assume some form of
correlation between the mass distribution and the light of cluster
galaxies \citep[e.g.,][]{broadhurst05}. We note that this process of using an
initial model to aid in the discovery of new arcs is not limited to
mass-follows-light models; fully parametric models such as \lenstool\
can provide good initial guesses. Following the recommendations of
\cite{johnson16}, each time a lensed source was
spectroscopically confirmed, we reassessed any images with no
spectroscopic confirmation that have been predicted by the lens model
before adding them back in as lensing constraints. 

Owing to the superior resolution of \hst\ and the broader wavelength coverage compared to the previous ground-based data, we discovered new arc candidates in most of the cluster fields. These
are presented in Table~\ref{tab.arcstable}, with  coordinates,
IDs, and where available, spectroscopic or photometric redshifts. 

\subsection{Lensing Analysis of Individual Clusters
}\label{s.clusters}
In this section, we provide details on the identification of lensed
sources and modeling choices for each cluster in this \hst\ program. 
The ability to construct a unique, high fidelity, lens model varies from cluster
to cluster, and is dominated by the availability of lensing constraints, availability of spectroscopic redshifts, and other factors that are less straightforward to quantify such as the distribution of lensing constraints in the field of view. 
Numerical estimators that are often used for assessing models, such as $\chi^2$, reduced $\chi^2$, and rms, may be good indicators of the statistical errors in the parameter space, but do not necessarily encompass the full uncertainty \citep[e.g.,][]{johnson16,cerny18}. 
We therefore resort to providing a subjective assessment of the reliability of each model as a well-understood cosmic telescope. We rate the lens models in three broad categories: Category $A$ denotes a high quality, robust model, whose magnification is relatively well-constrained; in category $B$ we place acceptable models, with somewhat higher systematic  uncertainties; and $C$ denotes poor models, that are under-constrained. 
 We indicate below the caveats and point to sources of uncertainty for 
individual cluster models if they significantly reduce the reliability of the model, i.e., fall in category $B$ or $C$. 

Table~\ref{tab.arcstable} lists the lensing constraints in each field, which are described in the subsections below. 
We tabulate the IDs, coordinates, and redshifts of the lensed galaxies, as well as individual emission knots within the lensed galaxies that were used as constraints. 
The IDs appear as $AB.X$ where $A$ is a number or a letter indicating the source ID (or system name);  for images with identified substructure, $B$ is a number or a letter
indicating the ID of the emission knot within the system; and $X$ is a number indicating the ID of the lensed image within 
the multiple image family. For example, 1a.3 would be the third image of emission knot $a$ in system \#1.  Images without clearly identified substructure would appear as $A.X$.
In a few cases, we retain the naming systems that were used in previous publications. 

\begin{figure*}
\epsscale{1.0}
\plotone{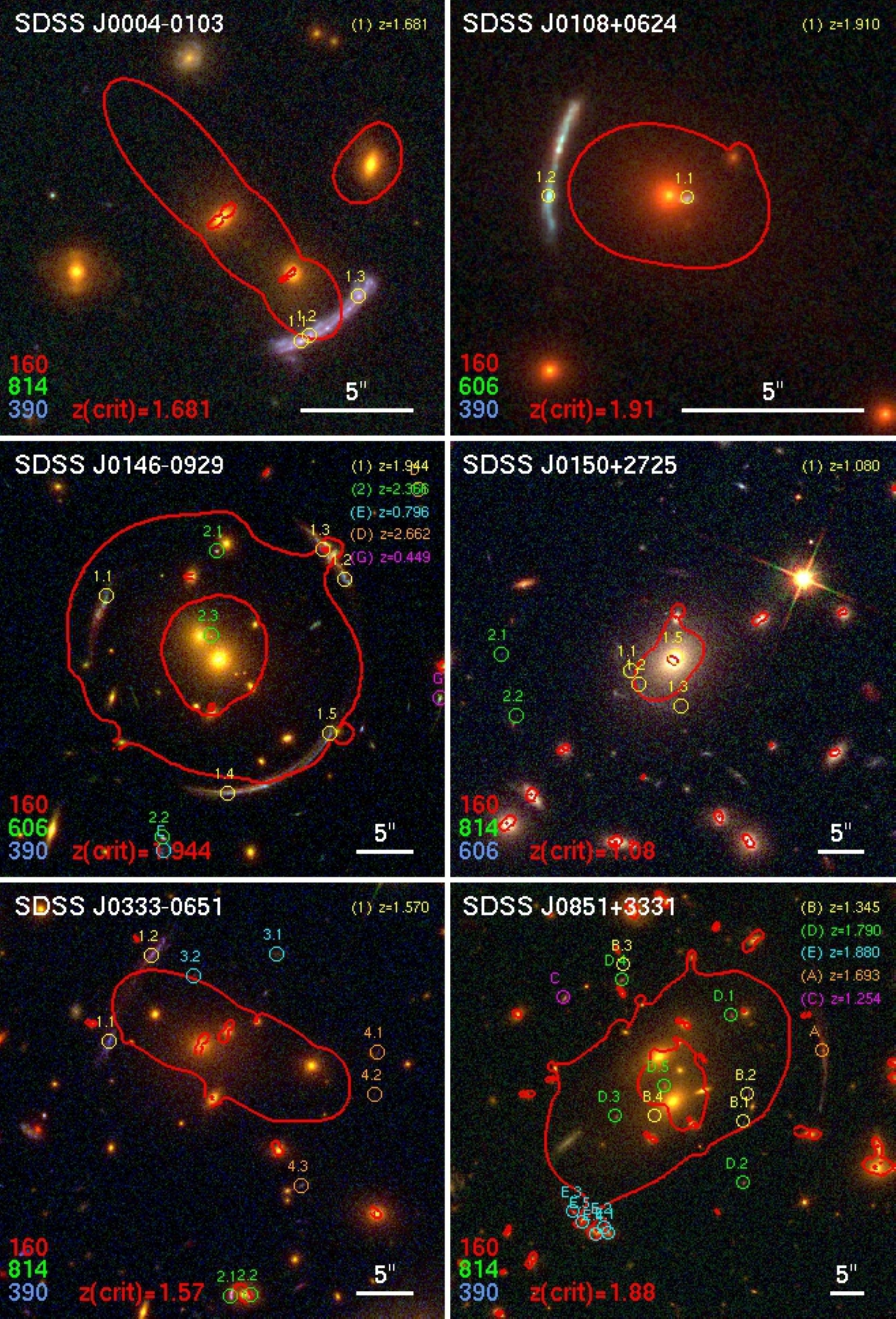}
\caption{Composite color images of the SGAS-HST clusters SDSS~J0004$-$0103,  
SDSS~J0108$+$0624,  
SDSS~J0146$-$0929,  
SDSS~J0150$+$2725,  
SDSS~J0333$-$0651, and  
SDSS~J0851$+$3331. North is up and East is left in all panels. The images are from the \hst\ bands that best show the lensed galaxies, with the filters noted in the bottom-left corner of each panel.} \label{fig.1}
\end{figure*}

In Figures~\ref{fig.1} through \ref{fig.7}, we present the 37 clusters in this program. 
The images are rendered from three of the \hst\ bands as noted in the bottom-left 
corner of each panel. The clusters are shown in order of R.A., and their names are shown on the top-left. 
The field of view of each panel is selected in order to best show the strong lensing region; we plot a 5\as\ scale bar
at the bottom-right corner of each panel. The critical curves are over-plotted in red, representing the loci of maximum
magnification for a source at redshift $z$, as indicated in the bottom-left of each panel. 
We plot the locations of the observed lensed galaxies. Each family is labeled in the 
same color; for clarity, we do not label multiple knots within the lensed images; they
are given in Table~\ref{tab.arcstable}.
Finally, spectroscopic redshifts 
 are listed at the top-right corner of each panel, with the same color code. For completeness,
we list spectroscopic redshifts of sources even if they fall outside of the 
field of view of the panel. 

\subsubsection{SDSS J0004$-$0103}  
The \hst\ imaging confirms the bright arc that was
previously discovered in ground-based data. 
The \hst\ resolution reveals the clumpy morphology of this arc. We
identify several distinct emission knots, and determine that the arc
consists of three merging partial images of the background source. We
use individual emission clumps as lensing constraints.
The cluster itself is diffuse, with no obvious dominant
galaxy at its core. We use the Gapper estimator to calcuate the
velocity dispersion from spectroscopic redshifts of nine cluster galaxies measured by 
\cite{carrasco17}, $\sigma_v = 556\pm178$ km~s$^{-1}$. This velocity dispersion is
consistent with a small group rather than a rich cluster. 
We use the velocity dispersion as an upper limit on the lensing mass
distribution when computing the lens model. 

Due to the limited lensing constraints and the uncertainty of where
the cluster center is, the lens model is not unique and has large
systematic uncertainty. We rank its fidelity for predicting the magnification anywhere in the field and its ability to constrain the mass distribution as $C$ (poor, under-constrained).   
At the same time, the model is successful at reconstructing the morphology and geometry of the lensed source and explaining the lensing configuration, and can be used to place limits on the magnification of the lensed galaxy. 

\subsubsection{SDSS J0108$+$0624}  
We confirm the identification of one bright arc, at a projected
angular distance $3\farcs3$ from the BCG, at $\zspec=1.91$ \citep{rigby18b}. The \hst\ data reveal a
demagnified counter image separated $0\farcs52$ from the BCG. We also
identify several bright emission knots, color variation along the
magnified arc, and clumpy morphology in both images, which helps constrain the lens model. 
No additional lensed galaxies were identified. The internal morphology and color 
variation along the giant arc indicate that this is a system of two images, 
with the giant arc being one complete image 
of the source, and the demagnified central image the second image. A third image is predicted 
behind the central cluster galaxy. We observe faint F390W emission at this location. However, since some
of the other cluster members show faint emission in this filter at their centers as well,  
the data cannot rule out that this emission is coming from the central cluster galaxy. 
This model is not as well constrained as others in this sample, since the constraints come from a single lensed galaxy with only two images. Since the lens is dominated by a central galaxy and the identification of the central image provides a good constraint on the center of the lens, the model is of sufficient fidelity for measuring the mass distribution in the core. However we place it in category $B$ due to the high uncertainty on the lensing magnification. 

\subsubsection{SDSS J0146$-$0929}  
The primary source lensed by SDSS~J0146$-$0929 appears as an
incomplete Einstein ring around the BCG, with mean projected radius of
$12\farcs1$. The \hst\ resolution reveals morphology and color
variation along the lensed galaxy. We identify image 1.1 (northeast of
the BCG) to be a complete image of the source. The arc that appears to
the south of the BCG is a merging pair of two partial images of the
source, which we label 1.4 and 1.5. The arc northwest of the BCG is
affected by the lensing potential of two interloper galaxies, 
likely cluster members, leading to a complex critical curve at this
location and breaking up this arc into two partial images 1.2 and 1.3 (the `main' arc) and
additional partial images buried in the light of the interlopers. 
We use several emission clumps in the source as constraints. 
The spectroscopic redshift of this galaxy was reported by \cite{stark13}.
We identify and spectroscopically confirm a second lensed galaxy with two secure images, 2.1 and 2.2. 
One image was predicted by the lens model near the center of the
cluster. We identify a likely candidate, but it is not used as a constraint.

We obtained \multislit\ spectroscopy from Magellan on 2013 Nov 6, using
IMACS on the Baade telescope (Table~\ref{tab.spec}).
From these data, we measured
the redshift of source~\#2, $\zspec=2.366$,  
from \ciiidoublet. The other background sources with spectroscopic redshifts that are listed in Table~\ref{tab.spec}
are likely not strongly lensed (i.e., we do not expect them to have
counter images).
The model is well-represented by a central cluster-scale halo and
cluster member galaxies. We leave the slope parameters of the two
galaxies near the northwest arc as free parameters, and fix their
positional parameters to their observed values.  
The lens model is well-constrained, and we classify it in category $A$. 

\subsubsection{SDSS J0150$+$2725}  
The main arc in this field is a low surface brightness galaxy at
$\zspec=1.08$, measured from the \OII\ line in spectroscopy from the 3.5 m telescope at the 
Apache Point Observatory (APO), obtained on 2012 Jan 20, with the APO/DIS instrument. 
 Although some candidates were suggested from ground data, a
careful inspection of the \hst\ images shows that their morphology or
color are not the same and thus we rule them out as strongly lensed systems. 
We identify a counter image of the giant arc very close to the central
galaxies, embedded in the BCG light. However, the distinct blue color
of this galaxy gives it a high contrast against the cluster galaxies,
which are mostly dropped out of our bluest band (F390W). 
We also identify two faint images of a candidate lensed galaxy. A
third image is predicted, but would be less magnified and fall under our
detection limit. 
The lensing constraints are sufficient for a robust model, however, 
due to the lack of spectroscopic redshift for source \#2 the slope of the mass distribution is 
not well constrained. We therefore classify the model in category $B$ (acceptable with higher uncertainty).

\subsubsection{SDSS J0333$-$0651}  
The giant arc in this field, source~\# 1 at $\zspec=1.57$, was spectroscopically confirmed with Magellan/FIRE observation obtained on 2013 Mar 01. It is likely a single, highly distorted and flexured lensed galaxy. It appears resolved in the \hst\ data, however, its substructure and color variation do not show lensing symmetry. 
We identify one secure lensed source, likely a galaxy-galaxy lensing event  forming a partial Einstein ring around an elliptical galaxy $22\farcs2$ south of the BCG (close to the bottom of Figure~\ref{fig.1}). We have no spectroscopic data for this source. We further identify two other candidate lensed systems, labeled 3 and 4. 

Due to the limited strong lensing constraints, the lens model for this cluster is less robust than other clusters in our program. 
We therefore classify it as $B+$. 
Spectroscopic redshifts of sources \#3 and \#4 would strengthen the model, as well as deeper observations to identify additional (fainter) images that are expected to occur for system \#3.
Alternatively, if no further constraints are revealed, future analysis may use the flexion (curved distortion due to the third derivatives of the lensing potential) of the main arc as a constraint, 
to improve the fidelity of the model \citep[e.g.,][]{cain16}. Flexion constraints are currently not implemented in the lens modeling algorithm we use.  

\subsubsection{SDSS J0851$+$3331}  
This cluster lenses a few background sources, some of which were identified
by \cite{bayliss11}. 
Source~A  at $z_{spec}=1.6926$ \citep[][labeled A1 and A2]{bayliss11}, appears as a highly
stretched and flexured arc, with significant color gradient, and is a single image. The two
images of source~B that were identified and measured by \citet{bayliss11}, labeled B.1, B.2 at $\zspec=1.3454$,
have matching morphology in the \hst\ imaging, with the expected 
lensing symmetry. We identify two new counter images for this source, B.3 and B.4, for
a total of four  images of the lensed source. 
Source~C is a single image. 
We identify two new lensed systems in the \hst\ data: source~D has
five images around the center of the cluster, with spectroscopic
redshift $z_{spec}=1.79$, measured from \hst-WFC3-IR grism
spectroscopy with \hst\ (GO-14622, PI: Whitaker).
Source~E has four secure images, and is brightest in the infrared bands.  The 
lensing configuration of the images of source~E is dominated by two cluster galaxies $\sim20''$
south of the BCG. A candidate partial fifth image is identified. This source was the science target of GO-14622, which measured a spectroscopic redshift $z_{spec}=1.88$, confirming the photometric redshift
that we initially inferred based on Spitzer observations and the \hst\ data presented here. 
An analysis of this source will be presented in a forthcoming publication (Akhshik, M. et al., in preparation). 
Given the number, quality, and distribution of constraints, the model is of high fidelity, and we classify it in category $A$.

\subsubsection{SDSS J0915$+$3826}  
This cluster was previously modeled by \cite{bayliss10} based on
ground-based data. That work identified two lensed sources, and
measured their redshifts: source A at $\zspec=1.501$ and source B,
a \lya -emitting galaxy at $\zspec=5.2$. Our new \hst\ imaging
confirm the interpretation of these images as lensed
galaxies. Furthermore, with the \hst\ resolution we pin down the
lensing morphology of the giant arc and determine that it is made of
three merging images of the source galaxy. Several emission knots are
clearly identified within the galaxy. For source B,  
\cite{bayliss10} were unable to identify a counter image, and therefore
this source was not used as a constraint. Our
\hst\ data help reveal two counter images with the same unique color
as source B, enabling its use as lensing constraint. 
Lensing potential from a nearby cluster-member galaxy
perturbs the critical curve at the location of B.1.
The model is of high fidelity, and we classify it in category $A$.

\subsubsection{SDSS~J0928$+$2031}  
The distribution of cluster member galaxies and the lensing evidence in this 
field suggest that SDSS~J0928$+$2031 is composed of two subhalos, each dominated by a 
bright elliptical galaxy. The northwest halo appears to be the more massive, based on 
the lensing evidence and its appearance in archival \chandra\ data.
This cluster was selected as a strong lens based on an identification of an elongated arc north 
of the BCG, labeled 1 in Figure~\ref{fig.2}. The giant arc likely forms along the critical curve for its redshift, with a bright knot falling just outside of the caustic and thus not multiply-imaged. We identify several additional lensing constraints near the northwest core of the cluster.   
Spectroscopy was obtained with the Magellan Clay telescope using FIRE on 2013 Feb 28, securing spectroscopic redshifts of the main arc at $\zspec=1.891$, and of two other sources  that are likely not multiply-imaged (Table~\ref{tab.spec}). We note a group of red galaxies near source C, which appear to be low-redshift dusty galaxies; while their FIRE spectra did not have high enough signal-to-noise for a high confidence redshift, their \spitzer\ luminosity and optical colors are similar to those of source C, and the spectra are consistent with $z\sim1.9$.  
We obtained \multislit\ spectroscopy from Gemini North as part of program GN15AQ38 (PI: Sharon), and measured spectroscopic redshifts of arc candidates 2 and 3, and cluster-member galaxies (see Tables~\ref{tab.arcstable}, \ref{tab.spec}).

The second bright galaxy in this cluster lies 46'' southeast of the BCG in projection. We identify a flexured galaxy at $\zspec=0.856$ northwest of the galaxy, in the direction of the BCG. No counter images are identified for this galaxy, and it is likely in the strong-weak lensing regime rather than strongly lensed. We identify a candidate radial arc, system 5, and use it as a constraint in the south halo.
The lack of strong lensing constraints prevent us from robustly modeling the mass distribution that may be attributed to a possible dark matter clump in the south region. The projected mass density implied from the model is consistent with the velocity dispersion, which we measured from 15 galaxies we observed with GMOS and four galaxies from SDSS-DR12. The model therefore provides a reliable description of the mass and magnification around the northwest core (category $A$), from multiple lensing constraints and source redshifts around its core; but it is underconstrained in the area surrounding the southeast core (category $C$).

\subsubsection{SDSS J0952$+$3434} 
A giant arc is observed south of the cluster core ($\zspec=2.1896$, \citealt{kubo10}). The giant arc does not show particular symmetry, and we do not robustly identify a counter image. We identify tentative faint multiple images of other sources that can be used to constrain the lens model. However, the low confidence with which they are identified, and the lack of confirmation of a counter image for the giant arc, results in a low-confidence lens model (category $B$). The lens model, the loose appearance of the galaxy distribution, and overall density of galaxies, are consistent with the low projected velocity dispersion we calculate for this cluster (Table~\ref{tab.clusterz}) from spectroscopic redshifts of eight cluster-member galaxies from Gemini/GMOS and two galaxies with spectra in SDSS-DR12 (Table~\ref{tab.spec}). 

\subsubsection{SDSS J0957$+$0509}  
The \hst\ data show that the giant arc of source~1 in SDSS~J0957$+$0509 
is composed of several merging images of a single background galaxy. 
Cluster-member galaxies that are in close projected proximity to the arc contribute to a complex lensing configuration. We identify five bright images
within the giant arc, and one faint demagnified image near the center of a galaxy $7\farcs8$ west of the BCG. The redshift of this source was measured by \cite{bayliss11}, $\zspec=1.821$. 
In addition to the giant arc, we identify two other lensed systems in the \hst\ data, \#2, \#3. Each of these systems 
has four visible images, with a fifth demagnified image predicted by the lens model near the center of the BCG. 
The lens model is reliable, and we classify it as $A-$. Securing spectroscopic redshifts for sources \#2 and \#3 would add confidence to the model that is otherwise well-constrained.

\subsubsection{SDSS J1002$+$2031}  
\begin{figure*}
\epsscale{1.0}
\plotone{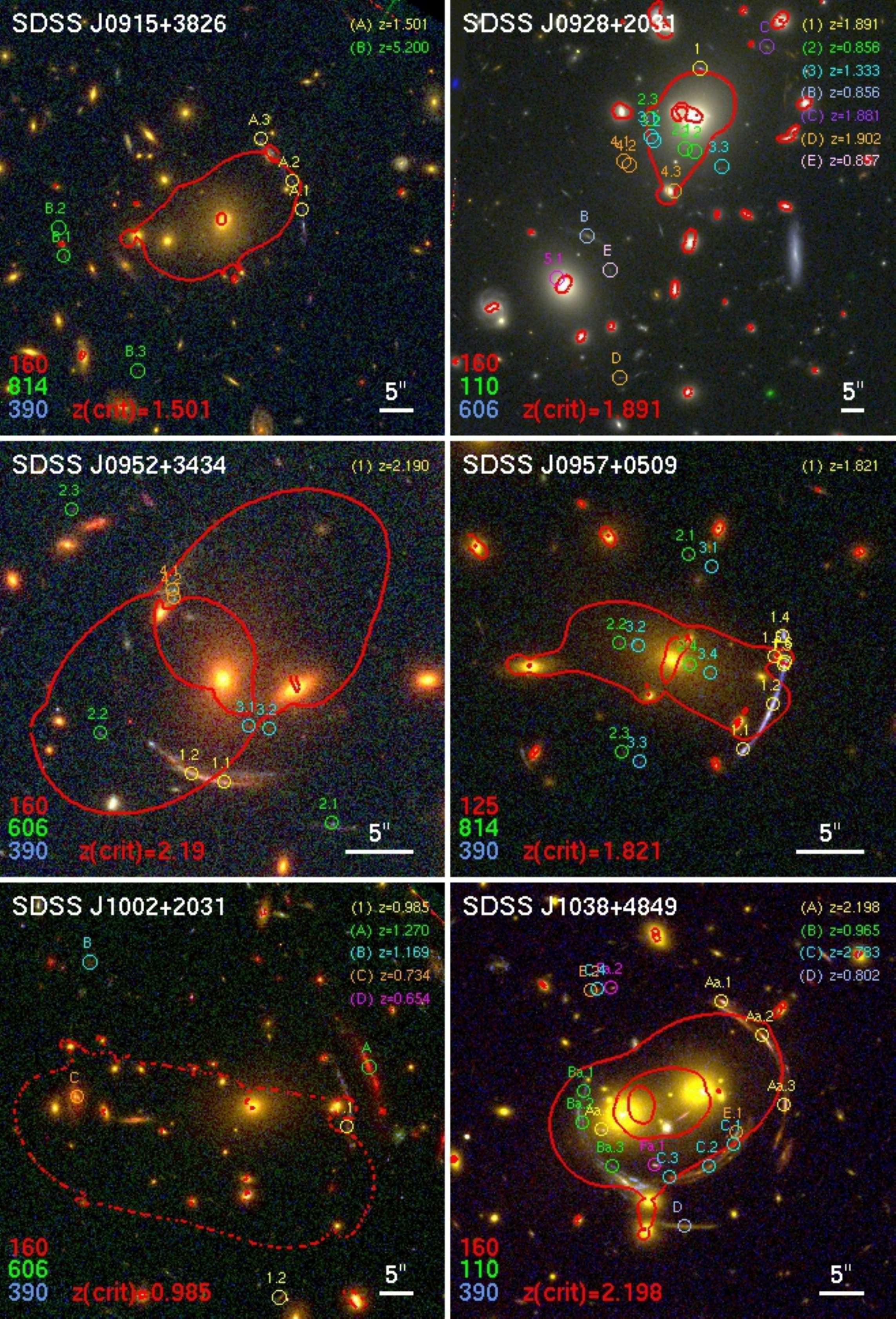}
\caption{Same as Figure~\ref{fig.1}, for the clusters 
SDSS~J0915$+$3826, 
SDSS~J0928$+$2031,  
SDSS~J0952$+$3434,  
SDSS~J0957$+$0509,  
SDSS~J1002$+$2031, and  
SDSS~J1038$+$4849.}\label{fig.2}
\end{figure*}

\begin{figure*}
\epsscale{1.0}
\plotone{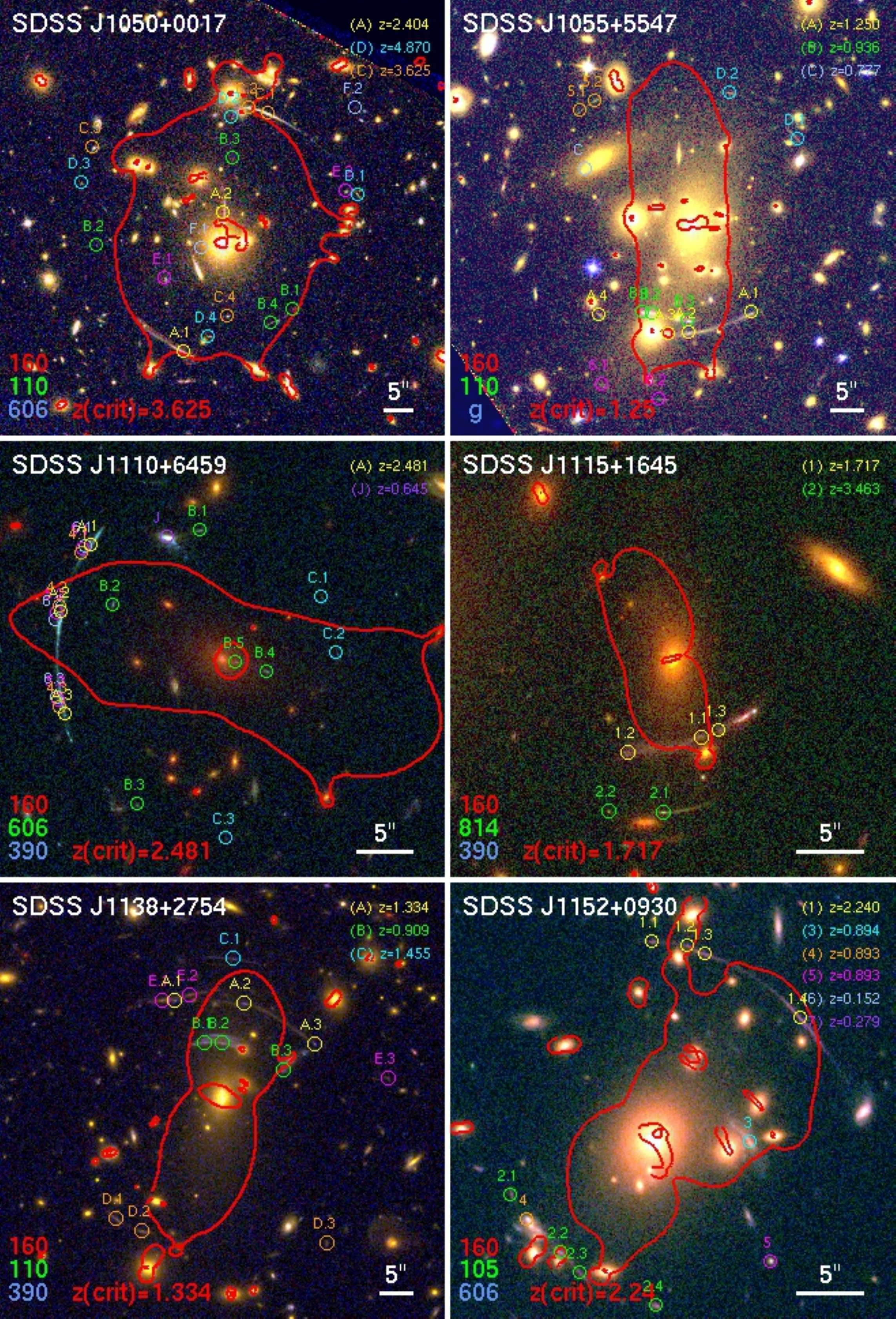}
\caption{Same as Figure~\ref{fig.1}, for the clusters 
SDSS~J1050$+$0017, 
SDSS~J1055$+$0017,
SDSS~J1110$+$6459, 
SDSS~J1115$+$1645, 
SDSS~J1138$+$2754, and
SDSS~J1152$+$0930. 
} \label{fig.3}
\end{figure*}

\begin{figure*}
\epsscale{1.0}
\plotone{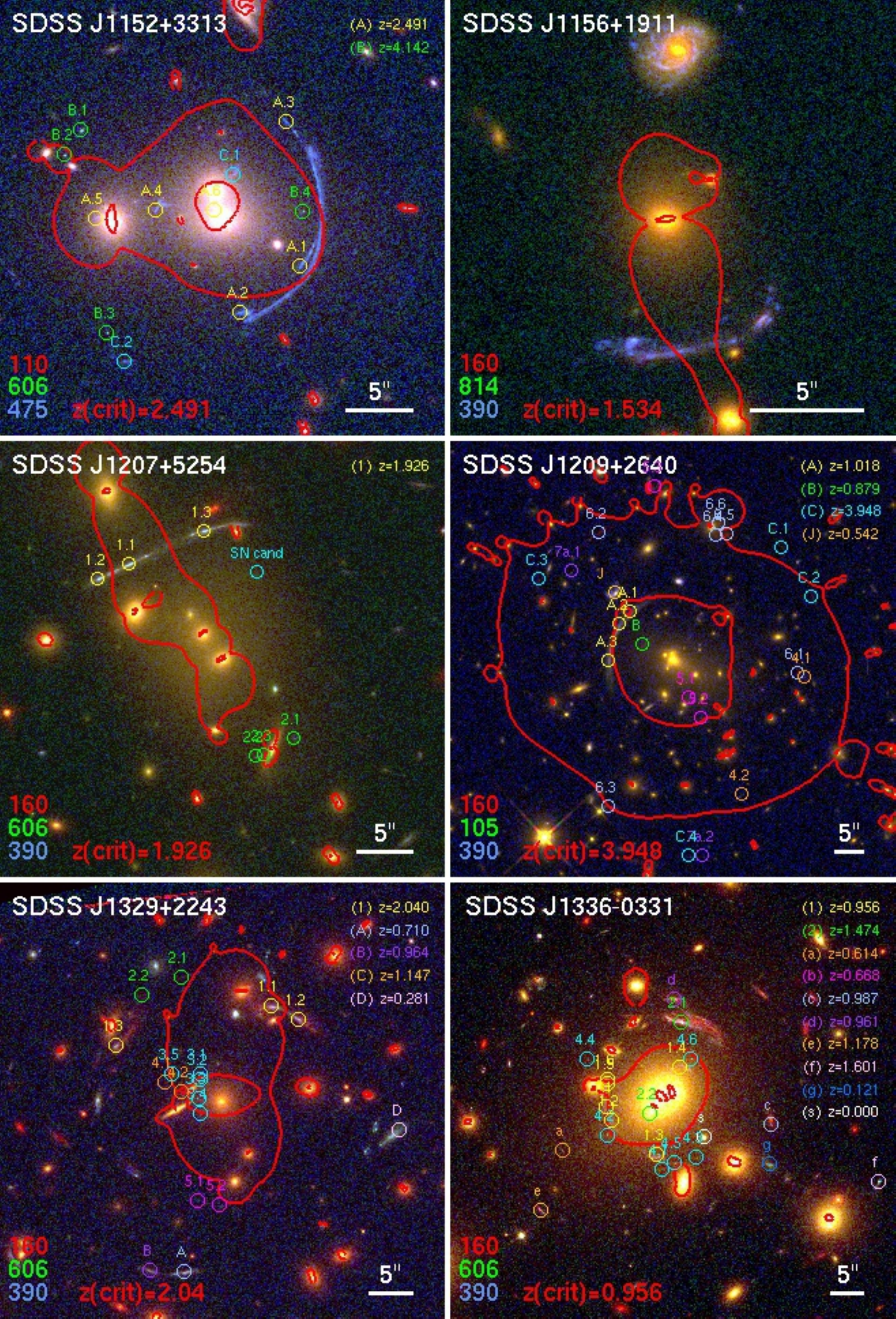}
\caption{Same as Figure~\ref{fig.1}, for the clusters 
SDSS~J1152$+$3313,
SDSS~J1156$+$1911, 
SDSS~J1207$+$5254, 
SDSS~J1209$+$2640, 
SDSS~J1329$+$2243, and
SDSS~J1336$-$0331.
} \label{fig.4}
\end{figure*}

\begin{figure*}
\epsscale{1.0}
\plotone{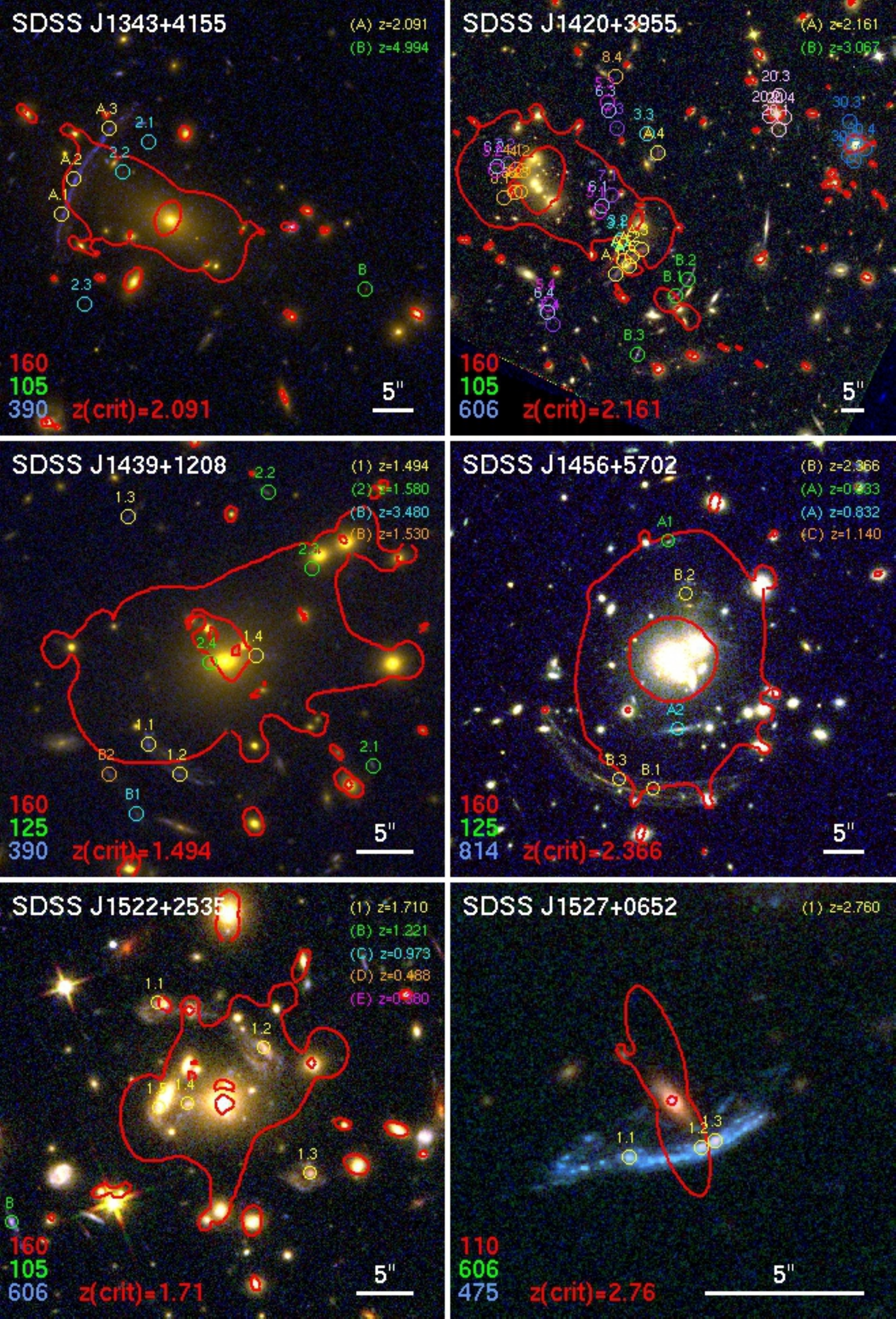}
\caption{Same as Figure~\ref{fig.1}, for the clusters 
SDSS~J1343$+$4155,
SDSS~J1420$+$3955, 
SDSS~J1439$+$1208, 
SDSS~J1456$+$5702, 
SDSS~J1522$+$2535, and
SDSS~J1527$+$0652.
} \label{fig.5}
\end{figure*}

\begin{figure*}
\epsscale{1.0}
\plotone{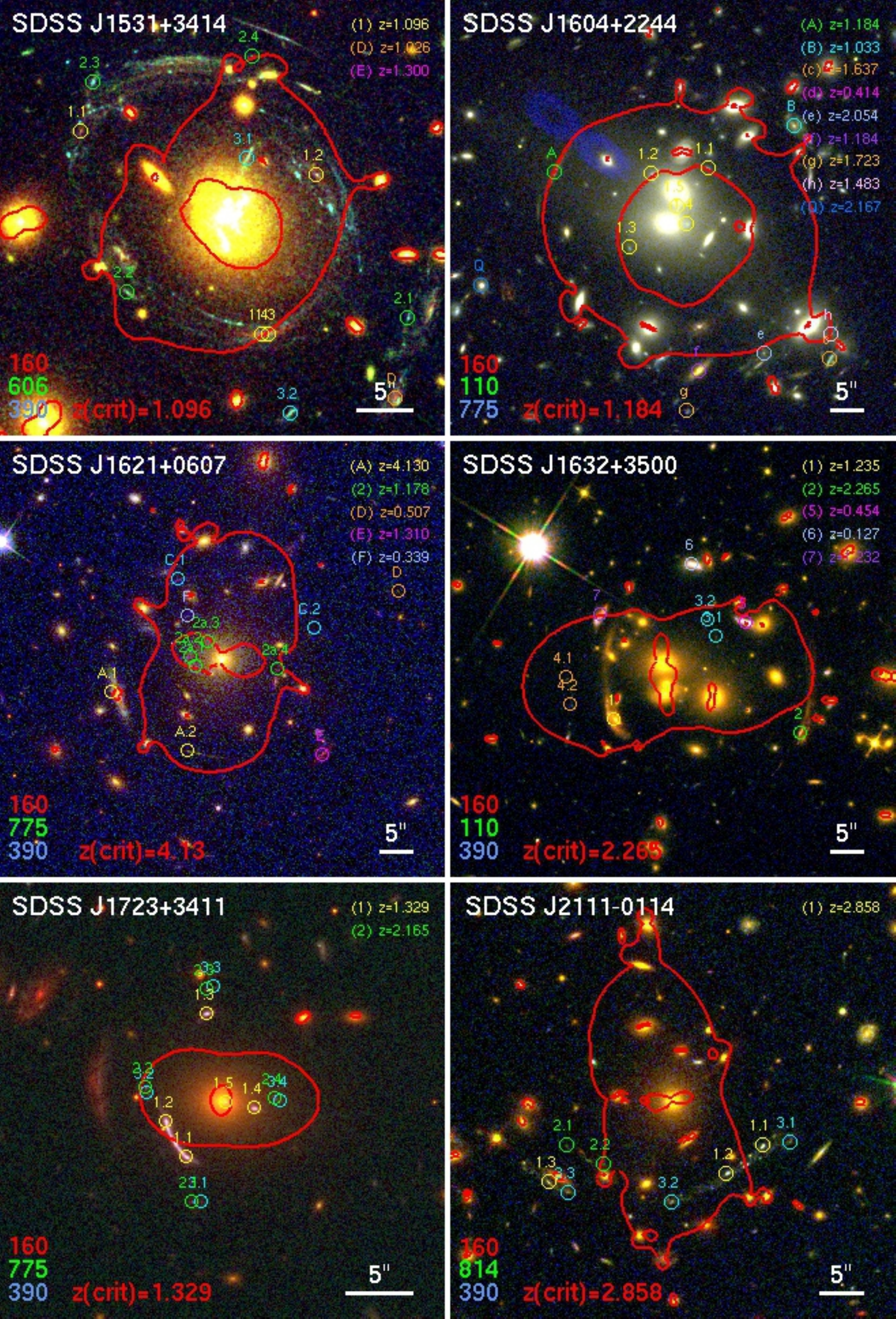}
\caption{Same as Figure~\ref{fig.1}, for the clusters 
SDSS~J1531$+$3414,
SDSS~J1604$+$2244, 
SDSS~J1621$+$0607, 
SDSS~J1632$+$3500, 
SDSS~J1723$+$3411, and
SDSS~J2111$-$0114.
SDSS~J2243$-$0935.
} \label{fig.6}
\end{figure*}

\begin{figure*}
\epsscale{1.0}
\plotone{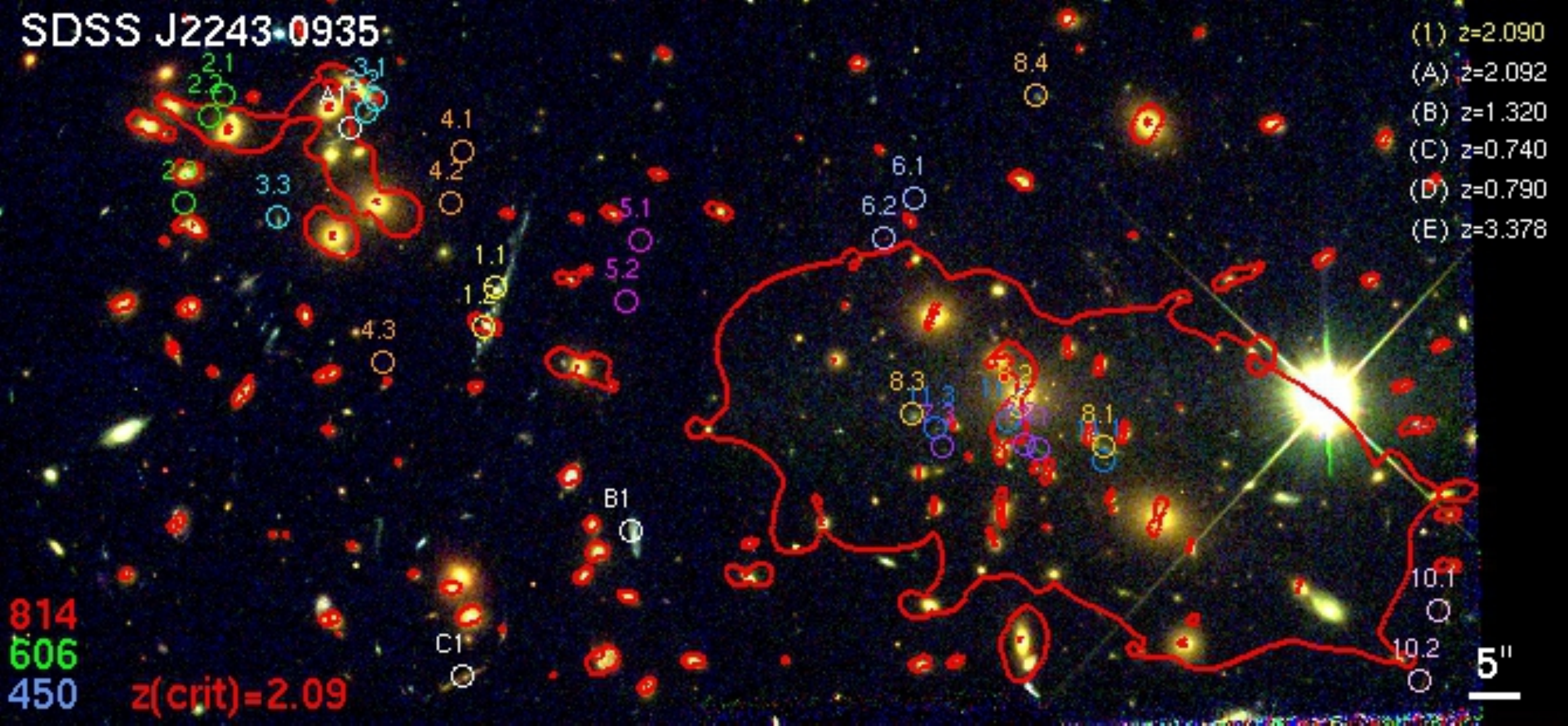}
\caption{Same as Figure~\ref{fig.1}, for the cluster
SDSS~J2243$-$0935.
} \label{fig.7}
\end{figure*}

Ground-based imaging of this cluster show an obvious giant arc, located $14\farcs0$ west of the BCG, labeled \#1 in Figure~\ref{fig.2}. A second elongated feature appears bright in red/infrared bands, $18\farcs1$ west of the BCG, labeled A in Figure~\ref{fig.2}. Using our \hst\ imaging data, we identify a tentative candidate counter image for the giant arc (labeled \#1.2). The red feature, which also looks like an elongated arc from the ground, appears as two galaxies in the \hst\ data. It does not have a counter image, and is likely a magnified single-image. We do not identify other strong-lensing constraints in the data. 
We observed this field using GMOS on Gemini North (GN11AQ19 and GN15BQ26) and obtained spectroscopic redshifts of background galaxies and cluster member galaxies. The spectroscopy places the blue giant arc at $z_{spec}= 0.985$, and galaxy~A that was initially identified as a red arc at $\zspec=1.27$. Another bright background galaxy, C, is found $26\farcs1$ east of the BCG, with spectroscopic redshift $\zspec=0.734$. These two spectroscopically-confirmed, singly-imaged galaxies help in constraining possible lens models, as we can rule out models that predict counter images for these galaxies.  
Nevertheless, we find that these constraints are not sufficient for a robust lens model. This model is therefore classified as $C-$ (poor, under-constrained).
The dashed critical curve that is shown in Figure~\ref{fig.2} represents a family of models that is capable of reproducing the lensing geometry of the blue arc, however, these models are likely incorrect because they predict counter images for galaxy~A. 

\subsubsection{SDSS J1038$+$4849}  
As can be seen in Figure~\ref{fig.2}, there are several giant arcs with large azimuthal coverage around the cluster core. \cite{bayliss11} measured spectroscopic redshifts of four unique sources (see Table~\ref{tab.arcstable}). The new \hst\ data improve upon archival data with the long wavelength baseline. Since the lensed images vary significantly in color and morphology, these new data are useful for a robust identification of the matched counterparts in the lensed images of each source, and in distinguishing the sources from each other. 

We follow the naming scheme of \cite{bayliss11} for the giant arcs, however, the exact locations of constraints within each arc are slightly different, owing to the superior \hst\ resolution. Source A, at $\zspec=2.198$, is identified northwest of the cluster core, as a giant arc of three merging images. The images are highly magnified in the tangential direction, and several emission clumps can be isolated in each. The source is likely a galaxy pair. We identify a fourth image of this source east of the cluster core. 

Source B, at $\zspec=0.9652$ is seen as a giant arc southeast of the cluster core, brightest in the UVIS bands. The lens model clarifies the lensing configuration of this system: the source galaxy appears to be a spiral galaxy with bright bulge and two loose spiral arms. The central part of the galaxy has one image, that appears slightly distorted in the tangential direction. This is the brightest part of this arc, labeled as image B1 in \cite{bayliss11}. In the source plane, some emission knots in one of the spiral arms cross the caustic, forming the elongated giant arc Ba.1, Ba.2, Ba.3.

Source C, at $\zspec=2.7830$, is lensed into a giant arc composed of three merging images south of the core, and a newly-identified counter image north-northeast of the core. The model predicts a demagnified central image for this source, 2-4 magnitudes fainter, in the light of the central galaxies. We do not identify this image in the current data depth.

Source D in \cite{bayliss11}, at $\zspec=0.802$, is a distorted singly imaged galaxy, and not used as a constraint in the lens model. 

We identify two new lensed galaxies. Source E is likely part of system C, however, since it has significantly bluer color than system C we leave its redshift as a free parameter. The model-predicted redshift for this source is consistent with that of source C. We identify two images of source F, south and north of the cluster core. A third image is predicted by the lens model to appear near the center of the cluster, predicted to be two magnitudes fainter. We do not identify this demagnified image. 

The model is well-constrained and is classified in category $A$. 
 
\subsubsection{SDSS J1050$+$0017}  
Our lens model of this cluster was presented in \cite{bayliss14.1050}, who measured the physical properties of one of the giant arcs (system C, at $\zspec=3.625$). We refer the reader to this publication for a full description of the lens model and spectroscopic confirmation of the lensed galaxies. We follow the naming scheme of \cite{bayliss14.1050} in the brief description below; see also Table~\ref{tab.arcstable} and Figure~\ref{fig.3}.

We identify several lensed sources in the field of SDSS~J1050$+$0017. 
Source A has two images: a giant tangential arc south of the cluster center, and a radial arc north of the BCG. We identify three images of source B, a faint tangential arc and a counter image. Source C has four images, two of them form the giant arc northwest of the cluster center, and the other two appear northeast and south of the BCG. 
Source D has four images around the cluster core, and we identify two images each for sources E and F. The model only uses as constraints the spectroscopically confirmed systems: A, C, and D. The model is well-constrained and is classified in category $A$. 

\subsubsection{SDSS J1055$+$5547}  
During the UVIS observation of SDSS~J1055$+$5547, the fine guidance sensors failed to lock on one of the two guide stars, and the \hst\ visit was executed with only one guide star and gyro control. As a result, the UVIS observation of this cluster failed. The WFC3-IR imaging in F110W and F160W were not significantly affected.
Our request for a re-observe of the failed visits was denied due to STScI's strict policy of not re-observing failed observations when more than 90\% of the targets in the program are complete. 
Since the UVIS data were critical for the science outcome of GO-13003, this cluster was excluded from the analysis of the lensed background sources. Nevertheless, we provide a lens model for this cluster, from a combination of the WFC3-IR data and archival Subaru imaging in $g$, $r$, and $i$ (see \citealt{oguri12} for details of the Subaru observation of this field). Figure~\ref{fig.3} is a color rendition made from the F160W and F110W in red and green, respectively, and Subaru $r$ band in blue. The \hst\ data are superior in resolution, and the Subaru data complement them with a broad wavelength coverage which is useful for a robust identification of lensed features. 
Furthermore, we identify the cluster member galaxies in this field from the Subaru data, using the $r$ and $i$ bands. 

The cluster lenses several background galaxies, four of them were identified and targeted for spectroscopy by \cite{bayliss11}. The lensed images of source A ($\zspec=1.250$) and source B ($\zspec=0.936$) appear in the south of the field. Nearby galaxies complicate the lensing configuration of A and B. Source C ($\zspec=0.777$) in \cite{bayliss11} is a single image of a background galaxy. Source D was targeted for spectroscopy, but did not yield a redshift measurement. We identify a candidate counter image for this system. We identify other lensing features around the cluster core, as listed in Table~\ref{tab.arcstable}.  The model is well-constrained and is classified in category $A$. 

\subsubsection{SDSS J1110$+$6459}  
The lensing analysis and spectroscopic observations of SDSS~J1110$+$6459 are
presented in Johnson et al. (2017a)  in more detail.
We identify four multiply imaged lensed galaxies, denoted A, 
B, C, D in Figure~\ref{fig.3}. 
The main arc, A, is a giant arc formed by a naked cusp configuration of three
merging images of the same source galaxy. 
In Johnson et al. (2017a) we measured a
source redshift $\zspec=2.4812$ from Gemini spectroscopy of
all three arcs, and identified  more than 
 $20$ emission knots in each of the three segments of
the arc, thanks to the high tangential magnification. 
In \cite{johnson17,johnson17L} and \cite{rigby17} we measured their 
size, luminosity, and star formation rate, and find that the typical source-plane
unlensed size is $<100$~pc.   
System B is a five-image set, comprised of three tangential and two radial arcs, with distinctive color
and morphology.
System C is detected with two secure images. A third candidate is predicted by the model but not unambiguously
confirmed. 
System D is identified as three separate emission knots near system
A (labeled D,E,F in \citealt{johnson17}). It is not spectroscopically confirmed, however, since its
deflection angle is slightly different than that of system A we conclude
that it is a separate galaxy from system A, that happens to be on the
same line of sight. The best-fit model redshift for this system is $z_{model}=2.37$. 

We refer the reader to \cite{johnson17} for the full details of
the lens model of  SDSS~J1110$+$6459. In short, this cluster was
modeled using the hybrid \lenstool\ method, with the cluster potential
modeled as a combination of mass components centered on grid nodes,
following the prescription of \cite{jullo09}. The halos of
cluster galaxies were added as described in Section~\ref{s.galaxies}.
The model is well-constrained and is classified in category $A$. 

We also note a ``jellyfish'' galaxy in this cluster at $\zspec=0.6447$, spectroscopically
confirmed at the cluster redshift near image B.1  \citep{johnson17}. Based on its
morphology, this galaxy is likely going through a rapid epoch of
star formation as it falls into the cluster \citep{sun10,ebeling14,mcpartland16}.

\subsubsection{SDSS J1115$+$1645}  
Two lensed galaxies are spectroscopically confirmed in the field of SDSS~J1115$+$1645, both south of the cluster BCG. The redshift of System~1 was measured by \cite{stark13}, $\zspec=1.7170$. 
The redshift of system 2 was measured by \cite{bayliss12}, $z_{spec}=3.4630$. 
 The SDSS~DR12 catalog provides spectroscopic redshift of another cluster galaxy at $\zspec=0.5350$, at a projected radius of $51\farcs2$ from the BCG. 

 We identify a foreground structure on the line of sight to the lensing cluster, likely a cluster or a group, at $\zspec \sim0.19$. The structure appears as a clear red sequence in the color-magnitude diagram of the field. Three galaxies in this cluster/group have spectroscopic redshifts in SDSS~DR12, $z_{SDSS}=0.19107$, $0.19372$, $0.19602$, within $\sim3'$ from the lens. The foreground cluster/group is likely boosting the lensing efficiency of SDSS J1115$+$1645, as seen in other strong-lensing selected clusters \citep{bayliss14LOS}. The lensing evidence suggests that the cluster at $z_{lens,spec}=0.537$ is responsible for the observed lensing configuration; however, due to the small number of constraints the current analysis cannot quantify the contribution of the intervening structure. The model is classified in category $B$, since although it has lensing constraints from two galaxies, they are both on the south side of the cluster leaving its north region under-constrained. 

\subsubsection{SDSS J1138$+$2754}   
Five lensed galaxies are identified in this field. All the spectroscopic redshifts were measured by \cite{bayliss11}. Source A has three images north of the BCG; all three images were  spectroscopically confirmed at $\zspec=1.334$. Source B, at $\zspec=0.909$, has three images, two of them form a thick arc north of the BCG.  Source C is likely a single image, also north of the BCG, with a tentative redshift $\zspec=1.455$. Two images of source D were also identified and targeted by \cite{bayliss11} but the spectra did not yield a redshift measurement. We identify a candidate third image for this system. The new \hst\ data help discriminate the different part of what looks like a single giant arc in the ground-based data. We identify a new lensed candidate, labeled E. 

We measure a photometric redshift for source D, $z_{phot}=2.94$ from the \hst\ and \spitzer\ photometry. Since the images of this galaxy are the only constraints south of the BCG, we use the photometric redshift as a prior in our analysis. 
The model is well-constrained and is classified as $A-$. A spectroscopic redshift measurement of source D would add confidence to model predictions south of the core.  

\subsubsection{SDSS J1152$+$0930}  
System 1 forms a faint thin giant arc northwest of the BCG, identified as source D in \citet{bayliss11}; we rename the lensed features in this field for clarity. The arc is fairly smooth in the \hst\ data with no prominent substructure. A faint elongated arc appears in the southwest of the cluster, which we interpret as four images of a lensed galaxy. 
Other faint arc-like features are observed in the data, however, we are unable to confirm additional multiple-image lensing constraints. We measure a redshift for system 1 of $\zspec=2.24$, from tentative \ciiidoublet\ lines in low signal-to-noise \multislit\ spectra using IMACS on the Magellan Baade telescope, observed in April 2014. The same observations result in spectroscopic redshift of two foreground sources at $\zspec=0.152$ (source 6)  and $\zspec=0.279$ (source 7). 
We confirm the redshifts of background sources that were formerly measured by \cite{bayliss11} for the galaxies labeled 3 (A2) and  4 (A3), however, the \hst\ data, as well as the lens model, indicate that these are single images.  Given the similar redshifts of galaxies 3, 4, 5, $\zspec \sim 0.893$, it is likely that they are part of a small background group. 
The lens model is based on the northwest and southeast arcs, with the redshift of the former used as a fixed constraint. Table~\ref{tab.arcstable} lists the positions that were used as constraints. However, due to the relative smooth appearance of the faint arcs, it is difficult to identify the exact locations of the counterparts of each multiply-imaged feature. 
The model is classified as $A-$. A spectroscopic redshift measurement of source \#2 would add confidence to the model in the southeast area.

\subsubsection{SDSS J1152$+$3313}  
\cite{bayliss11} measured the redshifts of two lensed galaxies in this cluster; source A ($\zspec=2.491$), which forms the giant arc in the west, and two images of source B ($\zspec=4.1422$). With the \hst\ data, the internal structure of the giant arc is resolved, and we identify several unique features in it that are used as lensing constraints. We find that the giant arc is made of three merging images of the background galaxy. In addition, we identify three counter images, for a total of six detectable images for this system. A seventh image is predicted by the lens model near the center of the southern central galaxy (near A.6); several features are detected in the galaxy light, but since they cannot be uniquely matched with features in the other arcs, we do not use this image as a constraint. 
For source B we identify four images, with a fifth, demagnified image predicted by the lens model to appear near the center of the south central galaxy, but is too faint to be uniquely identified in the galaxy light. 
We identify two images of a new lensed galaxy, labeled C, with two images, and a demagnified third predicted near the core of the north central galaxy. It was not targeted for spectroscopy and therefore its redshift is unknown.
The model is well-constrained and is classified in category $A$. 

\subsubsection{SDSS J1156$+$1911}  
In the field of SDSS J1156$+$1911 we detect one giant arc at
$\zspec=1.543$ (Stark et al. 2013). We assume a cluster redshift from the spectroscopic
measurement of its BCG, $z_{BCG,spec}=0.54547\pm0.00013$ from SDSS.  
The BCG redshift was also measured by \cite{stark13}. We do not detect a
counter image for the giant arc.  
The lens model and the color gradients within the arc indicate that
the east and west portions of the arc are not multiply imaged. The
source galaxy crosses two caustics in the source plane, resulting in
magnification and stretching of only a small part of the galaxy. 
The model is reliable for reconstructing the morphology of the lensed galaxy, and can explain its lensing geometry. Given the limited lensing constraints we classify it in category $B$. 

\subsubsection{SDSS~J1207$+$5254}  
SDSS J1207$+$5254, at $z=0.275$, was reported  by \cite{kubo10}. 
The cluster core contains several elliptical galaxies, and
not one dominant central galaxy. The small curvature of the $\zspec=1.926$
giant arc is a result of an elongated projected mass density distribution, caused by a
secondary halo which coincides with the bright elliptical galaxy to the northeast. 
The cluster was observed in our Gemini program, but  
 the pipeline failed to generate useable flat fields for most of the mask due to scattered light from the bright acquisition stars. 

Interestingly, an over-density of luminous red
galaxies is apparent in the WFC/IR data northwest of the cluster center, 
and based on the color and brightness it is likely a background structure. 

The giant arc in SDSS~J1207$+$5254 is a naked-cusp configuration of
three images, two of 
which are partial images. We identify 10 individual emission knots
that are used as constraints; no counter image is identified in the
data beyond the giant arc, as expected in such configuration. 
A secondary system of two high-confidence arcs is identified southwest
of the cluster center, with unknown redshift. A third image in this system is 
predicted by the lens model to be embedded in the light of a cluster-member galaxy, and is visually
confirmed; this third image is not used as a constraint in the model. 
The lens model is classified as $B+$, with higher uncertainties in the southwest region.

Comparing the new \hst\ data and $gri$ Gemini/GMOS
observations taken on 2008-12-23 as part of GN-2008A-Q-25, we found in the GMOS
data a point-like source that does not have a counterpart in the \hst\
image, at R.A., Decl.=[181.8982, 52.918553]. Since this source only
appears in the archival images we were unable to confirm it
spectroscopically. If extragalactic, this source could be a supernova (SN)
at the cluster redshift without an apparent host, i.e., an
intracluster SN. These have been shown to compose as much as $20\%$ of
cluster SNe \citep{galyam03,sand11,sharon10}.
Alternatively, the transient could be a supernova in the
foreground or behind the cluster, with an undetected faint host. The
magnification at the position of the transient depends on the source redshift, 
but given its close proximity to the cluster center (see Figure~\ref{fig.4}) it could be significant, 
ranging from 1.8 for a source at $z=0.4$, to 7 for a source at $z=1$, 
and $\sim17$ for a source at $z=2$. 
The transient appears in two of the three bands, with $g > 25.9  \pm 0.3$,
$r = 23.87 \pm 0.07$ and $i = 23.28 \pm 0.08$. The duration of the
GMOS observations was 10 minutes, during which the position of the
source remains stable. For a thorough discussion of SN alternatives
in a single-epoch discovery we refer the reader to \cite{galyam02} and \cite{sharon10}.  

\subsubsection{SDSS J1209$+$2640}  
The discovery of the lensing cluster SDSS~J1209$+$2640 was reported by \cite{ofek08}. This paper identified several lensed candidates, and measured the redshift of the brightest arc, A, $\zspec=1.018$. \cite{bayliss11} measured spectroscopic redshifts of source B, $\zspec=0.789$, and C, $\zspec=3.948$. With archival WFC2 data and the broad wavelength coverage made available by our new \hst\ data, we identify the counter images of C, and several new arcs. We keep the ID names of systems A, B, and C, and re-label the new systems below.

Source 4.1 is approximately the suggested candidate D in \cite{ofek08}. \cite{bayliss11}  obtained a spectrum of this arc, but was unable to secure spectroscopic redshift for it. We identify a candidate counter image for it in the south part of the field, which was labeled H2 in \cite{ofek08}.

Source 5 is identified as a radial pair southwest of the BCG, 5.1 and 5.2, with a counter image in the north part of the field. This system has unique colors and diffuse surface brightness that helps in its identification. 

Source 6, with six images, is bright in the WFC3-IR bands and virtually absent from the optical data. Since we also detect this source in our Spitzer imaging, we conclude that it is likely a $z\sim1-3$ dusty galaxy, which is supported by the lens model. A cluster galaxy in the north part of the cluster core complicates the lensing configuration and is responsible for three of the images. 

We identify two candidate images for source 7, one of which is the arc labeled I in \cite{ofek08}. 

The lensing signal of this cluster is modeled with one halo for the cluster and halos for the cluster galaxies. We let the parameters of the galaxy near the three north images of system 6 to be solved for by the lens model.  We note that to reproduce the lensing evidence this model requires external shear in addition to these components.   
The model is well constrained, and classified in category $A$.

Our Gemini/GMOS spectroscopy did not yield new redshifts for arc candidates in this field. We add 10 spectroscopic redshifts for cluster-member galaxies to the measurements of \cite{bayliss11} and update the velocity dispersion of this cluster (Table~\ref{tab.clusterz}). 
Finally, we note a galaxy that is likely undergoing ram pressure stripping, near the core of the cluster, at R.A, Decl. = [182.3517695, 26.68261135], labeled J. Its redshift, as well as that of a nearby galaxy, were measured by \cite{ofek08} as $\zspec=0.542$.

\subsubsection{SDSS J1329$+$2243}  
The main lensed system in this field has three images of a galaxy at $\zspec=2.04$, north of the cluster center. Spectroscopic confirmation of the three images was obtained with Gemini/GMOS-North \citep{bayliss14LOS}. 
From the same \multislit\ observations, we measure the spectroscopic redshifts of 
other sources in the field. 

The orientation of sources A and B, located south of the BCG, suggest that they may be multiple images of the same source. We rule out this scenario based on their different spectroscopic redshifts, $\zspec=0.710$ and $\zspec=0.964$, respectively, which also places them outside of the strong lensing region. Similarly, the spectroscopic redshifts of galaxies C and D confirm that they are single images.

We identify a few faint sources as multiple image candidates based on colors and morphology, as listed in Table~\ref{tab.arcstable} and plotted in Figure~\ref{fig.4}. We used all the knots in system~1 as constraints in the model, the two secure images of system 2, and the two secure radial images of system 4, the latter two systems with unknown redshift set as a free parameter. The lensing morphology of system 3 is complicated by several cluster galaxies in the core of the cluster. The five listed images have similar colors and surface brightnesses in all the bands in which they appear; however, we cannot definitively determine whether they are all images of the same source. The lens model is consistent with several images being produced at this region. Nevertheless, these images are not used as constraints in the model published here.  
Deeper imaging to secure the identification of system 3, and spectroscopic redshifts of sources 2, 4, 5 would strengthen the model and increase its fidelity. It is otherwise well-constrained, and we classify it as $A-$.

\subsubsection{SDSS J1336$-$0331}  
We identify two giant arcs in this field. Southeast of the BCG, we detect an elongated arc (source~1), with a counter image on the opposite side of the BCG. The giant arc is composed of at least two images,  with a cluster-member galaxy adding complexity to the lensing potential that results in partial arcs between this galaxy and the BCG. 
A bright lensed galaxy (source~2) appears north of the BCG, with significantly redder core. We identify and spectroscopically confirm a radial arc south of the BCG, making this a system of two images.  Two other faint galaxies are identified as multiply-imaged lensed systems and used as constraints. 
We observed SDSS~J1336$-$0331 with the Magellan Clay 6.5 telescope using the LDSS3 instrument. Long slit spectroscopy of the two bright components of source~1 south of the BCG confirm their spectroscopic redshift at $\zspec=0.96$.  IMACS \multislit\ mask spectroscopy observed with the Magellan 6.5m Baade telescope confirms the radial arc south of the BCG (labeled 2.2 in Figure~\ref{fig.4}) as a counter image of 2.1, the giant arc to the north, at $\zspec=1.47$.
The same spectrum reveals multiple emission lines from the cluster redshift, likely from the galaxies at the core of the cluster. This spectrum, as well as a prominent dust lane in the light of the BCG, indicates significant star formation at the core of this cluster. 
The model is well constrained, and classified in category $A$.

\subsubsection{SDSS J1343$+$4155}  
A giant arc appears at the east side of the cluster core, with $\zspec=2.091$ \citep{diehl09}. The \hst\ resolution reveals some substructure in the giant arc, which helps with its interpretation as three merging partial images of the background galaxy which crosses a naked caustic cusp in the source plane. There are no counter images identified, as expected in this configuration. The discovery and spectroscopy of a $\zspec=4.994$ lensed \Lya\ emitting galaxy in the field of SDSS~J1343$+$4155 was reported by \cite{bayliss10}. A candidate counter image that was suggested by \cite{bayliss10} appears to not have the exact same color and morphology as the spectroscopically confirmed image. We therefore do not use it as a lensing constraint. The model does not predict multiple images for the \Lya\ emitter. 
We identify three faint images of a candidate third lensed galaxy. The improvement upon the model presented in \cite{bayliss10} includes revised constraints: we use the identified emission knots in the giant arc, remove the candidate counter image of the \Lya\ emitter, and add candidate 2 with free redshift. In addition, we account for contributions from individual cluster galaxies, which was not implemented in \cite{bayliss10}.
Since all the lensing constraints appear in one side of the cluster core, degeneracies between model parameters increase the uncertainties, especially in regions which have no lensing constraints (i.e., the southwest part of the cluster, where the \Lya\ source appears).
We therefore classify it in category $B$.
 
\subsubsection{SDSS J1420$+$3955}  
The appearance of two giant arcs in ground-based optical data led to the selection of this cluster as a strong lens.  The \hst\ imaging reveal several other lensed galaxies, resulting in a total of 10 lensed systems in this field. The lensing evidence indicates that SDSS~J1420$+$3955 is a complex structure.
\cite{bayliss11} report the spectroscopic redshifts of two giant arcs, labeled A and B.  
The lensing configuration of arc A, at  $\zspec=2.161$, appears to be affected by a cluster galaxy,
forming at least five images or partial images of the
source around the cluster member, with a sixth complete image
$25\farcs7$ north of it. The unique colors and morphology make this a
confident identification. Arc B is at $\zspec=3.066$; the \hst\ imaging reveal multiple emission knots, and enables the identification of a counter image, B.3.3. 

Newly identified lensed features include a red giant arc (source 3) with a counter image, bright in the WFC3/IR imaging, as well as other lensed galaxies, some of which form radial arcs. We identify two radial images of source 4. A group of three sources 5, 6, and 7, have four identified images each. Source 8 forms three images near one of the cluster galaxies near the core, and a fourth image northwest of the core of the cluster. None of these newly identified sources have spectroscopic redshifts measured. 

We also identify strong lensing evidence around two bright cluster galaxies located 53\as\ and 68\as\ west of the BCG. We use these lensing features to constrain the contribution from these structures to the overall lensing mass of  SDSS~J1420$+$3955.

Overall, the cluster is constrained by images from ten lensed galaxies, providing ample positional constraints. However, the lack of spectroscopic redshifts for all but two galaxies is a limitation. A spectroscopic redshift of either of sources 5, 6, 7 or 8 would place the cluster in category $A$; we currently classify it as $A-$. 

\subsubsection{SDSS J1439$+$1208}  
The main arc in this field appears as a merging pair (source~1), with redshift $\zspec=1.494$, measured from Magellan/FIRE spectroscopy.
A preliminary lens model predicts the location and morphology of a counter image (1.3) and radial arc, readily discovered in the \hst\ imaging (1.4), with robust confirmation by colors and morphology. We further discover another set of two images (2.1, 2.2) and a candidate radial image (2.4). The lens model predicts a fourth image between images 2.1 and 2.2. We detect a likely candidate counter image in that location (2.3). 

We pursued spectroscopic confirmation of the lensed candidates, using the Magellan Baade 6.5-m telescope with the IMACS camera. 
On 2013 March 17 we obtained longslit spectra of B1, 1.2, and 2.2.
On 2014 April 26 we obtained \multislit\ spectroscopy, targeting the candidate counter image of the main arc, 1.3;  arc 2.1, and the candidate arc B2. 
These observations resulted in spectroscopic redshift of 2.1, $\zspec=1.580$ from \OII\ line. The slit on 1.3 was unfortunately placed on a chip gap, and did not yield data.
A comparison between the spectra of B1 and B2 leads to the conclusion that they are not images of the same source. We observe a strong emission line in the spectrum of B1 that is not observed in B2, placing it at $\zspec=3.48$ if it is coming from \lya. A low-confidence emission line in the MOS spectrum of B1 suggests that it may be at $\zspec=1.53$, if this line is \OII.  

The model is well constrained, and classified in category $A$.

\subsubsection{SDSS J1456$+$5702}  
The \hst\ imaging of SDSS J1456$+$5702 reveals the details of an extended, low-surface brightness arc. While giant arcs are often observed to be long and thin, this one has an unusual width of $3\farcs4$. A counter image is detected close to the BCG. Due to its low surface brightness, this was the only main arc in GO-13003 that lacked spectroscopic redshift prior to its \hst\ observation. 
We secured a spectroscopic redshift for the giant arc, $\zspec=2.366$, from S~II, O~I, Si~II, and C~II lines in a Blue Channel spectrum obtained at MMT, on   2014 May 04 (PI: Bayliss).  

The lensing morphology of this system is primarily of three images: the giant arc in the south, a counter image north of the BCG, and a demagnified image near the core. The morphology of the giant arc is slightly complicated by cluster-member galaxies, causing some parts of the central region of the giant arc to be highly magnified and with added multiplicity. 

\cite{bayliss11} measured spectroscopic redshifts of three other galaxies in this field (Table~\ref{tab.arcstable}), most notably two galaxies at similar redshift, A1 and A2, on opposite sides of the cluster core. The \hst\ imaging indicate that these galaxies are not images of the same source; the lens model is also consistent with these galaxies being single images. 
The model is well-constrained and classified in category $A$.

\subsubsection{SDSS J1522$+$2535}  
In this field, we find one highly-magnified galaxy, with five multiple images. The galaxy is clearly resolved, 
and displays a bright core (brighter in the infrared bands) as well as spiral structure. The proximity to cluster galaxies
complicates the lensing configuration. In particular, a cluster galaxy is superimposed near the core of image 1.1 but does not significantly distort it. Images 1.4 and 1.5 are more strongly affected by the BCG and other bright galaxies near it. Nevertheless, due to its 
unique colors and morphology, all five images of the lensed galaxy are robustly identified. Moreover, all the images are spectroscopically confirmed by our Gemini/GMOS spectroscopy (Table~\ref{tab.spec}),  at $z_{spec}=1.7096$. 
Some of the lens model parameters of the three galaxies in close proximity to the lensed images were left as free parameters.
The model is well-constrained and classified in category $A$.

\subsubsection{SDSS J1527$+$0652}  
SDSS J1527$+$0652 was reported in \cite{koester10}, as a highly
magnified galaxy at $\zspec=2.76$. 
The bottom-right panel in Figure~\ref{fig.5} zooms in on the arc, which appears $14\farcs5$ south of the core of a galaxy cluster that is outside of the figure.   
The foreground cluster does not have one obvious central galaxy. The added
resolution of the \hst\ observations reveal that the lensing geometry
is locally dominated by the lensing potential of an elliptical galaxy,
boosted by the cluster potential. The arc is composed of three partial
images; counter images are neither detected nor predicted by the lens
model. 
The arc at $\zspec=2.76$ is the most obvious lensing feature in this field, however, it is not sufficient for constraining the mass distribution of the cluster itself. To constrain the cluster halo, we rely on several faint multiply-imaged lensed galaxies. These galaxies are not spectroscopically confirmed. The  lens model is consistent with the velocity dispersion measured from spectroscopy of 14 cluster galaxies (Bayliss et al., 2011a), $\sigma_v = 923 \pm 233$ \kms. The conversion between velocity dispersion and the PIEMD 
$\sigma_v$ parameter is given in \cite{eliasdottir07}. 

Bordoloi et al. (in preparation) observed this galaxy with the Keck Cosmic Web Imager (KCWI)  on UT 2017 June 21, with a field of view of $16\farcs5\times20\farcs4$, resulting in IFU spectroscopy of 
 the main highly magnified galaxy and nearby galaxies. The observations indicate that some of the emission that is evident in our \hst\ imaging
 comes from a line-of-sight interloper at $\zspec=2.543$. 
Magellan/MagE observations \citep{rigby18b} reveal several intervening absorption systems (Rigby J., et al., in preparation), including one at $\zspec=2.543$, for which an emission-line component is seen in the KCWI data, that explains diffuse emission seen in the \hst\ imaging. 
 Rigby et al. (in preparation) also find several other intervening absorption systems at lower redshifts. 
 The KCWI data also measure the redshift of several nearby galaxies, finding a small group at $z\simeq0.43$, including the galaxy nearest to the main arc. 
We therefore employ an iterative approach in modeling the lens. We first compute a lens model for the cluster as explained above; we then treat the cluster as providing cosmic shear, and proceed to model the galaxy near the main arc. 
The large number of observed emission clumps constrain the lens model, which is composed of the lens galaxy as well as contribution from external shear.  

We also note a fainter arc at R.A., Decl. = [231.92365, 6.862152778] caused by lensing by an elliptical galaxy close to the
edge of our \hst\ field of view.  This lensed arc was not used to constrain the lens model. 

The lensing constraints available are not sufficient for a robust model of the galaxy cluster; we classify the cluster model in category $C$. Nevertheless, like other models with similar lensing configuration (e.g., SDSSJ~1004$-$0103), this model provides a reliable description of the lensing configuration of the main arc in the field. For its high magnification uncertainty, we classify the lens model that is used for this lensed source in category $B$. 

\subsubsection{SDSS J1531$+$3414}  
We refer the reader to \cite{sharon15b} for detailed description
of the lensing analysis of SDSS~J1531$+$3414. We provide a short description here. 
In the field of this cluster we confirm three strongly lensed galaxies. Source 1 with five images, source 2 with five images, and source 3 with a tangential and radial arc. \cite{bayliss11} observed this field with \multiobject\ spectroscopy, and secured a redshift for source 1 at $\zspec=1.096$. For the other two sources, only lower limits were measured, $z>1.49$ for both. Our lensing analysis in \cite{sharon15b} confirms that galaxies A5, A6, B2, and C1 in \cite{bayliss11} are single images. 
The model is well-constrained and classified in category $A$.
 
\subsubsection{SDSS J1604$+$2244}  
This cluster was identified as a strong lens based on a bright elongated arc, labeled A, at $\zspec= 1.184$, which is located $\sim18$'' west of the two bright galaxies that dominate the cluster center. 
We measured the redshift of this arc, as well as six cluster-member galaxies, using Gemini GMOS (program GN11AQ19). The cluster is at relatively low redshift compared to other clusters in this sample, $z_{cluster}=0.286$. 
The arc appears to be a single image. Although other arc-like features are observed in the \hst\ data, we are able to identify only one secure family of multiple images that can be used to constrain the lens model. The system, labeled as source~1, consists of five images with distinct morphology and color. We measure the photometric redshift of the three images that are not too blended in the light of the foreground galaxies, finding $\zphot = 0.53$. We compute a lens model with the photometric redshift used as prior, $0.5<z_1<0.8$. We require that the lens model does not produce counter images for singly-imaged galaxies. 
The model is classified in category $B$; a spectroscopic redshift of source 1 would improve its reliability.

\subsubsection{SDSS J1621$+$0607}  
Three arcs were identified by \cite{bayliss11}, who were able to measure spectroscopic redshifts for two of them. Source A, which has two images, is a \Lya\  emitter at $\zspec=4.131$. Source and B is at $\zspec=1.1778$. Source C was targeted for spectroscopy, but the observation did not yield a redshift. With the \hst\ data, we find that A.1 is buried in the light of a foreground galaxy; we do not identify a third image for this source. We find two counter images for arc B, and obtained spectroscopic confirmation with Magellan using LDSS3 longslit on 2013 May 2. 
From the lensing analysis, we conclude that the arc identified as B1 by Bayliss et al. (2011) is a merging pair of two partial images, whereas the other two images are complete. We re-label B1 as 2.1 and 2.2, and the complete images are 2.3 and 2.4. The region of highest surface brightness in 2.3 and 2.4 falls outside of the caustic, and thus is not seen in the merging pair 2.1--2.2. We use three distinct features in this galaxy as constraints --- the bright core of the galaxy in 2.3 and 2.4, a dark dust lane, and a blue region that appear clearly in all the images. There is no robust identification of a counter image for C; we therefore do not use it as a constraint in the lens model. Its elongation and curvature suggest that it is lensed. From the best-fit model, this arc could be a merging pair straddling the critical curve at $z\sim2.4$, with a counter image near R.A., Decl. = [245.3811204, 6.123205304]. 
We identify a few plausible candidate images within 2\as\ of this location. This interpretation is consistent with the conclusion of \cite{bayliss11}, that arc C is likely in the ``redshift desert'', based on the lack of emission lines in its Gemini/GMOS spectrum.
The lens model is consistent with the velocity dispersion that we measure from our IMACS MOS observation (Table~\ref{tab.clusterz}), supplemented with the redshifts reported by \cite{bayliss11}. 
The model is well-constrained and classified in category $A$.

\subsubsection{SDSS J1632$+$3500}  
This field displays two giant arcs, approximately 15\as\ east and west of the cluster core. Spectroscopic redshifts were obtained from Gemini North GMOS observations in 2012 Apr 15 for the two sources: source~1 at $\zspec=1.235$, and source~2 at $\zspec=2.265$. The spectrum of the bright region in the eastern giant arc indicates the possible presence of an AGN, inferred from \HeI\ emission. 
The same Gemini observations secure spectroscopic redshift of a star-forming galaxy that is likely a cluster member, projected to be near the cluster core,  with somewhat large velocity offset from the mean ($\zspec=0.454$ and $\zspec=0.466$, respectively), but within 2.5$\sigma$ of the projected velocity dispersion. Other background, foreground, and cluster members with spectroscopic redshift measurement are listed in tables~\ref{tab.arcstable} and \ref{tab.spec}.   
Our \hst\ imaging data confirm that the two giant arcs are extremely distorted lensed images, however, no counter-images are identified.  We used the extended shape of the giant arcs as constraints in the lens model. In both cases, the brightest part of the galaxy (i.e., the galaxy core) lies outside of the lensing caustic and thus are singly-imaged. Two other possible lensed systems are identified and their positions are used as constraints in the lens model; their redshifts are left as free parameters. 
The model is well-constrained, and labeled as $A-$. The lensing magnification of the giant arcs has high uncertainty due to its close proximity to the critical curve, where small variations in the model result in large change in magnification. 

\subsubsection{SDSS J1723$+$3411}  
The bright arc in this field is one of the brightest lensed galaxies known to date, at $\zspec=1.3294\pm0.0002$ 
\cite{kubo10}. The source is lensed into five images, two of which merge to form the bright arc, images 1.1 and 1.2. The third and forth images, 1.3 and 1.4, are easily identified, and the fifth is a demagnified image near the core of the BCG, 1.5. Since the BCG light drops out from our bluest \hst\ filter, the fifth image is is clearly identified $0\farcs3$ from the center of the galaxy.  
We identify four images of a likely galaxy pair, labeled as sources 2 and 3. 
A redshift for source 2 was measured from \hst\ grism data (GO-14230, PI: Rigby), $\zspec=2.165$  based on \OIII\ emission lines. We were unable to derive a redshift for source 3. The lens model published here assumes that source 3 is at the same redshift as source 2; we have computed models that leave this redshift as a free parameter, and these models are indistinguishable from the one with a fixed redshift for source 3. 
A  distorted red galaxy east of the images 2.2 and 3.2 is likely a single image.
The model is well-constrained, and labeled as $A$.

Interestingly, the \hst\ data reveal a dust lane in the light of the BCG, indicating possible recent star formation 
\citep[e.g.,][]{cooke16}.

\subsubsection{SDSS J2111$-$0114} 
Carrasco et al. 2016 obtained spectroscopy of 46 cluster members in this field from VLT.
In calculating the cluster redshift and velocity dispersion
(Table~\ref{tab.clusterz}), we add to these measurements six additional redshifts for cluster
members from \cite{bayliss11}.  
The spectroscopic redshift of the main arc, $\zspec=2.858$, is adopted from
\cite{bayliss11}. 
The strong lensing constraints in this field are the main arc and two other candidate systems, all of which are  in naked cusp configuration. Since all the lensing constraints  appear south of the BCG, this field is under-constrained on the north side. We therefore classify this model in category $B$. The lens model is in good agreement with the observed velocity dispersion.

\subsubsection{SDSS J2243$-$0935}  
The primary arc in this field is an elongated arc at $\zspec=2.09$ \citep{bayliss11}. \cite{rigby18b} report that the lensed source hosts a broad-line AGN, based on broad rest-frame UV emission lines. This interpretation is confirmed by the morphology of this galaxy in the \hst\ imaging, that shows a point source embedded in a spatially extended host (labeled 1.1 in Figure~\ref{fig.7}). 
This straight arc is located 52\as\ east of the core of the cluster, in the direction to a prominent group of cluster-member galaxies. This lensing morphology, showing an arc with minimal or no curvature, is typically found in cluster mergers or bimodal distribution when arcs form in the region between two dominant halos. We find that the morphology of this arc is also affected by a small cluster-member galaxy projected on the line of sight, resulting in multiple images of a few of the emission knots in the host galaxy. 

The  F390W imaging data reveal a faint point-source embedded in the light of the cluster galaxy, separated from its core, likely a counter image of the AGN nucleus (labeled 1.2). {One of the MagE slits in the observations of \cite{rigby18b} partially covers this source. A careful extraction and inspection of the two-dimensional spectrum at the position of the candidate counter image suggest faint emission from this region, however, the signal-to-noise is not sufficient to spectroscopically confirm it as a counter image of the nucleus.}
{The AGN (1.1) clearly appears as an X-ray point source in archival Chandra observations of this field, detected at 7.5 sigma above the background. The X-ray emission from the positions of the optical point source (1.2) is detected at 2.2 sigma above the background, confirming that it is likely a counter image of the AGN.}

We identify several other multiply-imaged lensed galaxies around the west (main) and east (secondary) halos of this cluster core that help constrain the lens model. However, spectroscopic observations with the IMACS instrument on the Magellan Baade telescope, targeting the lensed images, did not yield redshifts for these multiple image families. Source~3 appears close to a source that was observed by \citep{bayliss11} to be at $\zspec=2.092$, however, the color and morphology of this source are not consistent with the sources identified as multiply imaged, and we therefore conservatively leave the redshift of source~3 as a free parameter. 
Several instances of galaxy-galaxy lensing are identified within the \hst\ field of view, owing to the overall high projected mass density in this region of the sky. We note these identifications, as well as the lensing constraints, in Table~\ref{tab.spec}.

The lack of spectroscopic redshifts of multiply-imaged lensed galaxies leave this cluster under-constrained. The redshifts of all but source~1 are left as free parameters. However, to improve the accuracy of the lens model and avoid catastrophic solutions we place priors on the redshift parameter of the most prominent arcs in each subhalo, informed by their photometric redshifts.  
These are source \#2 near the east halo, with photometric redshifts measured from images 2.1 and 2.3, and source \#8 near the west halo, with photometric redshifts measured from images 8.1 and 8.3. We use the photometric redshift posterior distribution to inform our redshift priors, $2.90<z_2<3.10$ and $2.95<z_8<3.05$.

This cluster was independently identified via its X-ray signal by \cite{ebeling10} and is also known as MACS~J2243$-$09. \cite{schirmer11} use weak lensing to map the cosmic web around this cluster and report on the filamentary structure around it. They report a mass of $M_{200} = 1.31^{+0.25}_{-0.20}\times 10^{15} h^{-1}_{70}$\msun within $r_{200}=2.13^{+0.18}_{0.12} h^{-1}_{70}$~Mpc. Our strong lensing analysis estimates that the mass within 500 kpc is $M\sim7.5\times 10^{14}$\msun. Although highly uncertain due to the lack of spectroscopic redshifts, this mass estimate is consistent with the weak lensing measurement. We classify this model as $A-$. Its fidelity can be improved by securing spectroscopic redshifts of some of the lensed galaxies.

\section{Summary}\label{s.summary}
We present strong lens models for 37 galaxy clusters from the Sloan Giant Arcs Survey Large \hst\ program, GO-13003. 
The lens models are constrained by positions of multiply-imaged lensed galaxies that were uniquely identified in the \hst\ imaging. We measured spectroscopic redshifts for lensed galaxies from data obtained with Magellan, Gemini, APO, and \hst. The lensing constraints that were used in this work, including image positions and redshifts from this work and from the literature, are listed in the appendix. 

Magnification from strong lensing clusters of galaxies enables studies of the magnified sources behind them. The magnification boost increases the observed brightness of the background galaxies, leading to higher signal-to-noise per unit time of spectroscopic observations, allowing us to study galaxies that would otherwise require an unrealistic amount of observing time with current facilities. 
The lensing distortion increases the observed spatial resolution of the background source, which enables studies of star formation on spatial scales as small as a few tens of parsecs at cosmic noon. 

In order to take full advantage of the lensing magnification in analyzing the background galaxies, one needs to incorporate the information from lens models. Once the properties of the foreground lens are properly modeled, the derived lens model outputs are used in order to convert the observed measurements  of background sources to their intrinsic ones. The deflection maps are used to ray-trace the light from the image plane to the source plane, and thus reconstruct the undistorted image of the source galaxy. They are used to determine the morphology of the source and measure or constrain the spatial extent of substructure within the source.  The magnification, which may vary significantly along giant arc-like images in the image plane, linearly converts between lensed and unlensed luminosity, and properties that are derived from it (i.e., star formation rate, stellar mass). 

With this publication, we release the lens model outputs of each field. The availability of constraints in each field varies, and depends not only on the lensing strength of the foreground cluster, but also on the extent of followup observations, depth of the available data, spatial distribution of constraints, and in many cases, fortuitous alignment between the sources and the lens. A description of the models and an assessment of their reliability are given in this paper. 
The lensing outputs are specific to the redshifts in the system, but can be converted to any source redshift; the equations needed for this conversion can be found in \citet[][Section~4.3]{johnson14}.  We also provide a set of model outputs computed from the MCMC process, from which uncertainties can be estimated.

\begin{acknowledgements}
Support for \HST\ program GO-13003 was provided by NASA through a grant from the Space Telescope Science Institute, which is operated by the Association of Universities for Research in Astronomy, Inc., under NASA contract NAS 5--26555.         
Based on observations made with the NASA/ESA Hubble Space Telescope, obtained at the Space Telescope Science Institute, which is operated by the Association of Universities for Research in Astronomy, Inc., under NASA contract NAS 5-26555. These observations are associated with programs GO-13003, GO-14622, GO-14230, GO-14896.
Some of the data presented in this paper were obtained from the Multimission Archive at the Space Telescope Science Institute (MAST). These data are associated with programs 
GO-11974, GO-11100.
KS acknowledges support from the University of Michigan's President's Postdoctoral Fellowship.       
KEW gratefully acknowledge support by NASA through Hubble Fellowship
grant \#HF2-51368 awarded by the Space Telescope Science Institute, which is operated by the Association of Universities for Research in Astronomy, Inc., for NASA.
MBB acknowledge support by NASA through grant HST-GO-14896-01.
Based on observations obtained at the Gemini Observatory, which is operated by the 
Association of Universities for Research in Astronomy, Inc., under a cooperative agreement 
with the NSF on behalf of the Gemini partnership: the National Science Foundation 
(United States), the National Research Council (Canada), CONICYT (Chile), the Australian 
Research Council (Australia), Minist\'{e}rio da Ci\^{e}ncia, Tecnologia e Inova\c{c}\~{a}o 
(Brazil) and Ministerio de Ciencia, Tecnolog\'{i}a e Innovaci\'{o}n
Productiva (Argentina).
This paper includes data gathered with the 6.5 meter Magellan Telescopes located at Las Campanas Observatory, Chile.
Observations reported here were obtained at the MMT Observatory, a joint facility of the University of Arizona and the Smithsonian Institution.
This work makes use of the Matlab Astronomy Package \citep{ofek14}.  
\end{acknowledgements}

 
\newpage
\appendix
\section{Lensing Constraints}
\startlongtable
\begin{deluxetable*}{llllll} 
\tablecolumns{6} 
\tablecaption{List of lensing constraints\label{tab.arcstable}} 
\tablehead{\colhead{ID} &
            \colhead{R.A.}    & 
            \colhead{Decl.}    & 
            \colhead{$z_{spec}$}     & 
            \colhead{$z_{spec}$}       & 
            \colhead{Notes}       \\[-8pt]
            \colhead{} &
            \colhead{J2000}     & 
            \colhead{J2000}    & 
            \colhead{}       & 
            \colhead{Reference}       & 
            \colhead{}             }
\startdata 
\sidehead {{SDSS~J0004$-$0103}}\hline\\[-5pt]
1.1 & 1.215462 & -1.055917 & 1.681 &Rigby+18 & \\ 
1.2 & 1.215350 & -1.055859 &  & & \\ 
1.3 & 1.214755 & -1.055367 &  & & \\ 
1a.1 & 1.215583 & -1.055996 &  & & \\ 
1a.2 & 1.215140 & -1.055762 &  & & \\ 
1a.3 & 1.214867 & -1.055521 &  & & \\ 
1b.1 & 1.215821 & -1.055905 &  & & \\ 
1b.2 & 1.215276 & -1.055667 &  & & \\ 
1b.3 & 1.214801 & -1.055190 &  & & \\ 
1c.1 & 1.214795 & -1.055314 &  & & \\ 
1c.2 & 1.215224 & -1.055719 &  & & \\ 
1c.3 & 1.215689 & -1.055942 &  & & \\ 
1d.1 & 1.215385 & -1.055875 &  & & \\ 
1d.2 & 1.215428 & -1.055900 &  & & \\ 
 \hline\hline
\sidehead{SDSS J0108$+$0624} \hline\\[-5pt]
1.1 & 17.174978 & 6.412078 & 1.91  &Rigby+18 & \\ 
1.2 & 17.176051 & 6.412092 &  & & \\ 
1a.1 & 17.175900 & 6.412684 & & & \\ 
1a.2 & 17.175025 & 6.412060 &  & & \\ 
1b.1 & 17.175964 & 6.412493 &  & & \\ 
1b.2 & 17.174992 & 6.412063 &  & & \\ 
1c.1 & 17.174932 & 6.412091 &  & & \\ 
1c.2 & 17.176030 & 6.411886 &  & & \\ 
 \hline\hline
\sidehead{SDSS J0146$-$0929} \hline\\[-5pt]
1.1 & 26.736156 & -9.496396 & 1.9436 & This work, FIRE; Stark+13 & \\ 
1.2 & 26.730240 & -9.495975 &  & & \\ 
1.3 & 26.730774 & -9.495248 &  & & \\ 
1.4 & 26.733138 & -9.501209 &  & & \\ 
1.5 & 26.730610 & -9.499755 &  & & \\ 
11.1 & 26.736201 & -9.496491 &  & & \\ 
11.2 & 26.730257 & -9.495890 &  & & \\ 
11.3 & 26.730918 & -9.495014 &  & & \\ 
12.1 & 26.736238 & -9.496558 &  & & \\ 
12.2 & 26.730315 & -9.495800 &  & & \\ 
12.3 & 26.730955 & -9.494963 &  & & \\ 
12.4 & 26.733018 & -9.501201 &  & & \\ 
13.1 & 26.736386 & -9.496849 &  & & \\ 
13.2 & 26.730378 & -9.495644 &  & & \\ 
13.3 & 26.731003 & -9.494875 &  & & \\ 
14.1 & 26.736177 & -9.496564 &  & & \\ 
14.2 & 26.730148 & -9.496054 &  & & \\ 
\hline
2.1 & 26.733406 & -9.495280 & 2.3660 &This work, IMACS & \\ 
2.2 & 26.734766 & -9.502288 &  & & \\
2.3 & 26.733561 & -9.497364 & & & Candidate\\
21.1 & 26.733440 & -9.495278 &  & & \\ 
21.2 & 26.734715 & -9.502318 &  & & \\
\hline
E & 26.73471004 & -9.502603976 & 0.7962  &This work, IMACS & Single image, near 2.2\\ 
D & 26.72841422 &-9.493789352 &2.6623  &This work, IMACS & Single image\\ 
G & 26.72786667 &-9.498882957 & 0.4485  &This work, IMACS & Single image\\ \hline\hline
\sidehead{SDSS J0150$+$2725} \hline\\[-5pt]
1.1 & 27.505066 & 27.426420 & 1.0800 &This work, APO& 2012 Jan 20  \\ 
1.2 & 27.504752 & 27.426018 &  & & \\ 
1.3 & 27.503328 & 27.425350 &  & & \\ 
1.5 & 27.503476 & 27.426849 &  & & \\ 
11.1 & 27.505501 & 27.427167 &  & & \\ 
11.2 & 27.503379 & 27.426822 &  & & \\ 
\hline
2.1 & 27.509517 & 27.426914 & \nodata & & \\ 
2.2 & 27.508971 & 27.425031 &  & & \\ 
\hline\hline
\sidehead{SDSS J0333$-$0651} \hline\\[-5pt]
1.1 & 53.271663 & -6.856274 & 1.5700 &This work, FIRE &Possibly single image \\ 
1.2 & 53.270623 & -6.854172 &  & & \\ 
\hline
2.1 & 53.268675 & -6.862509 & \nodata & &Galaxy-galaxy lensing \\ 
2.2 & 53.268182 & -6.862460 &  & & \\ 
\hline
3.1 & 53.267530 & -6.854139 & \nodata & &Candidate \\ 
3.2 & 53.269571 & -6.854654 &  & &Candidate \\ 
\hline
4.1 & 53.265041 & -6.856527  & \nodata & &Candidate \\  
4.2 & 53.265125 & -6.857559  &  & &Candidate \\ 
4.3 & 53.266938 & -6.859791  &  & &Candidate \\  \hline\hline
\sidehead{SDSS J0851$+$3331} \hline\\[-5pt]
B.1 & 132.908460 & 33.517522 & 1.3454 & Bayliss+11& \\ 
B.2 & 132.908210 & 33.518601 &  & & \\ 
B.3 & 132.914270 & 33.523888 &  & & \\ 
B.4 & 132.912760 & 33.517752 &  & & Candidate radial arc\\ 
B1.1 & 132.908500 & 33.517410 &  & & \\ 
B1.2 & 132.908200 & 33.518789 &  & & \\ 
\hline
D.1   & 132.909040 & 33.521854 & 1.79 &GO14622, PI: Whitaker  & \\ 
D.2   & 132.908460 & 33.515032 &  & & \\ 
D.3   & 132.914680 & 33.517739 &  & & \\ 
D.4   & 132.914390 & 33.523319 &  & & \\ 
D.5   & 132.912310 & 33.518961 &  & & \\
D1.1 & 132.909056 &  33.521854&   & & \\
D1.2 & 132.908477 &  33.515017 &  & & \\
D1.3 & 132.914686 &  33.517705&   & & \\
D1.4 & 132.914439 &  33.523272&  & & \\
D1.5 & 132.912316 &  33.519003&    & & \\
\hline
E.1 & 132.915050 & 33.512978 & 1.88 &GO14622, PI: Whitaker & \\
E.2 & 132.915260 & 33.513197 &  & & \\ 
E.3 & 132.916770 & 33.513843 &  & & \\ 
E.4 & 132.915660 & 33.512927 &  & & \\
E.5 & 132.916314 & 33.513417 &  & & Candidate partial image\\
E1.1 & 132.915090 & 33.513014 &  & & \\ 
E1.2 & 132.915220 & 33.513155 &  & & \\ 
E1.3 & 132.916820 & 33.513869 &  & & \\ 
\hline
A    & 132.904550 & 33.520420 & 1.6926 &Bayliss+11 & Single image \\
\hline
C    & 132.917233 & 33.522550 & 1.2539  &Bayliss+11 &Single image \\ \hline\hline
\sidehead{SDSS J0915$+$3826} \hline\\[-5pt]
A.1 & 138.908560 & 38.450062 & 1.5010 & Bayliss+10,11  & \\ 
A.2 & 138.909080 & 38.451231 &  & & \\ 
A.3 & 138.910690 & 38.452942 &  & & \\ 
Aa.1 & 138.910780 & 38.452817 &  & & \\ 
Ab.2 & 138.908530 & 38.449497 &  & & \\ 
Ab.3 & 138.910140 & 38.452420 &  & & \\ 
Ac.1 & 138.908590 & 38.448827 &  & & \\ 
Ac.2 & 138.910780 & 38.452575 &  & & \\ 
Ac.3 & 138.909643 & 38.451724 &  & & \\  
\hline
B.1 & 138.920960 & 38.448191 & 5.2000 & Bayliss+10,11  & \\ 
B.2 & 138.921280 & 38.449322 &  & & \\ 
B.3 & 138.917120 & 38.443534 &  & & \\ 
 \hline\hline
\sidehead{SDSS J0928$+$2031} \hline\\[-5pt]
1 & 142.018470 & 20.532302 & 1.8910 & This work, FIRE & \\ 
\hline
2.1 & 142.019460 & 20.527312 & 0.8555 & This work, GMOS& \\ 
2.2 & 142.018810 & 20.527153 &  & & \\ 
2.3 & 142.021730 & 20.529145 &  & & \\ 
\hline
3.1 & 142.021720 & 20.528090 & 1.3327 & This work, GMOS& \\ 
3.2 & 142.021500 & 20.527861 &  & & \\ 
3.3 & 142.017040 & 20.526253 &  & & \\ 
\hline
4.1 & 142.023490 & 20.526590 & \nodata & &Candidate \\ 
4.2 & 142.023160 & 20.526368 &  & & \\ 
4.3 & 142.020190 & 20.524736 &  & & \\ 
\hline
5.1 & 142.027803 & 20.519392 &\nodata  & &Candidate radial arc near south halo \\ 
\hline
B & 142.025835 & 20.522019 & 0.8563 &This work, GMOS  &Single distorted image near south core\\ 
\hline
C & 142.014117 & 20.533593 & 1.8806 &This work, FIRE & Single image \\ 
\hline
D & 142.023744 & 20.513353 & 1.9020 &This work, FIRE & Single image\\ 
\hline
E & 142.024390 & 20.519945 & 0.8565 &This work, GMOS & Single image\\ 
\hline\hline
\sidehead{SDSS J0952$+$3434} \hline\\[-5pt]
1.1 & 148.167560 & 34.577362 & 2.1900 &Kubo+10 & \\ 
1.2 & 148.168350 & 34.577527 &  & & \\ 
\hline
2.1 & 148.164870 & 34.576523 & \nodata & & \\ 
2.2 & 148.170630 & 34.578348 &  & & \\ 
2.3 & 148.171320 & 34.582908 &  & & \\ 
\hline
3.1 & 148.166930 & 34.578496 & \nodata & & \\ 
3.2 & 148.166430 & 34.578438 &  & & \\ 
\hline
4.1 & 148.168820 & 34.581285 & \nodata & & \\ 
4.2 & 148.168810 & 34.581105 &  & & \\ 
\hline\hline
\sidehead{SDSS J0957$+$0509} \hline\\[-5pt]
1.1 & 149.411832 & 5.156988 & 1.8210 &Bayliss+11 & \\ 
1.2 & 149.411235 & 5.157918 &  & & \\ 
1.3 & 149.410998 & 5.158717 &  & & \\ 
1.4 & 149.411013 & 5.159312 &  & & \\ 
1.5 & 149.411191 & 5.158897 &  & & \\ 
1.6 & 149.410982 & 5.158763 &  & & \\ 
\hline
2.1 & 149.412957 & 5.160956 & \nodata & & \\ 
2.2 & 149.414378 & 5.159153 &  & & \\ 
2.3 & 149.414307 & 5.156947 &  & & \\
2.4 & 149.412930 & 5.158728 &  & & \\
\hline 
3.1 & 149.412476 & 5.160714 & \nodata  & & \\ 
3.2 & 149.413997 & 5.159112 &  & & \\ 
3.3 & 149.413951 & 5.156746 &  & & \\ 
3.4 & 149.412504 & 5.158557 &  & & \\ 
 \hline\hline
\sidehead{SDSS J1002$+$2031} \hline\\[-5pt]
1.1 & 150.607950 & 20.516451 & 0.985 &This work, GMOS & Giant arc\\ 
1.2 & 150.610894 & 20.509550 & \nodata & &Candidate counter image \\ 
\hline
A & 150.607010 & 20.518948 & 1.270 &This work, GMOS & Single image\\ 
B & 150.619170  & 20.523185 &  1.169 &This work, GMOS & Single image\\ 
C & 150.619715  & 20.517718 &  0.734 &This work, GMOS & Single image\\ 
D & 150.596254  & 20.514684 &  0.654 &This work, GMOS & Single image\\ 

 \hline\hline
\sidehead{SDSS J1038$+$4849} \hline\\[-5pt]
Aa.1 & 159.676917 & 48.825040 & 2.198 & Bayliss+11 & A in Bayliss+11 \\ 
Aa.2 & 159.675061 & 48.823981 &  & & \\ 
Aa.3 & 159.674008 & 48.821871 &  & & \\ 
Aa.4 & 159.682485 & 48.821117 &  & & \\ 
Ab.1 & 159.676659 & 48.824781 &  & & \\ 
Ab.2 & 159.676111 & 48.824502 &  & & \\ 
Ab.3 & 159.674327 & 48.820902 &  & & \\ 
Ab.4 & 159.682786 & 48.821117 &  & & \\ 
\hline
Ba.1 & 159.683319 & 48.822281 & 0.9652 & Bayliss+11 & B in Bayliss+11 \\ 
Ba.2 & 159.683427 & 48.821341 &  & & \\ 
Ba.3 & 159.682014 & 48.820000 &  & & \\ 
Bb.1 & 159.683632 & 48.821441 &  & & \\ 
Bb.2 & 159.683631 & 48.821830 &  & & \\ 
Bb.3 & 159.682515 & 48.820109 &  & & \\ 
\hline
C.1 & 159.676358 & 48.820663 & 2.7830 & Bayliss+11 & C in Bayliss+11 \\ 
C.2 & 159.677509 & 48.819994 &  & & \\ 
C.3 & 159.679320 & 48.819666 &  & & \\ 
C.4 & 159.682720 & 48.825398 &  & & \\ 
Ca.1 & 159.676575 & 48.820271 & & & \\ 
Ca.2 & 159.675840 & 48.820938 &  & & \\ 
Ca.4 & 159.682460 & 48.825365 &  & & \\ 
Cb.1 & 159.676223 & 48.820798 &  & & \\ 
Cb.2 & 159.678027 & 48.819816 &  & & \\ 
Cb.3 & 159.679071 & 48.819662 &  & & \\ 
\hline
E.1 & 159.676264 & 48.821019 & \nodata & & \\ 
E.2 & 159.683006 & 48.825364 &  & & \\ 
\hline
Fa.1 & 159.680013 & 48.820036 & \nodata & & \\ 
Fa.2 & 159.682045 & 48.825450 &  & & \\ 
Fb.1 & 159.680118 & 48.820050 & \nodata & & \\ 
Fb.2 & 159.681968 & 48.825473 &  & & \\ 
\hline
D   & 159.678666 & 48.818142 & 0.8020 & Bayliss+11 & Single image \\ 
 \hline\hline
\sidehead{SDSS J1050$+$0017} \hline\\[-5pt]
A.1 & 162.668160 & 0.280149 & 2.4040 &Bayliss+14 & \\ 
A.2 & 162.666250 & 0.286671 &  & & \\ 
\hline
B.1 & 162.663000 & 0.282145 & \nodata & & \\ 
B.2 & 162.672260 & 0.285115 &  & & \\ 
B.3 & 162.665830 & 0.289216 &  & & \\ 
B.4 & 162.664060 & 0.281483 &  & &   \\ 
\hline
D.1 & 162.659930 & 0.287493 &  4.8700 & Bayliss+14 & \\ 
D.2 & 162.665880 & 0.291141 &  & & \\ 
D.3 & 162.672950 & 0.288075 &  & & \\ 
D.4 & 162.666960 & 0.280862 &  & & \\ 
\hline
C.1 & 162.664160 & 0.291329 & 3.6250 & Bayliss+14 & \\ 
C.2 & 162.665070 & 0.291628 &  & & \\ 
C.3 & 162.672400 & 0.289762 &  & & \\ 
C.4 & 162.666086 & 0.281812 &  & & \\
\hline
E.1 & 162.669020 & 0.283582 &\nodata  & & \\ 
E.2 & 162.660530 & 0.287685 &  & & \\ 
\hline
F.1 & 162.667300 & 0.285000  &\nodata  & & \\ 
F.2 & 162.660050 & 0.291583  & & & \\ 
 \hline\hline
\sidehead{SDSS J1055$+$5547} \hline\\[-5pt]
A.1 & 163.764788 & 55.802890 & 1.2500 &Bayliss+11 & \\ 
A.2 & 163.769380 & 55.802066 &  & & \\ 
A.3 & 163.770840 & 55.802015 &  &&\\ 
A.4 & 163.775825 & 55.802779 &  & & \\ 
\hline
B.1 & 163.772724 & 55.802923 & 0.9360 &Bayliss+11 & \\ 
B.2 & 163.771971 & 55.802880 &  & & \\ 
B.3 & 163.769481 & 55.802571 &  & & \\ 
Ba.1 & 163.772510 & 55.802923 &  & & \\ 
Ba.2 & 163.772229 & 55.802893 &  & & \\ 
Ba.3 & 163.769323 & 55.802577 &  & & \\ 
\hline
D.1 & 163.761436 & 55.809967 &\nodata   & &  D in Bayliss+11 \\ 
D.2 & 163.766349 & 55.811872 & & &\\ 
\hline
5.1 & 163.777241 & 55.811158 & \nodata & &\\ 
5.2 & 163.776207 & 55.811533 &  & & \\ 
\hline
6.1 & 163.775678 & 55.799939 &  & & \\ 
6.2 & 163.771550 & 55.799350 &  & & \\ 
\hline
C   & 163.776906 & 55.808742 & 0.777 & Bayliss+11 & Single image\\\hline\hline
\sidehead{SDSS J1110$+$6459} \hline\\[-5pt]
A.1 & 167.581480 & 64.999411 & 2.4812 & Johnson+17, Stark+13 & \\ 
A.2 & 167.583170 & 64.997795 &  & & \\ 
A.3 & 167.582980 & 64.995273 &  & & \\ 
\hline
B.1 & 167.575130 & 64.999798 & \nodata & & \\ 
B.2 & 167.580210 & 64.997948 & & & \\ 
B.3 & 167.578800 & 64.993072 &  & & \\ 
B.4 & 167.571270 & 64.996332 &  & & \\ 
B.5 & 167.573070 & 64.996565 &  & & \\ 
\hline
C.1 & 167.568090 & 64.998147 & \nodata & & \\ 
C.2 & 167.567230 & 64.996781 & & & \\ 
C.3 & 167.573630 & 64.992252 &  & & Candidate \\ 
\hline
4.1 & 167.582030 & 64.999212 & \nodata & & \\ 
4.2 & 167.583280 & 64.997903 &  & & \\ 
4.3 & 167.583280 & 64.995456 &  & & \\ 
\hline
5.1 & 167.581910 & 64.999308 & \nodata & & \\ 
5.2 & 167.583400 & 64.997748 &  & & \\ 
5.3 & 167.583360 & 64.995603 &  & & \\ 
\hline
6.1 & 167.581760 & 64.999396 & \nodata & & \\ 
6.2 & 167.583470 & 64.997575 &  & & \\ 
6.3 & 167.583400 & 64.995701 &  & & \\
\hline
J     & 167.577020 & 64.999612 & 0.6447 2 & Johnson+17 & ``Jellyfish'' galaxy\\ \hline\hline
\sidehead{SDSS J1115$+$1645} \hline\\[-5pt]
1.1 & 168.767840 & 16.758996 & 1.7170 &Stark+13 & \\ 
1.2 & 168.769390 & 16.758679 &  & & \\ 
1.3 & 168.767450 & 16.759129 &  & & \\ 
\hline
2.1 & 168.768650 & 16.757464 & 3.4630 &Bayliss+12 & \\ 
2.2 & 168.769810 & 16.757478 &  & & \\ 
\hline\hline
\sidehead{SDSS J1138$+$2754} \hline\\[-5pt]
A.1 & 174.539460 & 27.912498 & 1.3340 &Bayliss+11 & \\ 
A.2 & 174.536280 & 27.912373 &  & & \\ 
A.3 & 174.533020 & 27.910710 &  & & \\ 
\hline
B.1 & 174.538084 & 27.910779 & 0.9090 & & \\ 
B.2 & 174.537286 & 27.910748 &  & & \\ 
B.3 & 174.534447 & 27.909638 &  & & \\ 
\hline
C.1 & 174.536764 & 27.914191 & 1.4550 &Bayliss+11 &Single image \\ 
\hline
D.1 & 174.542160 & 27.903572 & \nodata &This work &$z_{phot} = 3.2$ \\ 
D.2 & 174.540960 & 27.903079 & &&\\ 
D.3 & 174.532440 & 27.902590 &  & & \\ 
\hline
E.1 & 174.540057 & 27.912498 & \nodata & & \\ 
E.2 & 174.538768 & 27.912722 &  & & \\ 
E.3 & 174.529624 & 27.909298 &  & & Candidate\\ 
 \hline\hline
\sidehead{SDSS J1152$+$0930} \hline\\[-5pt]
1.1 & 178.197480 & 9.508144 & 2.2400 & This work, IMACS& D1,D2 in Bayliss+11\\ 
1.2 & 178.196754 & 9.508056 &  & & \\ 
1.3 & 178.196378 & 9.507894 &  & & \\ 
1.4 & 178.194424 & 9.506574 &  & & \\ 
\hline
2.1 & 178.200403 & 9.502987 & \nodata & &C1 in Bayliss+11 \\ 
2.2 & 178.199381 & 9.501804 &  & & \\ 
2.3 & 178.198991 & 9.501391 &  & & \\ 
2.4 & 178.197409 & 9.500721 &  & & \\ 
\hline
3 & 178.195447 & 9.504058 & 0.8945 &This work, IMACS; Bayliss+11 &Single image; A2 in Bayliss+11 \\ 
\hline
4 & 178.200080 & 9.502489 & 0.893 &This work, IMACS; Bayliss+11  &Single image; A3 in Bayliss+11  \\ 
\hline
5 & 178.195037 & 9.501626 & 0.893 &Bayliss+11             &Single image; A1 in Bayliss+11\\ 
\hline
6 & 178.195959 & 9.498743 & 0.1520 &This work, IMACS &Foreground  \\ 
\hline
7 & 178.196645 & 9.498887 & 0.2790 &This work, IMACS &Foreground  \\ 
\hline
\hline\hline
\sidehead{SDSS J1152$+$3313} \hline\\[-5pt]
A.1 & 177.998849 & 33.227292 & 2.4910 &Bayliss+11 & \\ 
A.2 & 178.000309 & 33.226344 &  & & \\ 
A.3 & 177.999189 & 33.230220 &  & & \\ 
A.4 & 178.002384 & 33.228415 &  & & \\ 
A.5 & 178.003857 & 33.228260 &  & & \\ 
A.6 & 178.000963 & 33.228431 &  & & \\ 
A1.1 & 177.998621 & 33.227558 &  & & \\ 
A1.2 & 178.000112 & 33.226334 &  & & \\ 
A1.3 & 177.998987 & 33.230023 &  & & \\ 
A1.4 & 178.002341 & 33.228479 &  & & \\ 
A1.5 & 178.003788 & 33.228307 &  & & \\ 
A2.1 & 177.998551 & 33.227699 &  & & \\ 
A2.2 & 178.000171 & 33.226303 &  & & \\ 
A2.3 & 177.998930 & 33.229947 &  & & \\ 
A3.2 & 178.000151 & 33.226193 &  & & \\ 
A3.3 & 177.998534 & 33.229282 &  & & \\ 
A3.4 & 178.002320 & 33.228598 &  & & \\ 
A3.6 & 178.001038 & 33.228632 &  & & \\ 
A4.2 & 177.999190 & 33.226691 &  & & \\ 
A4.3 & 177.998617 & 33.229571 &  & & \\ 
A4.4 & 178.002181 & 33.228533 &  & & \\ 
A4.5 & 178.003671 & 33.228277 &  & & \\ 
A4.6 & 178.001066 & 33.228574 &  & & \\ 
\hline
B.1 & 178.004190 & 33.230054 & 4.1422 &Bayliss+11 & \\ 
B.2 & 178.004590 & 33.229554 &  & & \\ 
B.3 & 178.003560 & 33.225932 &  & & \\ 
B.4 & 177.998780 & 33.228408 &  & & \\ 
\hline
C.1 & 178.000480 & 33.229159 & \nodata & & \\ 
C.2 & 178.003140 & 33.225349 &  & & \\ 
 \hline\hline
\sidehead{SDSS J1156$+$1911} \hline\\[-5pt]
1 & 179.022440 & 19.185323 & 1.5430 & Stark+13& \\ 
\hline\hline
\sidehead{SDSS~J1207$+$5254} \hline\\[-5pt]
1.1 & 181.903370 & 52.918765 &  1.9260 &Kubo+10 & \\
1.2 & 181.904630 & 52.918406 &  & & \\ 
1.3 & 181.900360 & 52.919571 &  & & \\ 
1a.1 & 181.904920 & 52.918342 &  & & \\
1a.2 & 181.902730 & 52.918963 &  & & \\ 
1a.3 & 181.900840 & 52.919494 &  & & \\ 
1b.1 & 181.903940 & 52.918579 &  & & \\ 
1b.2 & 181.904210 & 52.918498 &  & & \\ 
1c.1 & 181.902350 & 52.918905 &  & & \\ 
1c.2 & 181.901000 & 52.919288 &  & & \\ 
\hline
2.1 & 181.896690 & 52.914499 & \nodata & & \\ 
2.2 & 181.898240 & 52.914044 &  & & \\ 
2.3 & 181.897894 & 52.914102 & & &Candidate \\ 
\hline
SN cand& 181.8982  & 52.918553  & & &Location of SN candidate, \\ &&&&&discovered in archival GMOS imaging \\ \hline\hline
\sidehead{SDSS J1209$+$2640} \hline\\[-5pt]
A.1 & 182.350930 & 26.681689 & 1.0180 & Ofek+08 & \\ 
A.2 & 182.351500 & 26.681164 &  & & \\ 
A.3 & 182.352090 & 26.679416 &  & & \\ 
\hline
B   & 182.35032  & 26.680163 & 0.879  & Bayliss+11 & \\ 
\hline
C.1 & 182.342960 & 26.684707 & 3.9480 & Bayliss+11 & \\ 
C.2 & 182.341400 & 26.682441 &  & & \\ 
C.3 & 182.355790 & 26.683225 &  & & \\ 
C.4 & 182.347856 & 26.670245 &  & & \\ 
\hline
4.1 & 182.341782 & 26.678621 & \nodata & & \\ 
4.2 & 182.345070 & 26.673153 &  & & Candidate\\ 
\hline
5.1 & 182.347890 & 26.677651 & \nodata & & \\ 
5.2 & 182.347220 & 26.676739 &  & & \\ 
5.3 & 182.349667 & 26.687590 &  & & \\ 
\hline
6.1 & 182.342125 & 26.678861 &\nodata  & & IR (dusty)\\ 
6.2 & 182.352617 & 26.685458 &  & & \\ 
6.3 & 182.352073 & 26.672528 &  & & \\ 
6.4 & 182.346474 & 26.685323 &  & & \\ 
6.5 & 182.345869 & 26.685350 &  & & \\ 
6.6 & 182.346284 & 26.685887 &  & & \\ 
\hline
7a.1 & 182.354066 & 26.683616 &\nodata  & & \\ 
7a.2 & 182.347138 & 26.670215 &  & & Candidate \\ 
7b.1 & 182.353591 & 26.684219 &  & &\\ 
7b.2 & 182.347467 & 26.670170 &  & & Candidate\\ 
\hline
J       & 182.351763 & 26.682600 & 0.542 &  Ofek+08 & ``Jellyfish'' galaxy \\  \hline\hline
\sidehead{SDSS J1329$+$2243} \hline\\[-5pt]
1.1 & 202.392260 & 22.723947 & 2.0400 &Bayliss+14 &  Knots in source 1  \\ 
1.2 & 202.391340 & 22.723539 &  & & \\ 
1.3 & 202.397396 & 22.722757 &  & & \\ 
1b.1 & 202.391650 & 22.723738 &  & & \\ 
1b.2 & 202.391889 & 22.723873 &  & & \\ 
1b.3 & 202.397483 & 22.722747 &  & & \\ 
1c.1 & 202.391092 & 22.723370 &  & & \\ 
1c.2 & 202.392442 & 22.723829 &  & & \\ 
1c.3 & 202.397387 & 22.722657 &  & & \\ 
1d.1 & 202.390930 & 22.723279 &  & & \\ 
1d.2 & 202.392627 & 22.723718 &  & & \\ 
1d.3 & 202.397329 & 22.722648 &  & & \\ 
1e.2 & 202.392077 & 22.723914 &  & & \\ 
1e.1 & 202.391450 & 22.723595 &  & & \\ 
1e.3 & 202.397453 & 22.722719 &  & & \\ 
\hline
2.1 & 202.395249 & 22.724825 & \nodata & & Tangential arc, candidate\\ 
2.2 & 202.396536 & 22.724304 &  & & \\ 
\hline
3.1 & 202.394619 & 22.721897 & \nodata  & & Candidate system near radial critical curve\\ 
3.2 & 202.394611 & 22.721712 &  & & \\ 
3.3 & 202.394645 & 22.721121 &  & & \\ 
3.4 & 202.394589 & 22.720651 &  & & \\ 
3.5 & 202.395540 & 22.721882 &  & & \\ 
\hline
4.1 & 202.395765 & 22.721646 & \nodata & & Candidate \\ 
4.2 & 202.395232 & 22.721315 &  & & \\ 
\hline
5.1 & 202.394698 & 22.718009 & \nodata & & Tangential arc, candidate\\ 
5.2 & 202.393956 & 22.717865 &  & & \\ 
\hline
A & 202.395142 & 22.715818 & 0.7096 &This work, GMOS  & Single image\\ 
\hline
B & 202.396303 & 22.715868 & 0.9644 &This work, GMOS  & Single image\\ 
\hline
C & 202.396603 & 22.710030 & 1.1472 &This work, GMOS  & Single image\\ 
\hline
D & 202.388014 & 22.720156 & 0.2811 &This work, GMOS  & Single image\\ 
 \hline\hline
\sidehead{SDSS J1336$-$0331} \hline\\[-5pt]
1.1 & 204.002570 & -3.525374 & 0.9556 & This work, LDSS3 & \\ 
1.2 & 204.002340 & -3.525918 &  & &\\ 
1.3 & 204.000510 & -3.527313 &  & & \\ 
1.4 & 203.999560 & -3.523771 &  & & \\ 
1.5 & 204.002510 & -3.524433 &  & & \\ 
1.6 & 204.002520 & -3.524291 &  & & \\ 
\hline
2.1 & 203.999500 & -3.521916 & 1.4737 & This work, IMACS  & \\ 
2.2 & 204.000790 & -3.525640 &  & This work, IMACS & Spectroscopically confirmed radial arc\\ 
\hline
4.1 & 204.000320 & -3.527927 & \nodata & & \\ 
4.2 & 204.002540 & -3.526535 &  & & \\ 
4.3 & 203.998900 & -3.527469 &  & & \\ 
4.4 & 204.003330 & -3.523445 &  & & \\ 
4.5 & 203.999790 & -3.527670 &  & & \\ 
4.6 & 203.999120 & -3.523427 &  & & \\ 
\hline
a & 204.004370 & -3.527186 & 0.6139 & This work, IMACS & Possible redshift\\ 
b & 203.988837 & -3.528258 & 0.6685 & This work, IMACS & Single image\\
c & 203.995852 & -3.526103 & 0.9865 & This work, IMACS & Single image\\
d & 203.999791 & -3.520988 & 0.9613 & This work, IMACS & Single image\\
e & 204.005241 & -3.529637 & 1.1783 & This work, IMACS & Single image\\
f & 203.991477 & -3.528430 & 1.6015 & This work, IMACS & Single image\\
g & 203.995908 & -3.527730 & 0.1206 & This work, IMACS & Foreground\\
s & 203.998599 & -3.526622 & 0.00     & This work, IMACS & Star\\
 \hline\hline
\sidehead{SDSS J1343$+$4155} \hline\\[-5pt]
A.1 & 205.891790 & 41.917796 &2.0910 &Bayliss+11, Diehl+09 & \\ 
A.2 & 205.891220 & 41.919003 &  & & \\ 
A.3 & 205.889630 & 41.920730 &  & & \\ 
Aa.1 & 205.889900 & 41.920332 & & &\\ 
Aa.2 & 205.890240 & 41.920017 &  & & \\ 
Aa.3 & 205.891980 & 41.916926 &  & & \\ 
\hline
B & 205.877930 & 41.915293 & 4.994 &Bayliss+11 & \\ 
\hline
2.1 & 205.887850 & 41.920262 & \nodata & & \\ 
2.2 & 205.889010 & 41.919262 &  & & \\ 
2.3 & 205.890780 & 41.914773 &  & & \\ 
\hline\hline
\sidehead{SDSS J1420$+$3955} \hline\\[-5pt]
A.1 & 215.161652 & 39.914131 & 2.1610 &Bayliss+11 &  \\ 
A.2 & 215.160340 & 39.915191 &  & & \\ 
A.3 & 215.159600 & 39.915621 &  & & \\ 
A.4 & 215.158280 & 39.921504 &  & & \\ 
A.5 & 215.160952 & 39.914990 &  & & \\ 
A.6 & 215.160582 & 39.914590 &  & & \\ 
\hline
B.1 & 215.156830 & 39.912824 & 3.0665 &Bayliss+11 & \\ 
B.2 & 215.155840 & 39.913817 &  & & \\ 
B.3 & 215.159930 & 39.909231 &  & & \\ 
Bb.1 & 215.155430 & 39.914457 &  & & \\ 
Bb.2 & 215.157200 & 39.912639 &  & & \\ 
Bb.3 & 215.159780 & 39.909520 &  & & \\ 
\hline
3.1 & 215.161360 & 39.916025 & \nodata & & \\ 
3.2 & 215.161250 & 39.916211 &  & & \\ 
3.3 & 215.159150 & 39.922727 &  & & \\ 
\hline
4.1 & 215.169607 & 39.920562 & \nodata  & & \\ 
4.2 & 215.169066 & 39.920456 &  & & \\ 
4b.1 & 215.169749 & 39.920578 &  & & \\ 
4b.2 & 215.168985 & 39.920438 &  & & \\ 
\hline
5.1 & 215.162934 & 39.917972 &\nodata  & & \\ 
5.2 & 215.171229 & 39.920415 &  & & \\ 
5.3 & 215.162289 & 39.924584 &  & & \\ 
5.4 & 215.166954 & 39.912265 & & & Candidate\\
\hline
6.1 & 215.162750 & 39.918301 &\nodata  & & \\ 
6.2 & 215.171055 & 39.920716 &  & & \\ 
6.3 & 215.162135 & 39.924116 &  & & \\ 
6.4 & 215.167052 & 39.911817 &  & & Candidate\\
\hline
7.1 & 215.161918 & 39.918919 &\nodata  & & \\ 
7.3 & 215.161482 & 39.923071 &  & & \\ 
7.2 & 215.170257 & 39.920849 & & & Candidate\\
7.4 & 215.166625 & 39.911040 & & & Candidate\\
\hline
8.1 & 215.170597 & 39.918805 &\nodata  & & \\ 
8.2 & 215.169685 & 39.919134 &  & & \\ 
8.3 & 215.169165 & 39.919150 &  & & \\ 
8.4 & 215.161679 & 39.926218 &  & & \\ 
\hline
20.1 & 215.148556 & 39.922919 &\nodata  & &Near NW halo1 \\ 
20.2 & 215.149334 & 39.923745 &  & & \\ 
20.3 & 215.148650 & 39.925028 &  & & \\ 
20.4 & 215.148153 & 39.923660 &  & & \\ 
\hline
30.1 & 215.142593 & 39.921047 &\nodata  & &Near NW halo2 \\ 
30.2 & 215.143058 & 39.921337 &  & & \\ 
30.3 & 215.142942 & 39.923436 &  & & \\ 
30.4 & 215.141806 & 39.921806 &  & & \\

 \hline\hline
\sidehead{SDSS J1439$+$1208} \hline\\[-5pt]
1.1 & 219.792668 & 12.138371 & 1.4940 &This work, FIRE &2013 Mar 01 \\ 
1.2 & 219.791888 & 12.137640 &  & & \\ 
1.3 & 219.793188 & 12.143946 &  &  & \\ 
1.4 & 219.789968 & 12.140544 &  & & \\ 
1a.1 & 219.792718 & 12.138600 &  & & \\ 
1a.2 & 219.791448 & 12.137471 &  & & \\ 
1a.3 & 219.793018 & 12.144041 &  & & \\ 
1a.4 & 219.789758 & 12.140576 &  & & \\ 
1b.1 & 219.792568 & 12.138661 &  & & \\ 
1b.2 & 219.791118 & 12.137460 &  & & \\ 
1b.3 & 219.792838 & 12.144199 &  & & \\ 
1b.4 & 219.789508 & 12.140580 &  & & \\ 
1c.1 & 219.791888 & 12.137807 &  & & \\ 
1c.2 & 219.792268 & 12.138138 &  & & \\ 
\hline
2.1 & 219.787058 & 12.137849 & 1.5800 &This work, IMACS & \\ 
2.2 & 219.789678 & 12.144543 &  & & \\ 
2.3 & 219.788568 & 12.142683 &  & & \\ 
2.4 & 219.791131 & 12.140390 &  & & Radial arc candidate\\
\hline
B1 & 219.7929577 &  12.13667192 &3.48  &This work, IMACS & \\
\hline
B2 & 219.7936562 &  12.13763932 &1.53  &This work, IMACS & Possible redshift, single image\\  \hline\hline
\sidehead{SDSS J1456$+$5702} \hline\\[-5pt]
B.1 & 224.004820 & 57.034582 & 2.3660 &This work, MMT & Knots in giant arc \\ 
B.2 & 224.002780 & 57.041188 &  & & \\ 
B.3 & 224.006960 & 57.034895 &  & & \\ 
Ba.1 & 224.008300 & 57.034957 &  & & \\ 
Ba.2 & 224.002640 & 57.040816 &  & & \\ 
Bb.1 & 224.009840 & 57.036207 &  & & \\ 
Bb.2 & 224.002320 & 57.041565 &  & & \\ 
Bc.1 & 224.000690 & 57.034937 &  & & \\ 
Bc.2 & 224.002700 & 57.041638 &  & & \\ 
Bd.1 & 223.999750 & 57.035006 &  & & \\ 
Bd.2 & 224.003040 & 57.041559 &  & & \\ 
Be.1 & 223.999940 & 57.034689 &  & & \\ 
Be.2 & 224.003390 & 57.041169 &  & & \\ 
Bf.1 & 224.010680 & 57.036451 &  & & \\ 
Bf.2 & 224.001640 & 57.041280 &  & & \\ 
\hline
A1 & 224.003905 & 57.042969 & 0.8331&Bayliss+11 & Single image\\ 
\hline
A2 & 224.003257 & 57.036603 & 0.8324 &Bayliss+11 &Single image \\ 
\hline
C & 224.022021 & 57.034751 & 1.1400 &Bayliss+11 &Single image \\ 
 \hline\hline
\sidehead{SDSS J1522$+$2535} \hline\\[-5pt]
1.1 & 230.721557 & 25.593341 & 1.7096 & This work, GMOS & Substructure in source 1\\ 
1.2 & 230.718683 & 25.592248 &  & & \\ 
1.3 & 230.717426 & 25.589195 &  & & \\ 
1.4 & 230.720750 & 25.590884 &  & & \\ 
1.5 & 230.721538 & 25.590773 &  & & \\ 
1a.1 & 230.721580 & 25.593523 &  & & \\ 
1a.2 & 230.718270 & 25.591888 &  & & \\ 
1a.3 & 230.717480 & 25.589447 &  & & \\ 
1a.4 & 230.720770 & 25.590667 &  & & \\ 
1b.1 & 230.721730 & 25.593499 &  & & \\ 
1b.2 & 230.718350 & 25.591777 &  & & \\ 
1b.3 & 230.717670 & 25.589403 &  & & \\ 
1b.4 & 230.720970 & 25.590605 &  & & \\ 
1c.1 & 230.721800 & 25.592924 &  & & \\ 
1c.2 & 230.719150 & 25.592023 &  & & \\ 
1c.3 & 230.717670 & 25.588817 &  & & \\ 
1d.1 & 230.721910 & 25.593010 &  & & \\ 
1d.2 & 230.718990 & 25.591908 &  & & \\ 
1d.3 & 230.717830 & 25.588910 &  & & \\ 
\hline
B &230.725525  & 25.5879675    & 1.221       & This work, GMOS &  Possible redshift, single image  \\ 
C &230.727051 &  25.58725537 &  0.973161  & This work, GMOS &  Single image  \\ 
D & 230.7270299& 25.57886337 &  0.487957  & This work, GMOS &Foreground galaxy  \\ 
E  &230.715947 &  25.57524935 &  0.379631  & This work, GMOS & Foreground galaxy  \\   \hline\hline
\sidehead{SDSS J1527$+$0652} \hline\\[-5pt]
1.1 & 231.938559 & 6.872048 & 2.7600 & Koester+10& \\ 
1.2 & 231.937925 & 6.872142 &  & & \\ 
1.3 & 231.937803 & 6.872197 &  & & \\ 
1b.1 & 231.938366 & 6.872013 &  & & \\ 
1b.2 & 231.938053 & 6.872052 &  & & \\ 
1c.1 & 231.938274 & 6.872022 &  & & \\ 
1c.2 & 231.938162 & 6.872030 &  & & \\ 
1c.3 & 231.937547 & 6.872303 &  & & \\ 
1d.1 & 231.938452 & 6.871979 &  & & \\ 
1d.2 & 231.937931 & 6.872059 &  & & \\ 
\hline
2.1 & 231.934215 & 6.877017 & \nodata &Candidate & \\ 
2.2 & 231.934355 & 6.876537 &  & & \\ 
2.3 & 231.934124 & 6.875461 &  & & \\ 
\hline 
3.1 & 231.939062 & 6.876848 &\nodata  &Candidate & \\ 
3.2 & 231.939158 & 6.875564 &  & & \\ 
3.3 & 231.939727 & 6.874700 &  & & \\ 
\hline 
4.1 & 231.944022 & 6.883274 &\nodata  &Candidate & \\ 
4.2 & 231.945432 & 6.879834 &  & & \\ 
4.3 & 231.947061 & 6.877657 &  & & \\ 
\hline 
5.1 & 231.951761 & 6.879191 &\nodata  &Candidate & \\ 
5.2 & 231.951803 & 6.882945 &  & & \\ 
5.3 & 231.947973 & 6.886712 &  & & \\ 
\hline 
6.1 & 231.942672 & 6.881764 &\nodata  &Candidate & \\ 
6.2 & 231.943085 & 6.880734 &  & & \\ 
 \hline\hline
\sidehead{SDSS J1531$+$3414} \hline\\[-5pt]
    1.1 & 232.79846  & 34.242414 &   1.096 &Bayliss+11 &  \\
    1.2 & 232.79147  & 34.241359 &     & & \\
    1.3 & 232.79290  & 34.237469 &     & & \\
    1.4 & 232.79312  & 34.237454 &     & & \\
    12.1 & 232.79883  & 34.241860 &     & & \\
    12.2 & 232.79204  & 34.241349 &     & & \\
    13.1 & 232.79863  & 34.241870 &     & & \\
    13.2 & 232.79214  & 34.241676 &     & & \\
    14.1 & 232.79861  & 34.242029 &     & & \\
    14.2 & 232.79197  & 34.241605 &     & & \\
    15.1 & 232.79850  & 34.242660 &    & &  \\
    15.2 & 232.79106  & 34.240519 &     & & \\
    16.1 & 232.79821  & 34.242889 &    & &  \\
    16.2 & 232.79379  & 34.237723 &     & & \\
    17.1 & 232.79725  & 34.242966 &     & & \\
    17.2 & 232.79311  & 34.242933 &     & & \\
    17.3 & 232.79574  & 34.238452 &     & & \\
    17.4 & 232.79018  & 34.239033 &     & & \\
\hline 
    2.1 & 232.78877  & 34.237850 &     & Bayliss+11 & limit: $z>1.49$  \\
    2.2 & 232.79710  & 34.238497 &     & & \\
    2.3 & 232.79809  & 34.243633 &     & & \\
    2.4 & 232.79337  & 34.244258 &     & & \\
    2a.1 & 232.78812  & 34.238060 &  &\\
    2a.2 & 232.79726  & 34.239392 &     & & \\
    2b.1 & 232.78891  & 34.237212 &     & & \\
    2b.2 & 232.79780  & 34.239135 &     & & \\
    2b.3 & 232.79825  & 34.243064 &     & & \\
\hline 
    3.1 & 232.79355  & 34.241752 &   &Bayliss+11 & limit:  $>1.49$   \\
    3.2 & 232.79224  & 34.235538 &     & & \\
\hline 
    D & 232.7891558 & 34.2359015  &   1.026 &Bayliss+11 & Single image  \\
\hline
    E & 232.7845467 & 34.2413218 &   1.3 &Bayliss+11 &  Single image\\ \hline\hline
\sidehead{SDSS J1604$+$2244} \hline\\[-5pt]
1.1 & 241.040867 & 22.739529 & \nodata & & \\ 
1.2 & 241.043363 & 22.739298 &  & & \\ 
1.3 & 241.044362 & 22.736277 &  & & \\ 
1.4 & 241.041803 & 22.737227 &  & & \\ 
1.5 & 241.042247 & 22.737956 &  & & \\ 
\hline
A & 241.047645 & 22.739371 & 1.1841 &This work, GMOS & Single image \\ 
\hline
B & 241.037062 & 22.741253 & 1.0333 &This work, GMOS  & Single image \\ 
\hline
c & 241.035457 & 22.731761 & 1.6370 &This work, GMOS  & \\ 
\hline
d & 241.047417 & 22.747950 & 0.4141 &This work, GMOS  & \\ 
\hline
e & 241.038404 & 22.731987 & 2.0540 &This work, GMOS  & Low-confidence redshift\\ 
\hline
f & 241.041255 & 22.731276 & 1.1843 &This work, GMOS & Low-confidence redshift \\ 
\hline
g & 241.041780 & 22.729624 & 1.7232 &This work, GMOS  & Low-confidence redshift \\ 
\hline
h & 241.035411 & 22.732763 & 1.4830 &This work, GMOS & Low-confidence redshift \\ 
\hline
Q & 241.050890 & 22.734746 & 2.1670 &SDSS DR12  & QSO \\ 
 \hline\hline
\sidehead{SDSS J1621$+$0607} \hline\\[-5pt]
A.1 & 245.389450 & 6.120616 & 4.1300 &Bayliss+11 & \Lya\ emitter; A1 in Bayliss+11\\ 
A.2 & 245.386300 & 6.118230 &  & &\Lya\ emitter;  A2 in Bayliss+11\\ 
\hline
2a.1 & 245.385990 & 6.121664 &1.1780  &Bayliss+11 & B1 in Bayliss+11\\ 
2a.2 & 245.386190 & 6.122059 &  & & \\ 
2a.3 & 245.385460 & 6.122681 & & & \\ 
2a.4 & 245.382610 & 6.121549 &  & & \\ 
2b.3 & 245.385240 & 6.122616 & 1.1780   &This work, LDSS3 & Falls outside of the caustic of 2a.1, 2a.2\\ 
2b.4 & 245.382740 & 6.121098 &  & & \\ 
2c.1 & 245.386100 & 6.121659 &  & & \\ 
2c.2 & 245.386260 & 6.121953 &  & & \\ 
2c.3 & 245.385360 & 6.122681 &  & & \\ 
2d.1 & 245.386200 & 6.121664 &  & & \\ 
2d.2 & 245.386310 & 6.121843 &  & & \\ 
2d.3 & 245.385288 & 6.122697 &  & & \\ 
2d.4 & 245.382710 & 6.121224 &  & & \\ 
\hline
C.1 &245.3867270 & 6.125210 &\nodata &Bayliss+11 &\\
C.2  &245.3811204 & 6.123205 & & & Predicted location of possible counter image\\
\hline
D &245.377640  &6.124750 &0.5066 & This work, IMACS & Single image\\
\hline
E &245.380753 &6.118075 &1.31  & This work, IMACS & Single image; low confidence redshift\\
\hline
F &245.386307 &6.123700 &0.3386 & This work, IMACS & Single image\\ \hline\hline
\sidehead{SDSS J1632$+$3500} \hline\\[-5pt]
1 & 248.047637 & 35.007370 & 1.2350 &This work, GMOS & \\ 
\hline
2 & 248.038365 & 35.006802 & 2.2650 &This work, GMOS & \\ 
\hline
3.1 & 248.042579 & 35.010737 &\nodata & &Candidate \\ 
3.2 & 248.043008 & 35.011403 &  & & \\ 
\hline
4.1 & 248.050018 & 35.009041 &  &\nodata &Candidate \\ 
4.2 & 248.049811 & 35.007939 &  & & \\ 
\hline
5 & 248.041096 & 35.011300 & 0.4538 &This work, GMOS &Cluster member, star forming \\ 
\hline
6 & 248.043778 & 35.013704 & 0.1271 &This work, GMOS &Foreground \\ 
\hline
7 & 248.048366 & 35.011598 & 1.2318 &This work, GMOS &Background, single image \\ 
 \hline\hline
\sidehead{SDSS J1723$+$3411} \hline\\[-5pt]
1.1 & 260.901550 & 34.198336 & 1.3294 & Kubo+10 & \\ 
1.2 & 260.902050 & 34.199054  &  & & \\ 
1.3 & 260.901040 & 34.201255 &  & & \\ 
1.4 & 260.899850 & 34.199329 &  & & \\ 
1.5 & 260.900572 & 34.199449  &  & & \\ 
1a.1 & 260.901820 & 34.198623 &  & & \\ 
1a.2 & 260.901890 & 34.198710 &  & & \\ 
\hline
2.1 &  260.901400 & 34.197421 & 2.1650 & This work, \hst-GO14230& PI: Rigby\\ 
2.2 & 260.902560 & 34.199755 &  & & \\ 
2.3 &  260.901040 & 34.201732 &  & & \\ 
2.4 & 260.899370 & 34.199521 &  & & \\ 
\hline
3.1 & 260.901180 & 34.197414 &\nodata  & &Assumed to be at same $z$ as source 2 \\ 
3.2 & 260.902500 & 34.199631 &  & & \\ 
3.3 & 260.900860 & 34.201795 &  & & \\ 
3.4 & 260.899230 & 34.199477 &  & & \\ 
 \hline\hline
\sidehead{SDSS J2111$-$0114} \hline\\[-5pt]
1.1 & 317.827850 & -1.241202 & 2.8580  &Bayliss+11  & A in Bayliss+11\\ 
1.2 & 317.828960 & -1.242054 &  & & \\ 
1.3 & 317.834400 & -1.242342 &  & & \\ 
1a.1 & 317.828530 & -1.241737 &  & & \\ 
1a.2 & 317.828260 & -1.241515 &  & & \\ 
1b.1 & 317.828340 & -1.241774 &  & & \\ 
1b.2 & 317.828450 & -1.241857 &  & & \\ 
1c.1 & 317.828910 & -1.242246 &  & & \\ 
1c.2 & 317.827940 & -1.241592 &  & & \\ 
1d.1 & 317.828700 & -1.242008 &  & & \\ 
1d.2 & 317.828120 & -1.241579 &  & & \\ 
\hline
2.1 & 317.833862 & -1.241214 & \nodata & & \\ 
2.2 & 317.832739 & -1.241775 &  & & \\ 
\hline
3.1 & 317.827023 & -1.241099 & \nodata & & \\ 
3.2 & 317.830653 & -1.242959 &  & & \\ 
3.3 & 317.833819 & -1.242636 &  & & \\ 
 \hline\hline
\sidehead{SDSS J2243$-$0935} \hline\\[-5pt]
1.1 & 340.850935 & -9.585705 & 2.09 &A2 in Bayliss+11; Rigby+18 & AGN core\\ 
1.2 & 340.851211 & -9.586760 &  & & \\ 
101.1 & 340.851430 & -9.587189 &  & &clumps in the host galaxy \\ 
101.2 & 340.851337 & -9.586573 &  & & \\ 
101.3 & 340.851173 & -9.586377 &  & & \\ 
102.1 & 340.851279 & -9.587084 &  & & \\ 
102.2 & 340.851068 & -9.586766 &  & & \\ 
102.3 & 340.851025 & -9.586260 &  & & \\ 
\hline
2.1 & 340.858480 & -9.580426 & \nodata & &  $z_{phot}\sim 3.0$ used as prior\\ 
2.2 & 340.858820 & -9.581002 &  & & \\ 
2.3 & 340.859550 & -9.583349 &  & & $z_{phot}\sim 3.0$ \\ 
\hline
3.1 & 340.854230 & -9.580533 & \nodata   & & \\ 
3.2 & 340.854490 & -9.580891 &  & & \\ 
3.3 & 340.856975 & -9.583722 &  & & \\ 
\hline
4.1 & 340.851790 & -9.581950 & \nodata   & & \\ 
4.2 & 340.852140 & -9.583397 &  & & \\ 
4.3 & 340.854020 & -9.587714 &  & & \\ 
\hline
5.1 & 340.846880 & -9.584385 & \nodata   & & \\ 
5.2 & 340.847240 & -9.586046 &  & & \\ 
\hline
6.1 & 340.839270 & -9.583212 & \nodata   & & \\ 
6.2 & 340.840110 & -9.584314 &  & & \\ 
\hline
7.1 & 340.835850 & -9.590120 & \nodata   & & \\ 
7.2 & 340.836190 & -9.589966 &  & & \\ 
7.3 & 340.838472 & -9.590060 &  & & \\ 
\hline
8.1 & 340.834010 & -9.590052 & \nodata & & $z_{phot}\sim 3.0$  used as prior\\ 
8.2 & 340.836370 & -9.588974 &    & & \\ 
8.3 & 340.839336 & -9.589144 &  & & $z_{phot}\sim 3.0$ \\ 
8.4 & 340.835870 & -9.580444 &  & & \\ 
8a.1 & 340.833738 & -9.590145 &   & & \\ 
8a.3 & 340.839226 & -9.589118 &  & & \\ 
8a.4 & 340.835763 & -9.580485 &  & & \\ 
\hline
10.1 & 340.824680 & -9.594526 & \nodata   & & \\ 
10.2 & 340.825210 & -9.596464 &  & & \\ 
\hline
11.1 & 340.833981 & -9.590439 & \nodata   & & \\ 
11.2 & 340.836653 & -9.589322 &  & & \\ 
11.3 & 340.838693 & -9.589496 &  & & \\ 
\hline
A1    & 340.854971 & -9.581328 & 2.092 & Bayliss+11 & \\
B1    & 340.847112 & -9.592340 &1.3202 & Bayliss+11 & Single image\\
C1   & 340.851833 & -9.596300 & 0.7403 & Bayliss+11 & Single image\\
D & 340.852299 &-9.604503912& 0.79&This work, IMACS & Galaxy-Galaxy lensing \\
E & 340.8573348 &-9.57659422& 3.378& This work, IMACS & Galaxy-Galaxy lensing\\ \hline\hline
\enddata 
 \tablecomments{Lensing constraints and other background or foreground
   galaxies with spectroscopic redshifts.
   The IDs of images of lensed
   galaxies are labeled as $AB.X$ where $A$ is a number or a letter indicating the source ID (or system name);  $B$ is a number or a letter
indicating the ID of the emission knot within the system; and $X$ is a number indicating the ID of the lensed image within 
the multiple image family.
References are: \citet{rigby18b,stark13,bayliss12,bayliss10,bayliss11,bayliss14.1050, kubo10,johnson17,ofek08,diehl09}.}
\end{deluxetable*} 

\section{Spectroscopy Catalog}
\startlongtable
\begin{deluxetable*}{rllll} 
 \tablecolumns{5} 
\tablecaption{List of spectroscopic redshifts\label{tab.spec}} 
\tablehead{\colhead{Cluster} &
            \colhead{R.A.}    & 
            \colhead{Decl.}    & 
            \colhead{z}     & 
            \colhead{Ref}       \\
            \colhead{} &
            \colhead{J2000}     & 
            \colhead{J2000}    & 
            \colhead{}       & 
            \colhead{}             }
\startdata 
SDSS~J0928$+$2031\\
& 09:28:06.668 & +20:32:06.46 & 0.1932 & Gemini/GMOS-GN15AQ38\\ 
& 09:28:04.818 & +20:31:46.84 & 0.1872 & Gemini/GMOS-GN15AQ38\\ 
& 09:28:03.098 & +20:31:25.64 & 0.1988 & Gemini/GMOS-GN15AQ38\\ 
& 09:28:04.928 & +20:31:29.24 & 0.1851 & Gemini/GMOS-GN15AQ38\\ 
& 09:28:10.617 & +20:31:52.38 & 0.1893 & Gemini/GMOS-GN15AQ38\\ 
& 09:28:10.325 & +20:31:42.66 & 0.1935 & Gemini/GMOS-GN15AQ38\\ 
& 09:28:07.084 & +20:31:20.02 & 0.1920 & Gemini/GMOS-GN15AQ38\\ 
& 09:28:14.500 & +20:31:44.04 & 0.1885 & Gemini/GMOS-GN15AQ38\\ 
& 09:28:06.301 & +20:30:57.09 & 0.1915 & Gemini/GMOS-GN15AQ38\\ 
& 09:28:02.236 & +20:30:34.59 & 0.1893 & Gemini/GMOS-GN15AQ38\\ 
& 09:28:04.629 & +20:30:44.78 & 0.1906 & Gemini/GMOS-GN15AQ38\\ 
& 09:28:13.988 & +20:31:24.52 & 0.1949 & Gemini/GMOS-GN15AQ38\\ 
& 09:28:08.272 & +20:30:51.62 & 0.1954 & Gemini/GMOS-GN15AQ38\\ 
& 09:28:07.468 & +20:30:22.92 & 0.1904 & Gemini/GMOS-GN15AQ38\\ 
& 09:28:04.962 & +20:30:06.18 & 0.1978 & SDSS DR12\\ 
& 09:28:04.534 & +20:31:45.15 & 0.1920 & SDSS DR12\\ 
& 09:28:13.826 & +20:30:43.65 & 0.1897 & SDSS DR12\\ 
& 09:28:22.205 & +20:29:10.13 & 0.1972 & SDSS DR12\\ 
& 09:27:47.657 & +20:33:34.87 & 0.1935 & SDSS DR12\\ 

SDSS~J0952$+$3434\\
& 09:52:40.222 & +34:34:46.07 & 0.3597 & SDSS DR12\\ 
& 09:52:41.616 & +34:34:14.08 & 0.3570 & SDSS DR12\\ 
& 09:52:41.693 & +34:36:03.10 & 0.3529 & Gemini/GMOS-GN15BQ26\\ 
& 09:52:38.978 & +34:35:54.77 & 0.3594 & Gemini/GMOS-GN15BQ26\\ 
& 09:52:35.313 & +34:35:39.75 & 0.3580 & Gemini/GMOS-GN15BQ26\\ 
& 09:52:37.398 & +34:34:57.64 & 0.3553 & Gemini/GMOS-GN15BQ26\\ 
& 09:52:40.756 & +34:34:14.84 & 0.3616 & Gemini/GMOS-GN15BQ26\\ 
& 09:52:43.880 & +34:34:17.73 & 0.3542 & Gemini/GMOS-GN15BQ26\\ 
& 09:52:40.352 & +34:34:06.46 & 0.3532 & Gemini/GMOS-GN15BQ26\\ 
& 09:52:46.453 & +34:33:52.97 & 0.3594 & Gemini/GMOS-GN15BQ26\\ 

SDSS~J1002$+$ 2031\\
& 10:02:26.832 & +20:31:01.92 & 0.3205 & SDSS-DR12\\ 
& 10:02:14.136 & +20:32:16.44 & 0.3196 & SDSS-DR12\\ 
& 10:02:28.776 & +20:29:05.64 & 0.3168 & SDSS-DR12\\ 
& 10:02:44.880 & +20:29:29.40 & 0.3085 & SDSS-DR12\\ 
& 10:02:45.072 & +20:31:02.28 & 0.3136 & SDSS-DR12\\ 
& 10:02:11.136 & +20:32:15.00 & 0.3187 & SDSS-DR12\\ 
& 10:02:38.496 & +20:28:01.20 & 0.3215 & SDSS-DR12\\ 
& 10:02:44.568 & +20:31:33.24 & 0.3305 & SDSS-DR12\\ 
& 10:02:42.696 & +20:30:36.72 & 0.3194 & SDSS-DR12\\ 
& 10:02:28.392 & +20:31:01.01 & 0.3194 & Gemini/GMOS-GN15BQ26\\ 
& 10:02:28.422 & +20:30:06.54 & 0.3152 & Gemini/GMOS-GN15BQ26\\ 
& 10:02:28.289 & +20:31:07.35 & 0.3245 & Gemini/GMOS-GN15BQ26\\ 
& 10:02:25.264 & +20:30:56.94 & 0.3175 & Gemini/GMOS-GN15BQ26\\ 
& 10:02:28.000 & +20:30:34.04 & 0.3217 & Gemini/GMOS-GN15BQ26\\ 
& 10:02:26.400 & +20:30:30.59 & 0.3191 & Gemini/GMOS-GN15BQ26\\ 
& 10:02:30.695 & +20:30:50.14 & 0.3188 & Gemini/GMOS-GN15BQ26\\ 
& 10:02:28.309 & +20:31:43.74 & 0.3206 & Gemini/GMOS-GN15BQ26\\ 
& 10:02:25.981 & +20:31:48.60 & 0.3230 & Gemini/GMOS-GN15BQ26\\ 
& 10:02:28.419 & +20:31:11.49 & 0.3157 & Gemini/GMOS-GN15BQ26\\ 
& 10:02:26.737 & +20:31:30.17 & 0.3093 & Gemini/GMOS-GN15BQ26\\ 
& 10:02:23.262 & +20:30:39.73 & 0.3235 & Gemini/GMOS-GN11AQ19\\ 
& 10:02:25.964 & +20:30:44.26 & 0.3153 & Gemini/GMOS-GN11AQ19\\ 
& 10:02:16.879 & +20:30:54.20 & 0.3275 & Gemini/GMOS-GN11AQ19\\ 
& 10:02:30.479 & +20:30:16.83 & 0.3126 & Gemini/GMOS-GN11AQ19\\ 
& 10:02:25.463 & +20:30:20.32 & 0.3223 & Gemini/GMOS-GN11AQ19\\ 
& 10:02:24.617 & +20:31:28.95 & 0.3226 & Gemini/GMOS-GN11AQ19\\ 
& 10:02:23.503 & +20:30:34.47 & 0.3238 & Gemini/GMOS-GN11AQ19\\ 
& 10:02:27.478 & +20:30:47.82 & 0.3197 & Gemini/GMOS-GN11AQ19\\ 
& 10:02:26.909 & +20:30:51.14 & 0.3143 & Gemini/GMOS-GN11AQ19\\ 
& 10:02:23.774 & +20:30:55.46 & 0.3195 & Gemini/GMOS-GN11AQ19\\ 
& 10:02:25.971 & +20:31:01.83 & 0.3114 & Gemini/GMOS-GN11AQ19\\ 
& 10:02:28.756 & +20:31:10.58 & 0.3139 & Gemini/GMOS-GN11AQ19\\ 
& 10:02:25.012 & +20:32:09.34 & 0.3157 & Gemini/GMOS-GN11AQ19\\ 
& 10:02:25.394 & +20:31:15.98 & 0.3189 & Gemini/GMOS-GN11AQ19\\ 
& 10:02:30.050 & +20:31:23.90 & 0.3151 & Gemini/GMOS-GN11AQ19\\ 
& 10:02:32.663 & +20:31:26.31 & 0.3228 & Gemini/GMOS-GN11AQ19\\ 
& 10:02:22.687 & +20:32:24.04 & 0.3120 & Gemini/GMOS-GN11AQ19\\ 

SDSS~J1522$+$2535 &&&&\\
& 15:22:53.524 & +25:36:40.27 & 0.6042 & SDSS-DR12\\ 
& 15:22:50.681 & +25:33:34.58 & 0.5917 & SDSS-DR12\\ 
& 15:22:45.102 & +25:36:56.98 & 0.6013 & Gemini/GMOS MOS\\ 
& 15:22:51.614 & +25:35:24.96 & 0.6046 & Gemini/GMOS MOS\\ 
& 15:22:51.900 & +25:35:04.95 & 0.5885 & Gemini/GMOS MOS\\ 
& 15:22:52.528 & +25:34:36.23 & 0.6071 & Gemini/GMOS MOS\\ 
& 15:22:52.686 & +25:34:46.52 & 0.5995 & Gemini/GMOS MOS\\ 
& 15:22:52.802 & +25:36:01.37 & 0.6023 & Gemini/GMOS MOS\\ 
& 15:22:53.304 & +25:35:54.81 & 0.5928 & Gemini/GMOS MOS\\ 
& 15:22:53.536 & +25:36:40.13 & 0.6043 & Gemini/GMOS MOS\\ 
& 15:22:54.162 & +25:34:25.91 & 0.6031 & Gemini/GMOS MOS\\ 
& 15:22:54.285 & +25:34:24.15 & 0.6060 & Gemini/GMOS MOS\\ 
& 15:22:54.313 & +25:35:24.39 & 0.6090 & Gemini/GMOS MOS\\ 
& 15:22:54.880 & +25:35:17.59 & 0.6088 & Gemini/GMOS MOS\\ 

SDSS~J1621$+$0607 &&&&\\
& 16:21:32.746 & +06:07:10.97 & 0.3382 & Bayliss+11\\ 
& 16:21:32.731 & +06:07:13.82 & 0.3390 & Bayliss+11\\ 
& 16:21:28.469 & +06:06:52.97 & 0.3437 & Bayliss+11\\ 
& 16:21:33.158 & +06:07:26.67 & 0.3408 & Bayliss+11\\ 
& 16:21:33.965 & +06:07:23.35 & 0.3420 & Bayliss+11\\ 
& 16:21:31.694 & +06:07:44.14 & 0.3406 & Bayliss+11\\ 
& 16:21:31.966 & +06:07:48.24 & 0.3436 & Bayliss+11\\ 
& 16:21:35.638 & +06:06:28.89 & 0.3367 & Bayliss+11\\ 
& 16:21:33.394 & +06:07:13.93 & 0.3505 & Bayliss+11\\ 
& 16:21:32.830 & +06:07:24.31 & 0.3391 & Bayliss+11\\ 
& 16:21:32.698 & +06:07:28.43 & 0.3522 & Bayliss+11\\ 
& 16:21:16.227 & +06:10:59.42 & 0.3446 & Magallan/IMACS MOS\\ 
& 16:21:16.797 & +06:18:32.11 & 0.3037 & Magallan/IMACS MOS\\ 
& 16:21:18.122 & +06:13:58.32 & 0.3185 & Magallan/IMACS MOS\\ 
& 16:21:19.303 & +06:14:39.07 & 0.3008 & Magallan/IMACS MOS\\ 
& 16:21:22.008 & +06:08:20.09 & 0.3000 & Magallan/IMACS MOS\\ 
& 16:21:22.420 & +06:01:29.46 & 0.3397 & Magallan/IMACS MOS\\ 
& 16:21:25.778 & +06:02:09.85 & 0.3412 & Magallan/IMACS MOS\\ 
& 16:21:26.176 & +06:08:35.22 & 0.3421 & Magallan/IMACS MOS\\ 
& 16:21:36.517 & +06:04:07.66 & 0.3214 & Magallan/IMACS MOS\\ 
& 16:21:36.970 & +06:06:58.79 & 0.3433 & Magallan/IMACS MOS\\ 
& 16:21:39.339 & +06:02:01.51 & 0.3415 & Magallan/IMACS MOS\\ 
& 16:21:39.669 & +05:59:52.43 & 0.3161 & Magallan/IMACS MOS\\ 
& 16:21:41.159 & +06:02:16.98 & 0.3446 & Magallan/IMACS MOS\\ 
& 16:21:43.658 & +06:07:30.44 & 0.3412 & Magallan/IMACS MOS\\ 
& 16:21:43.672 & +06:03:43.78 & 0.3001 & Magallan/IMACS MOS\\ 
& 16:21:43.947 & +06:00:40.68 & 0.3402 & Magallan/IMACS MOS\\ 
& 16:21:46.199 & +06:02:55.47 & 0.3460 & Magallan/IMACS MOS\\ 
& 16:21:46.831 & +06:03:13.22 & 0.3436 & Magallan/IMACS MOS\\ 
& 16:21:49.248 & +06:02:57.76 & 0.3439 & Magallan/IMACS MOS\\ 
& 16:21:53.388 & +06:04:56.83 & 0.3385 & Magallan/IMACS MOS\\ 
& 16:21:55.331 & +06:06:01.89 & 0.3418 & Magallan/IMACS MOS\\ 
& 16:21:56.464 & +06:14:45.48 & 0.3318 & Magallan/IMACS MOS\\ 
& 16:21:56.533 & +06:07:44.87 & 0.3963 & Magallan/IMACS MOS\\ 
& 16:21:59.376 & +06:16:36.90 & 0.3461 & Magallan/IMACS MOS\\ 
& 16:22:02.466 & +05:59:54.13 & 0.3446 & Magallan/IMACS MOS\\ 
& 16:22:11.728 & +05:59:43.15 & 0.3490 & Magallan/IMACS MOS\\ 
& 16:22:19.007 & +06:14:34.83 & 0.3412 & Magallan/IMACS MOS\\ 
& 16:21:32.746 & +06:07:10.97 & 0.3412 & Magallan/IMACS MOS\\ 
& 16:21:32.731 & +06:07:13.82 & 0.3436 & Magallan/IMACS MOS\\ 
& 16:21:28.469 & +06:06:52.97 & 0.3412 & Magallan/IMACS MOS\\ 
& 16:21:33.158 & +06:07:26.67 & 0.3392 & Magallan/IMACS MOS\\ 
& 16:21:33.965 & +06:07:23.35 & 0.3434 & Magallan/IMACS MOS\\ 
& 16:21:31.694 & +06:07:44.14 & 0.3392 & Magallan/IMACS MOS\\ 
& 16:21:31.966 & +06:07:48.24 & 0.3441 & Magallan/IMACS MOS\\ 
& 16:21:35.638 & +06:06:28.89 & 0.3435 & Magallan/IMACS MOS\\ 
& 16:21:33.394 & +06:07:13.93 & 0.3320 & Magallan/IMACS MOS\\ 
& 16:21:32.830 & +06:07:24.31 & 0.3326 & Magallan/IMACS MOS\\ 
& 16:21:32.698 & +06:07:28.43 & 0.3443 & Magallan/IMACS MOS\\ 

SDSS~J1632$+$3500 &&&&\\
& 16:32:07.962 & +35:00:46.950 &  0.467806  & Gemini/GMOS-GN11AQ19 \\
& 16:32:08.188 & +35:00:12.340 &  0.459047  & Gemini/GMOS-GN11AQ19 \\
& 16:32:12.515 & +34:59:54.832 &  0.464104  & Gemini/GMOS-GN11AQ19 \\
& 16:32:07.433 & +35:00:25.044 &  0.471901  & Gemini/GMOS-GN11AQ19 \\
& 16:32:04.769 & +35:01:01.796 &  0.465734  & Gemini/GMOS-GN11AQ19 \\
& 16:32:09.859 & +35:00:40.741 &  0.453848  & Gemini/GMOS-GN11AQ19; star forming \\
& 16:32:14.773 & +35:00:24.579 &  0.471658  & Gemini/GMOS-GN11AQ19 \\
& 16:32:02.585 & +35:01:43.225 &  0.469637  & Gemini/GMOS-GN11AQ19 \\
& 16:32:16.188 & +35:00:41.964 &  0.463060  & Gemini/GMOS-GN11AQ19 \\
& 16:31:55.872 & +35:00:10.084 &  0.471787  & SDSS DR12 \\
& 16:32:08.117 & +35:00:32.674 &  0.470646  & SDSS DR12 \\
\enddata 
 \tablecomments{Spectroscopic redshifts of cluster-member galaxies.}
\end{deluxetable*} 

\end{document}